\documentclass[fleqn,usenatbib]{mnras}

\usepackage[T1]{fontenc}
\usepackage{ae,aecompl}
\usepackage{multirow}

\usepackage{graphicx}	
\usepackage{amsmath}	
\usepackage{amssymb}	

\newcommand{\CII}{[C\,{\sc ii}]}
\newcommand{\hii}{\ion{H}{II}}

\newcommand{\kms}{km~s$^{-1}$} 
 
\newcommand{\trot}     {$T_{\rm rot}$} 
\newcommand{\Nmeth} {$N_{\rm CH_3OH}$}
\newcommand{\Nethy} {$N_{\rm C_2H}$}
\newcommand{\Ncycloprop} {$N_{\rm c-C_3H_2}$} 
\newcommand{\Nform} {$N_{\rm H_2CO}$} 
\newcommand{\Nhcn} {$N_{\rm HCN}$}

\usepackage{newtxtext,newtxmath}

\title[Dark cloud-type chemistry in PDRs]{Dark cloud-type chemistry in PDRs with moderate UV field\thanks{This work is based on observations carried out under the project 027-18 with the IRAM~30~m Telescope. IRAM is supported by INSU/CNRS (France), MPG (Germany), and IGN (Spain).}}

\author[M. S. Kirsanova et al.]{
Maria~S. Kirsanova,$^{1,2}$\thanks{E-mail: kirsanova@inasan.ru}
Anna~F. Punanova,$^{2}$
Dmitry~A. Semenov,$^{3,4}$
Anton~I. Vasyunin$^{2}$
\\
$^{1}$Institute of Astronomy, Russian Academy of Sciences, 119017, 48 Pyatnitskaya Str., Moscow, Russia\\
$^{2}$Institute of Natural Sciences and Mathematics, Ural Federal University,
19 Mira Str., 620002 Yekaterinburg, Russia\\
$^{3}$Max-Planck-Institut f{\"u}r Astronomie, K{\"o}nigstuhl 17, 69117 Heidelberg, Germany\\
$^{4}$Department of Chemistry, Ludwig Maximilian University, Butenandtstra{\ss}e 5--13, 81377 Munich, Germany
}

\date{Accepted 2021 August 6. Received 2021 July 22; in original form 2020 Deecmber 8}

\pubyear{2021}

\begin{document}
\label{firstpage}
\pagerange{\pageref{firstpage}--\pageref{lastpage}}
\maketitle

\begin{abstract}
We present a study of emission lines of small hydrocarbons C$_2$H and $c$-C$_3$H$_2$, and COMs precursors H$_2$CO and CH$_3$OH in order to better understand the possible chemical link between the molecular abundances and UV radiation field in photodissociation regions (PDRs). We study two PDRs around extended and compact \hii{} regions with $G \leq 50$~Habings in the S235 star-forming complex. We find the highest abundances of both hydrocarbons on the edges of molecular clumps, while $c$-C$_3$H$_2$ is also abundant in the low-density expanding PDR around compact \hii{} region S235\,A. We see the highest methanol column density towards the positions with the UV~field $G\approx 20-30$~Habings and explain them by reactive desorption from the dust grains. The $N_{\rm C_2H}/N_{\rm CH_3OH}$ ratio is lower by a factor of few or the order of magnitude in comparison with the Horsehead and Orion Bar PDRs. The ratio is similar to the value observed in hot corinos in the Perseus cloud. We conclude that ion-molecular and grain surface chemical routes rule the molecular abundances in the PDRs, and the PDRs inherit molecular abundances from the previous dark stage of molecular cloud evolution in spite of massive stars already emitting in optics.
\end{abstract}




\section{Introduction}\label{sec:intro}

Irradiated edges of molecular clouds reveal rich chemical compositions ranging from diatomic species H$_2$ \citep[e. g.][]{Marconi1998, 2000A&A...364..301W, 2016ApJ...819..136A} and CO~\citep[e. g. recent works by][]{Goicoechea_2016, Joblin_2018} to polyatomic carbon chains \citep[e. g.][]{1985ApJ...298L..61M, 1988A&A...206..108C, 2008A&A...478L..19A} and complex organic molecules \citep[COMs, see e. g.][]{2003A&A...412..157J,Leurini_2010,2014FaDi..168..103G,Cuadrado_2017}. The very detection of COMs in highly-irradiated gas is surprising given the current understanding of their formation. The accepted mechanisms of COMs formation include either their synthesis on water ice on relatively cold grains ($T_{\rm dust} = 15-25$~K) followed by desorption \citep[see, e. g.][]{Garrod_2006}, or gas-phase synthesis from precursors formed on cold grains with $T_{\rm dust} =10$~K~\citep[see, e. g.][]{Vasyunin_2013, Vasyunin_2017}. None of those two mechanisms shall be efficient under the conditions of highly-irradiated medium of PDRs such as the Orion Bar PDR with $T_{\rm gas} > 100$~K \citep{Joblin_2018} and $T_{\rm dust} =50-70$~K \citep{Arab_2012, Salgado_2016}. Furthermore, it is difficult to reconcile with an idea that these COMs have been formed during the earlier, cold and ``UV dark'' protostellar stage because COMs and their precursors are fragile to the UV radiation \citep[][]{Oberg_2016}.

\citet{Cuadrado_2017} have proposed that photodesorption of ices is responsible for appearance of COMs in the gas phase of the Orion Bar PDR irradiated by high UV~field with $G=10^4-10^5$ (in the units of the Habing UV radiation field). \citet{Guzman_2013} came to the same conclusion about the role of the photodesorption of ices by studying abundances of molecular precursors for COMs such as H$_2$CO and CH$_3$OH in the Horsehead PDR, where the UV~field is moderate with $G=100$~Habings. In contrast, \citet{Esplugues_2016} concluded that the non-thermal reactive desorption upon surface recombination reactions is responsible for the presence of the simple COMs in the PDRs with moderate UV~fields.

On the other hand, the Horsehead PDR has surprised observers by showing high abundances of the small hydrocarbons -- precursors for long hydrocarbon chains. For example, \citet{Pety_2005} and \citet{LeGal_17} have found that abundances of small hydrocarbons, such as C$_2$H, c-C$_3$H$_2$, C$_4$H cannot be explained with the chemical models including gas-phase and dust surface chemistry. They proposed that UV~photons are able to destroy polycyclic aromatic hydrocarbons (PAHs) in the transition zone between \hii{} region and PDR, driving the observed high abundances of the small hydrocarbons. However, \citet{Murga_2020} explored photodissociation of PAHs in the Orion Bar and Horsehead PDRs and haven't found significant contribution of that chemical channel in comparison with low- and high-temperature gas-phase chemical reactions.

While the PDRs in the Orion region are among the most studied objects of that type, their main parameters (UV~field $G$ and gas number density $n$) cover only a limited parameter space. Thus we need to increase the number of PDRs with known chemical composition covering a broader range of $G$ and $n$ in order to understand the formation and survival of molecules in the irradiated gas. The first aim of this study is to specify which kind of the desorption processes prevails in the PDRs with moderate UV~field. The second aim of our study is to explore whether the enhanced abundances of small hydrocarbons are common for the  PDRs with moderate UV~field. To achieve these science goals, we explore molecular abundances in the three PDRs with moderate UV~fields against those found in Horsehead, focusing on precursors of COMs (H$_2$CO and CH$_3$OH) and long carbon chains (C$_2$H and C$_3$H$_2$). We obtain spatial distributions of their abundances, covering a wide area around the sources of the UV~emission and away from them, providing a benchmark for a detailed comparison with astrochemical models in the future.

\section{Target regions}

\begin{figure*}
    \centering
    \includegraphics[height=11.2cm]{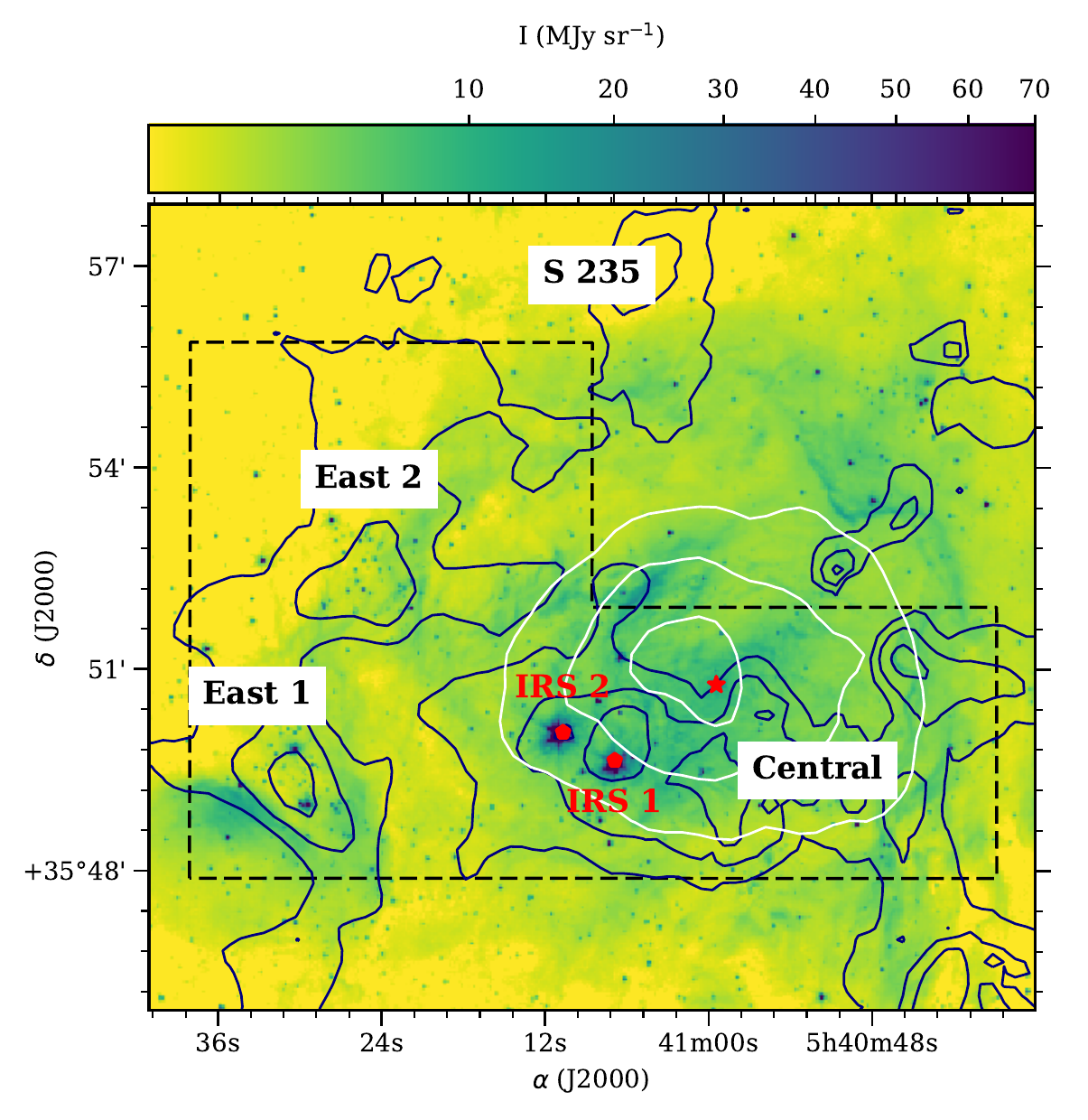}
    \includegraphics[height=11.2cm]{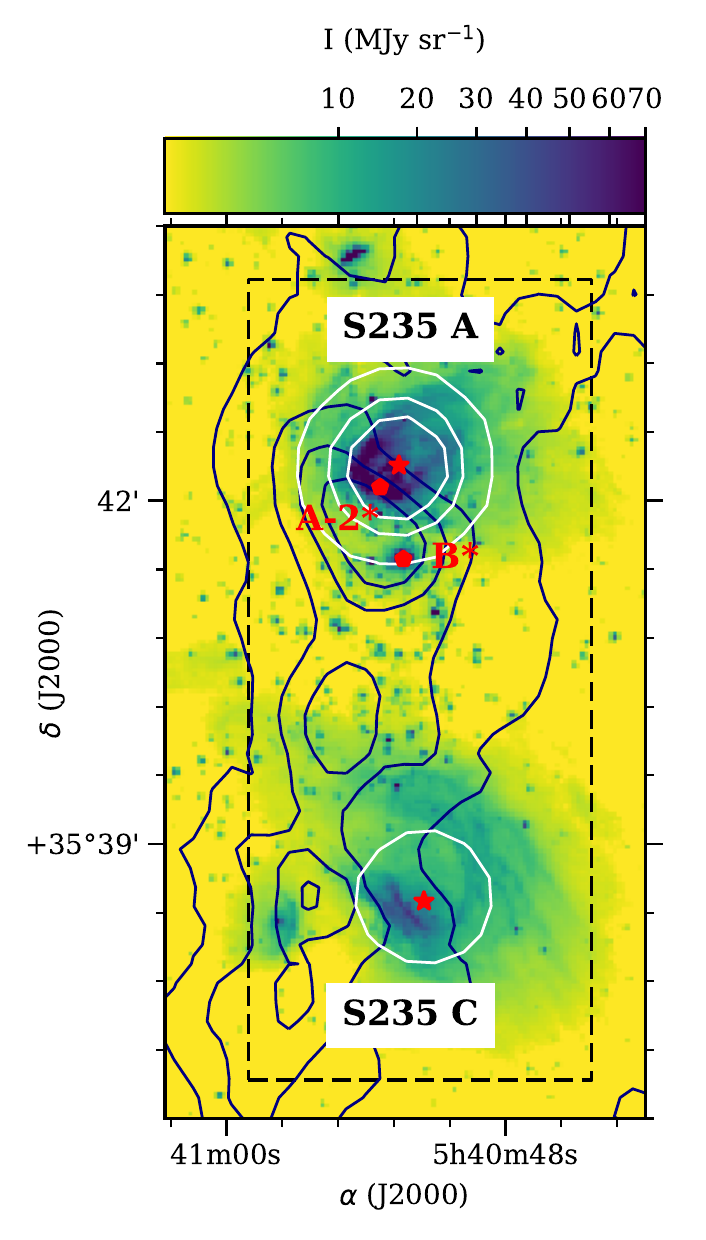}
    \caption{The 3.6~\micron{} {\it Spitzer} image of the S235 (left) and S235\,ABC (right) regions \citep[][]{2004sptz.prop..201F}. White contours show 1.4~GHz continuum emission from NVSS~survey~\citep[][]{Condon_1998} at 0.01, 0.5 and 0.1~Jy/beam. Names of the \hii{} regions as well as dense molecular clumps with the embedded young stellar clusters are given in bold face in white boxes. Black contours show levels of hydrogen column density: 1, 2, 3 and $4\cdot 10^{22}$~cm$^{-2}$ based on CO observations by \citet{Bieging_2016}. The ionizing sources of the \hii{} regions are shown by the red stars. Red diamonds show other bright infrared sources: IRS~1, IRS~2 \citep{Evans_1981}, S235~B$^{*}$~\citep{Boley_2009} and S235\,A-2$^{*}$ \citep{Kirsanova_2020_PDR}. Black dashed line show the area mapped by IRAM~30-m telescope.
    }
    \label{fig:general_view_spitzer}
\end{figure*}

We selected irradiated molecular clouds bordering the extended \hii{} region Sh2-235 \citep[S235 hereafter,][]{Sharpless_1959} and the two compact \hii{} regions S235\,A and S235\,C \citep{Israel_1978}. The regions belong to the same giant molecular cloud G174+2.5, studied in CO emission lines by \citet{Evans_1981, Heyer_1996, Kirsanova_2008, Bieging_2016, Ladeyschikov_2016}. The distance to S235\,A is 1.56~kpc according to the maser parallax measurements by \citet{Burns_2015}. S235 is ionized by an O9.5~V star BD+35~1201~\citep[see ][]{Georgelin_1973, Cruz-Gonzales_1974, Avedisova_1984}, while S235\,A and S235\,C are ionized by B0.5~\citep{Evans_1981} and B0.3-B3~\citep{Thompson_1983} stars, respectively. S235 and S235\,ABC are regions of active star formation \citep[][]{Allen_2005, Klein_2005, Camargo_2011, Dewangan_2011}; rich young stellar clusters surround S235 and also S235\,A~\citep{Kirsanova_2008}. There are two bright infrared sources IRS\,1 and IRS~2 in the Central clump around S235, they are young stellar objects with optical emission lines showing an outflow activity~\citep{Alvarez_2004}. The area containing the compact \hii{} regions together with the  Herbig~Be-type star S235~B$^*$~\citep{Boley_2009} will be called S235\,ABC hereafter, while the northern part of the area will be called S235\,AB. Fig.~\ref{fig:general_view_spitzer} shows the 3.6~\micron{} {\it Spitzer} image of the target PDRs.

\citet{Anderson_2019} studied the \CII{} at 158~\micron{} line emission in S235 and found that the \hii{} region expands to the observer: the front neutral wall of the \hii{} region is moving relative to the rear and side walls. \citet{Kirsanova_2020} combined optical and infrared data and showed that the material in the rear wall has a higher column density than in the front wall. The Central clump, situated behind the \hii{} region \citep{Camargo_2011, Anderson_2019} in the rear neutral wall, is irradiated by higher UV-flux than the East~1 clump in the side neutral wall. While S235\,A is deeply embedded into neutral gas of the molecular cloud ($A_{\rm V} = 4$ and $6-12$~mag towards S235 and S235\,A, respectively), it is also expanding to the observer as \citep{Kirsanova_2020_PDR}. 

\section{Observations}

We mapped irradiated molecular clouds near the \hii{} regions with the IRAM~30~m antenna on August 19--21 2018. The observed areas are shown in Fig.~\ref{fig:general_view_spitzer}. The on-the-fly maps were obtained with the EMIR~090 (3~mm band) and EMIR~150 (2~mm band) heterodyne receivers\footnote{\url{http://www.iram.es/IRAMES/mainWiki/EmirforAstronomers}} simultaneously in position switching mode. We tested five reference positions via frequency switching observations. The chosen reference position (J2000 $\alpha$=05$\rm ^h$41$\rm ^m$18.663$\rm ^s$, $\delta$=+36$^\circ$01\arcmin20.93\arcsec) was proved to have no contamination of any of the observed lines down to the rms of 0.05-0.1~K in the units of the antenna temperature $T_{\rm A}$, which is sufficient for the sensitivity of our data. We used the FTS~50 backend with the spectral resolution of 50~kHz, the corresponding velocity resolutions were $\simeq$0.17 for the 3~mm and $\simeq$0.10~km~s$^{-1}$ for the 2~mm bands. The beam sizes were $\simeq$28\arcsec{} for the 3~mm and $\simeq$17\arcsec{} for the 2~mm bands. The system temperatures were 96-105~K at 3~mm and 156-164~K at 2~mm. Typical weather conditions corresponded to pwv=8--15~mm. The beam sizes, spectral resolutions, and sensitivities, are given in Table~\ref{tab:observations}. Forward efficiencies were 0.95 and 0.93 and beam efficiencies were 0.81 and 0.73 for the observations at 3 and 2~mm, respectively. Sky calibrations were obtained every 10~min. Pointing and focus were checked by observing QSO~B0316+413, QSO~B0430+052, and QSO~B0439+360 every two (pointing) and six (focus) hours. The observed transitions and frequencies are given in Table~\ref{tab:detectedlines}. 

\begin{table}
\begin{tabular}{cccccc}
\hline
\hline
Molecule     & Frequency & $HPBW$    & $\Delta \varv_{\rm res}$ & rms  & $T_{\rm sys}$ \\
             & (GHz)     & (\arcsec) & (km~s$^{-1}$)            & (K)  & (K) \\
\hline
CH$_3$OH     & 84.5  & 29.3  & 0.18  & 0.04 & 105\\
c-C$_3$H$_2$ & 85.3  & 29.0  & 0.18  & 0.04 & 105\\
      C$_2$H & 87.3  & 28.3  & 0.17  & 0.05 & 96\\
         HCN & 88.6  & 27.9  & 0.17  & 0.04 & 96\\
     H$_2$CO & 140.8 & 17.6  & 0.11  & 0.13 & 156\\
    CH$_3$OH & 143.9 & 17.2  & 0.10  & 0.12 & 164\\
	CH$_3$OH & 145.1 & 17.0  & 0.10  & 0.12 & 164\\
c-C$_3$H$_2$ & 145.1 & 17.0  & 0.10  & 0.12 & 164\\
\hline
\end{tabular}
\caption{Observation parameters.}
\label{tab:observations}
\end{table}

\section{Methods}

\subsection{Production of the spectral data cubes}

The spectral data cubes were produced with the \texttt{CLASS} package\footnote{Continuum and Line Analysis Single-Dish Software \url{http://www.iram.fr/IRAMFR/GILDAS}}. All spectra were smoothed by a factor of two using Hanning window, to the 100~kHz resolution (0.35 and 0.20~km~s$^{-1}$ at 3 and 2~mm), to improve the sensitivity. With the typical line widths of $\sim$2~km~s$^{-1}$, the lines remain well resolved. We produced two sets of the data cubes: (1) data convolved to the native resolution with the pixel sizes of 1/3 of the beam sizes, for the integrated intensity maps (see Fig.~\ref{fig:mom0}), and (2) data convolved to the same largest beam of 29.3$^{\arcsec}$ with the pixel size of 9\arcsec{} to calculate the column densities and to allow their pixel-by-pixel comparison. The convolution and regridding were done with the \texttt{CLASS} methods \texttt{table} and \texttt{xy\_map}. Since the combined map was observed as three separate smaller maps, the sensitivity across the combined map is not uniform. Fig.~\ref{fig:noisemaps} in Appendix~\ref{app:noise} shows the sensitivity maps at 2 and 3~mm. The intensity scale was converted to the main-beam temperature scale according to the beam efficiency values\footnote{Kramer, C., Penalver, J., \& Greve, A. 2013: \url{http://www.iram.es/IRAMES/mainWiki/CalibrationPapers?action=AttachFile\&do=view\&target=eb2013-v8.2.pdf}} (given above). 
\subsection{Archival data}

We use maps of CO and $^{13}$CO emission from \citet{Bieging_2016} and pre-calculated gas column density on the basis of these data from~\citet{Kirsanova_2020_PDR}. The spatial resolution of that data is 38\arcsec{}. \citet{Kirsanova_2020} and \citet{Kirsanova_2020_PDR} produced maps of dust temperature in the S235 and S235\,ABC observed area, respectively. The S235 maps were calculated using {\it AKARI} data at 65, 90,  140 and 160~\micron, obtained with a spatial resolution of 61\arcsec. The S235\,ABC maps are based on Bolocam galactic plane survey at 350~\micron{} and 1.1~mm as well as SCUBA-2 data at 450 and 850~\micron{} and have a spatial resolution of 33\arcsec.

\subsection{Column density and abundance calculations}\label{sec:columndensity}

Calculations of molecular column densities were done in approximation of the local thermodynamic equilibrium (LTE) as we observed/detected only one transition of each molecule in many of the directions in our maps (except methanol, see Sec.~\ref{sec:calcmeth}). Nevertheless, we believe that the LTE approximation is reliable for the observed area due to two considerations. First, we observed molecular clouds, which are dense enough to excite e.g. the CS(2--1) and line emission observed by \citet{Kirsanova_2008} and avoided the area with the rarefied gas to the north of BD+35~1201. The detection of this line means that the gas number density is at least $\approx 10^4$~cm$^{-3}$ or higher \citep[see discussion of the effective excitation density by][]{2015PASP..127..299S}, and this line serves as an approximate density indicator. The effective excitation density is consistent with gas number densities determined by \citet{Kirsanova_2014} using ammonia inverse transitions. Bearing in mind moderate optical depth of the ammonia lines, we expect that actual gas number density in the studied regions can be higher. We admit that the gas might be close to the verge of LTE and non-LTE state, however, the LTE analysis is the best we can do with only one or two transitions of each molecule. Second, \citet{Bieging_2016} analysed the CO and $^{13}$CO line emission in S235 and found that the molecular gas is close to the LTE conditions. 

Using the emission measure of S235 $EM\sim 10^5$~pc~cm$^{-6}$, $T_{\rm e} =7280$~K \citep{Kirsanova_2020} and the Eq.~4.60 from \citet{2016era..book.....C}, we find the background brightness temperature of S235, S235\,A and C much smaller ($\ll 1$~K) than the spectral lines brightness temperature at 2 and 3~mm. By neglecting the background temperature and using Rayleigh-Jeans approximation, we calculate molecular column densities using the following Eq.~\citep[see e.~g.][]{Goldsmith_1999, Mangum_2015,Kalenskii_2016}:

\begin{equation}
{\rm ln} N = {\rm ln} \frac{N_{\rm u}}{g_{\rm u}} + Q(T) + \frac{E_{\rm u}}{{\rm k} T}.
\label{eq:Ntot}
\end{equation}

Here, the $N_{\rm u}/g_{\rm u}$ ratio is the number density of the upper level of the transition divided by the upper level degeneracy:

\begin{equation}
\frac{N_{\rm u}}{g_{\rm u}} = \frac{8 \pi {\rm k} \nu_{\rm ul}^2}{{\rm hc^3} A_{\rm ul} g_{\rm u}} W {\rm (cm^{-2})}.
    \label{eq:Nup}
\end{equation}

The $Q(T)$ value is the rotation-spin partition function at the excitation temperature $T$, $E_{\rm u}/k$ is the energy of the upper level of the transition (K), $A_{\rm ul}$ is the Einstein coefficient (s$^{-1}$), $\nu_{\rm ul}$ is the frequency of the transition from the upper (u) to low (l) level (Hz), and $W$ is the velocity-integrated line intensity (K~cm~s$^{-1}$). All the spectroscopic parameters are given in Table~\ref{tab:detectedlines}.

We applied pre-calculated values of $Q(T)$ from CDMS catalogue for particular excitation temperature. The upper level degeneracy, the ratio of $Q(T)/g_{\rm u}$ is the same in JPL and CDMS, while specific values of $Q(T)$ and $g_{\rm u}$ could differ by a common factor. The pre-calculated $Q(T)$ values are given in Appendix~\ref{app:Qsprot}.

\begin{table*}
	\centering
	\caption{Detected lines and their spectroscopic parameters. The line strength number $S_{\rm ul}$ is applied only for the transitions with hyperfine splitting. All spectroscopic constants were taken from CDMS~\citep{Muller_2001}. $^{*}$ Energy of the $ortho$ c-C$_3$H$_2$ ground state level $1_{0,1}$ is 2.3~K. $^{**}$ Energy of the $ortho$ H$_2$CO ground state level $1_{1,1}$ is 15.2~K.}
	\label{tab:detectedlines}
	\begin{tabular}{ccccccc}
		\hline
		Molecule & Transition & Frequency & $A_{\rm ul}$  & $E_{\rm u}/{\rm k}$ & $g_{\rm u}$ & $S_{\rm ul}$\\
		         &  & (MHz)     & (s$^{-1}$)    & (K)                 & --          & --    \\
		\hline
		\multirow{6}{*}{C$_2$H} & 1$_{3/2, 1}$ -- 0$_{1/2, 1}$ & 87284.156 & $6.60\cdot10^{-6}$ & 4.2 & 3 & 0.17 \\
		                        & 1$_{3/2, 2}$ -- 0$_{1/2, 1}$ & 87316.925 & $6.48\cdot10^{-5}$ & 4.2 & 5 & 1.66\\
		                        & 1$_{3/2, 1}$ -- 0$_{1/2, 0}$ & 87328.624 & $3.22\cdot10^{-5}$ & 4.2 & 3 & 0.83\\
		                        & 1$_{1/2, 1}$ -- 0$_{1/2, 0}$ & 87402.004 & $3.23\cdot10^{-5}$ & 4.2 & 3 & 0.83\\
		                        & 1$_{1/2, 0}$ -- 0$_{1/2, 1}$ & 87407.165 & $1.30\cdot10^{-5}$ & 4.2 & 1 & 0.33\\
		                        & 1$_{1/2, 1}$ -- 0$_{1/2, 0}$ & 87446.512 & $6.63\cdot10^{-6}$ & 4.2 & 3 & 0.17\\
		\hline
		\multirow{2}{*}{$ortho$ $c$-C$_3$H$_2$} & 2$_{1, 2}$ -- 1$_{0, 1}$ & 85338.896 & $2.61\cdot10^{-5}$ & 6.4$^{*}$ & 5 & --\\
		                             & 3$_{1, 2}$ -- 2$_{2, 1}$ & 145089.606 & $6.77\cdot10^{-5}$ & 16.0$^{*}$ & 7 & --\\
		\hline
        $ortho$ H$_2$CO                    & 2$_{1,2}$  -- 1$_{1,1}$  & 140839.517& $5.30\cdot10^{-5}$ & 21.9$^{**}$ & 15 & --\\
        \hline
		\multirow{10}{*}{CH$_3$OH}  & 5$_{-1}$  -- 4$_{0}$ E   & 84521.172  & $1.97\cdot10^{-6}$ & 40.4 & 44 & --\\
		                            & 3$_{1}$  -- 2$_{1}$ A$^+$   & 143865.790 & $1.08\cdot10^{-5}$ & 28.3 & 28 & --\\
		                            & 3$_{0}$  -- 2$_{0}$ E    & 145093.750 & $1.23\cdot10^{-5}$ & 27.1 & 28 & --\\
		                            & 3$_{-1}$  -- 2$_{-1}$ E  & 145097.435 & $1.10\cdot10^{-5}$ & 19.5 & 28 & --\\
		                            & 3$_{0}$  -- 2$_{0}$ A$^+$   & 145103.185 & $1.23\cdot10^{-5} $& 13.9 & 28 & --\\
		                            & 3$_{2}$  -- 2$_{2}$ E    & 145126.191 & $1.81\cdot10^{-5} $& 36.2 & 28 & --\\
		                            & 3$_{-2}$  -- 2$_{-2}$ E  & 145126.386 & $1.81\cdot10^{-5} $& 39.8 & 28 & --\\
                                    & 3$_{1}$  -- 2$_{1}$ E    & 145131.864 & $3.02\cdot10^{-5} $& 34.9 & 28 & --\\
                                    	\hline
       \multirow{3}{*}{HCN}        & 1$_{1}$ -- 0$_{1}$ & 88630.416 & $1.02\cdot10^{-3}$ & 4.25 & 3 & 1.00 \\
                                   & 1$_{2}$ -- 0$_{1}$ & 88631.848 & $1.70\cdot10^{-3}$ & 4.25 & 5 & 1.66 \\
                                   & 1$_{0}$ -- 0$_{1}$ & 88633.936 & $3.40\cdot10^{-4}$ & 4.25 & 1 & 0.33 \\
		\hline
    \end{tabular}
\end{table*}

\subsubsection{CH$_3$OH}\label{sec:calcmeth}

We used five lines from the CH$_3$OH~$J=3_K-2_K$ series at 143-145~GHz and one line of the $J=5_K-4_K$ series at 84~GHz to calculate the excitation temperature (henceforth \trot) and column density \Nmeth{} from Eqs.~\ref{eq:Ntot}~and~\ref{eq:Nup}. The blended lines of E-methanol $3_2-2_2$ and $3_{-2}-2_{-2}$ were not included in the analysis. By combining the $N_{\rm u}/g_{\rm u}$ values, we produced the rotation diagram in each pixel of the map and estimated \trot{} with the generalized least squares method. Examples of the rotational diagrams and spatial distribution of the \trot{} values are shown in Appendix.~\ref{app:rotdiag}. For lines which were not detected in a particular pixel (i. e. $S/N$ ratio for the integrated intensity $W$ is less than 3), we set $N_{\rm u} =0$ with the corresponding uncertainty given by the noise level of the spectrum and number of the spectral channels occupied by the line. The integrated intensities of the $J=3_K-2_K$ and $J=5_{-1}-4_0$ lines ($W$) were calculated by \texttt{gauss} fit method in \texttt{CLASS} together with their uncertainties ($\delta W$) simultaneously. This is useful since the uncertainty of the ${\rm ln} (N_{\rm u} / g_{\rm u})$ value in Eq.~\ref{eq:Nup}, given by the error propagation method, corresponds to $\delta W/W$. The error propagation was also used to calculate the uncertainty of the total column density $N$ according to Eq.~\ref{eq:Ntot}. Since we used the mixture of the A and E-methanol lines in the rotation diagrams, the $Q(T)$ value, given in Table~\ref{tab:partition_functions}, was also selected for the mixture.

The CH$_3$OH(5$_{-1}$-4$_0$)~E line can be excited in the maser regime~\citep[see e.g. ][]{Cragg_2005, Kalenskii_2016}. We compared the line widths and systematic velocities of this line to those of the (3$_K$-2$_K$) line and found them similar in the majority of the observed areas. Therefore we did not expect the maser amplification of the CH$_3$OH(5$_{-1}$-4$_0$)~E line and used this line together with the (3$_K$-2$_K$) lines to calculate \Nmeth{}. In those pixels where the (5$_{-1}$-4$_0$)~E line apparently had the maser amplification, we obtained negative $T_{\rm rot}$ from the rotational diagrams. One example of such empty pixel between S235\,A$^*$ and S235~B$^*$ can be found below in Fig.~\ref{fig:abundratio_S235A}.

\subsubsection{H$_2$CO and c-C$_3$H$_2$}

Only one line of the $ortho$-formaldehyde at 140~GHz and two lines of $ortho$ c-C$_3$H$_2$ at 85 and 145~GHz were observed in this study. To calculate \Nform{} and \Ncycloprop{}, we adapted $T$=10~K as the excitation temperature. With single H$_2$CO line, excitation temperature can not be measured, and with two c-C$_3$H$_2$ lines can not be estimated reliably since rotation diagrams give reliable results when they include many lines \citep{Goldsmith_1999}. Also, the number of pixels with the detected 145~GHz line was much less than the number of pixels with the detected 85~GHz line. We chose not to use the methanol \trot{} values, since the excitation temperatures of these molecules are different in PDRs \citep[see, e. g. analysis by][where they found the values of about 12--13~K for H$_2$CO in the Orion Bar PDR]{Cuadrado_2017}. The corresponding $ortho$ partition functions are given in Table~\ref{tab:partition_functions}. Table~\ref{tab:detectedlines} shows absolute values of $E_{\rm u}$, therefore we subtracted the energy of the ground level of $15.2$~K (for H$_2$CO) and $2.3$~K (for c-C$_3$H$_2$) from the corresponding $E_{\rm u}$ value. We assume that the ortho-to-para ratio for both molecules is defined by the statistical weights 3:1 while it can be as low as 1.5:1 depending on physical conditions in molecular clouds \citep[see e. g.][]{1984A&A...137..211K,1993ApJS...89..123M}.

\subsubsection{C$_2$H and HCN}\label{sec:c2h_hcn}

To fit the hyperfine structures of HCN and C$_2$H, we used the \texttt{hfs} method of \texttt{CLASS}. The method computes line profiles with the assumptions of Gaussian velocity distributions and equal excitation temperatures for all hyperfine components. It varies four parameters (integrated intensity $T_{\rm ant}\times\tau$, central velocity of the main hyperfine component $V_{\rm LSR}$, line width $\Delta\varv$, and optical depth $\tau$) and finds the best fit. We started with this method but the {\it hfs} method could not fit the spectra of the lines in the majority of pixels due to asymmetric shape of the lines in many pixels, where one or several velocity components present \citep[see, e. g.][]{2013MNRAS.436.3186P}. However, the ratio of the ethynyl hyperfine components at 87328.624 to 87316.925~MHz agrees with the theoretical value of 0.5 in the majority of pixels in the map, therefore, we considered the lines as optically thin. The emission of ethynyl in other PDRs is also optically thin \citep[e.~g.][]{Cuadrado_2015,Buslaeva_2021}. Since optical depth of HCN can be significant \citep[e.~g.][]{2011A&A...534A..77P, 2017MNRAS.466...49J}, our approach allows obtaining only lower limit of the \Nhcn. To be conservative, we used the same value of the excitation temperature $T=10$~K to calculate column densities of ethynyl \Nethy{} and hydrogen cyanide \Nhcn.

The values of \Nethy{} and \Nhcn{} were calculated for the integrated line intensities over the whole multiplet, the $Q$ for these molecules is approximated by the formula for linear molecules:
\begin{equation}
    Q(T)=\frac{{\rm k}T}{{\rm h}B_0} + \frac{1}{3},
\end{equation}
where $B_0~=~43.674$~GHz and $44.315$~GHz are the rotational constants for C$_2$H and HCN, respectively.

\subsection{Molecular abundances}\label{sec:abund}

We smoothed the column density maps of the studied molecules to the resolution of 38\arcsec, limited by the CO data described above. Following \citet{Bieging_2016}, we assumed the ${\rm CO}/{\rm H_2}=10^{-4}$ and ${\rm CO}/{\rm ^{13}CO}=80$ and also corrected the column density values by a factor of 1.4 for helium. To obtain the abundances relative to the total number of hydrogen nuclei $x$, we divided the column densities for each pixel by an additional factor of 2: $x = N/N_{\rm H_{2}}/2$.

\subsection{UV~field}\label{sec:marion}

We used an approach and parameters described in~\citet{Kirsanova_2020} to calculate the intensity of the UV~flux in units of Habing field from the maps of dust temperature. The approach is based on geometrical considerations introduced by \citet{tielensbook}, Eq.~5.44, and also takes dust extinction into account. In order to check results of this approach, we calculated the UV~field intensity for S235 with MARION model~\citet{Kirsanova_2009} for the East~1 dense clump, which is located in the side wall of S235, using spectrum for O~9.5-type star and parameters of the ionized gas obtained by \citet{Kirsanova_2020}. 
The results of these two approaches agrees within factor of 2 or 3 in various parts of the clump. Therefore, we use the approach for all the observed regions.

\section{Results}\label{sec:results}

\begin{figure}
    \centering
    \includegraphics[width=\columnwidth]{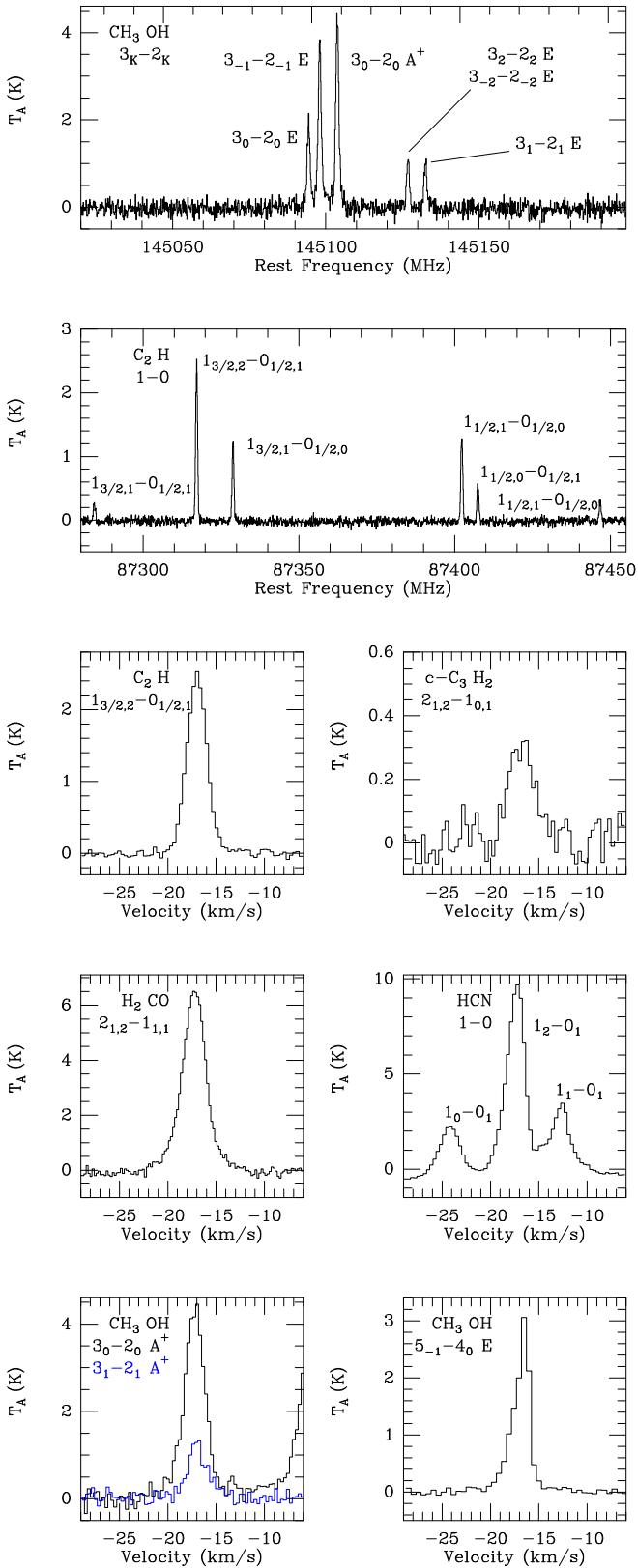}
    \caption{The spectra of the observed lines at the peak of methanol $3_K-2_K$ emission in S235\,ABC area. Two top panels show the methanol $3_K-2_K$ series at 145~GHz and all six hyperfine components of the C$_2$H(1--0) line. All other panels show particular lines with the same spectral scale.}
    \label{fig:spectra}
\end{figure}

We summarize the detected molecular emission lines in Table~\ref{tab:detectedlines}. Examples of the observed spectra at the peak of the methanol $3_K-2_K$ emission in S235\,ABC are shown in Fig.~\ref{fig:spectra}. The methanol spectra have single-component shape with blue wing at the peak. The shape of the $5_{-1}-4_0$~E methanol line is clearly deviates from the Gauss here: the red part of the line has a narrow shape typical for the maser excitation, but the blue part has a shape close to Gaussian. The central $1_2-0_1$ component of the HCN multiplet has similar skewed profile. The brightness ratio of the HCN hyperfine components is not 1:5:3, as expected in theory, but 1.0:3.8:1.3; the $1_1-0_1$ component is diminished. The ratio of the $1_2-0_1$ to the $1_0-0_1$ component corresponds to the optical depth of 0.7, but we obtained negative value for the depth using the $1_0-0_1$ and $1_0-0_1$ components. The ratio of the C$_2$H(1--0) hyperfine components in the same direction is in agreement with the theoretical values from Table~\ref{tab:detectedlines} within 10\%, therefore they are optically thin. The lines of H$_2$CO, C$_2$H and c-C$_3$H$_2$ molecules have Gaussian shape.

\subsection{Molecular emission maps}

\subsubsection{C$_2$H and $c$-C$_3$H$_2$}

Maps of the integrated emission in C$_2$H(1--0) and $c$-C$_3$H$_2$(2$_{1,2}$--1$_{0,1}$) transitions are shown in Fig.~\ref{fig:mom0}. The emission of the hydrocarbons can be clearly seen in S235 and S235\,ABC, the emission occupies the entire observed area. The line emission follows the distribution of molecular column density, calculated from the CO lines. The C$_2$H and $c$-C$_3$H$_2$ molecules have different dipole moments ($D=0.77$ and 3.43~Debyes, respectively) and they are expected to emit brightly from regions with different gas number density. However, the peaks of the $c$-C$_3$H$_2$(2$_{1,2}$--1$_{0,1}$) line emission are observed almost in the same directions where the C$_2$H(1--0) peaks exist. The spatial distribution of $c$-C$_3$H$_2$ emission differs from the emission produced by other molecules with high dipole moment such as HCN and H$_2$CO. The difference between high-D c-C$_3$H$_2$ vs HCN and H$_2$CO is likely due to different excitation parameters and abundance distributions. The brightest C$_2$H(1--0) lines have integrated intensities of $W=2.5$~K~\kms{}, they are observed in the East~1 clump in S235 along the chain of Class~0/I young stellar objects (YSO) \citep[][magenta crosses in Fig.~\ref{fig:mom0}]{Dewangan_2011}. Another two bright spots in the C$_2$H(1--0) map of S235 are observed towards infrared point sources IRS~1 and IRS~2 in the Central clump. In the S235\,ABC map, the C$_2$H(1--0) emission peak has an integrated intensity of $W=5$~K~\kms{} and we observe it in the area occupied by the Class~0/I~YSOs between the bright IR~sources S235\,A-2$^*$ and S235~B$^*$. It is also projected onto the S235\,AB-MIR source~\citep[not shown here, see][]{2006A&A...453..911F} and the peak of the SCUBA-850~\micron{} emission \citep[bottom of Fig.~\ref{fig:mom0}, see also][]{Klein_2005}. The emission of the hydrocarbons is widespread in the entire mapped area in S235 but highly concentrated in the dense molecular gas around S235\,A. We assign this behaviour to different density, geometry and probably also to the UV~radiation (see below).

Bright $c$-C$_3$H$_2$(2$_{1,2}$--1$_{0,1}$) line emission with $W=1.2$~K~\kms{} is concentrated on the western borders of the East~1 and East~2 clumps, which are illuminated by the ionizing star of S235. In the East~2 clump, the area with bright $c$-C$_3$H$_2$ emission coincides with the location of the Class~0/I~YSOs from \citet{Dewangan_2011}. The peak of the $c$-C$_3$H$_2$(2$_{1,2}$--1$_{0,1}$) emission in S235\,ABC is observed around young massive stars S235\,A$^*$ and S235\,A-2$^*$ and it is distributed wider in comparison with C$_2$H(1--0). The peak intensities of the $c$-C$_3$H$_2$(2$_{1,2}$--1$_{0,1}$) line emission are the same in the S235 and S235\,ABC regions, while the C$_2$H(1--0) line, as well as all other lines considered in this study, are almost a factor of two brighter in S235\,ABC than in S235.

\subsubsection{HCN, H$_2$CO and CH$_3$OH}

The HCN(1--0) line is the brightest detected line in the present study with the peak integrated intensity of $W \geq 20$~K~\kms{}. Spatial distribution of the HCN(1--0) and H$_2$CO(2$_{1,2}$--1$_{1,1}$) molecular emission is similar in both observed areas. The line emission peaks are located towards the mm-continuum peaks found by \citet{Klein_2005}, see Fig.~\ref{fig:mom0}. The brightest HCN(1--0) lines with $W\approx 22$~K~\kms{} are found in the East~1 clump towards the chain of the Class~0/I~YSOs. In the East~2 clump, the bright HCN(1--0) emission appears on the opposite side from the ionizing star and does not coincide with the location of the Class~0/I~YSOs. The bright H$_2$CO(2$_{1,2}$--1$_{1,1}$) line emission with $W=10$~K~\kms{} appears along the same chain of YSOs and also on the shielded side of the East~2 clump. Area around and to the south of the YSO IRS~1 in S235 also contains bright HCN(1--0) and H$_2$CO(2$_{1,2}$--1$_{1,1}$) emission. In the S235\,ABC area, the peak of the HCN(1--0) emission is found between the Class~0/I~YSOs and 235~B$^*$. The peak is shifted deeper into the molecular cloud from the ionizing star 235~A$^*$ to S235~B$^*$ in comparison with the emission peaks of the small hydrocarbons. There are secondary peaks of the HCN(1--0) and H$_2$CO(2$_{1,2}$--1$_{1,1}$) emission located to the south of S235~B, those also coincide with the mm-continuum emission peaks.

The methanol emission at 3 and 2~mm is concentrated around the mm-continuum peaks. The brightest methanol emission in the S235 area is found around the East~1 clump with $W=2$ and 9~K~\kms{} for the CH$_3$OH(5$_{-1}$--4$_0$)~E and CH$_3$OH(3$_{K}$--2$_K$), see Fig.~\ref{fig:spectra} and \ref{fig:mom0}, respectively. The CH$_3$OH(5$_{-1}$--4$_0$)~E emission peak is observed in the northern part of the clump similar to the HCN(1--0) emission. The bright CH$_3$OH(3$_K$--2$_K$) emission is distributed over the entire clump and follows the shape of the mm-continuum emission and the chain of the Class~0/I~YSOs. In the S235\,ABC area, the bright methanol emission in both transitions is observed between S235\,A$^*$ and S235~B$^*$ towards the mm-continuum peak. Also, the secondary methanol peaks are observed to the south of S235~B$^*$ similar to HCN(1--0) and H$_2$CO(2$_{1,2}$--1$_{1,1}$). The shape of the 5$_{-1}$-4$_0$~E line in Fig.~\ref{fig:spectra} is typical for a marginally resolved maser source~\citep[see, e. g.][]{Kurtz_2004}.

\begin{figure*}
    \includegraphics[height=10.0cm]{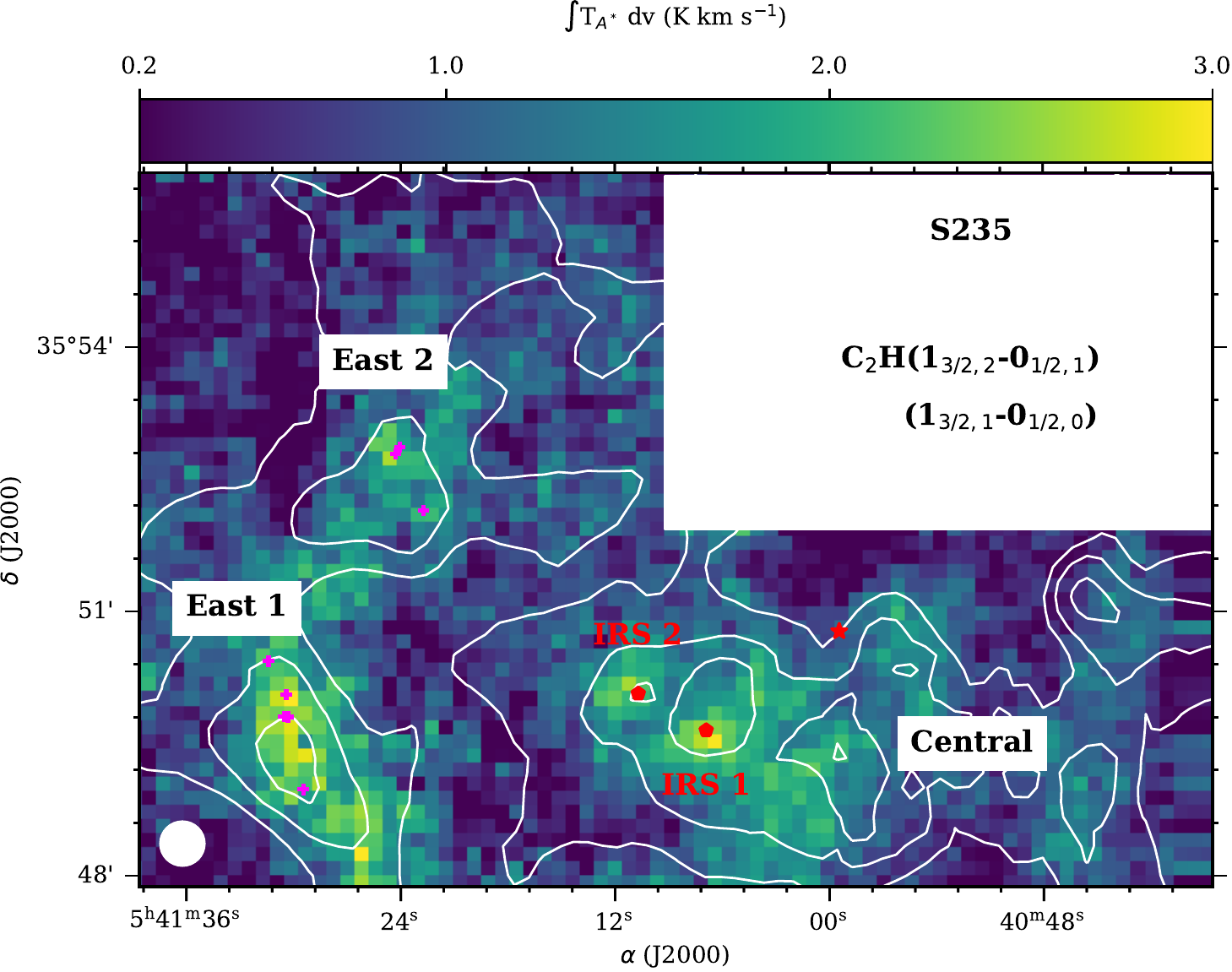}
    \includegraphics[height=10.0cm]{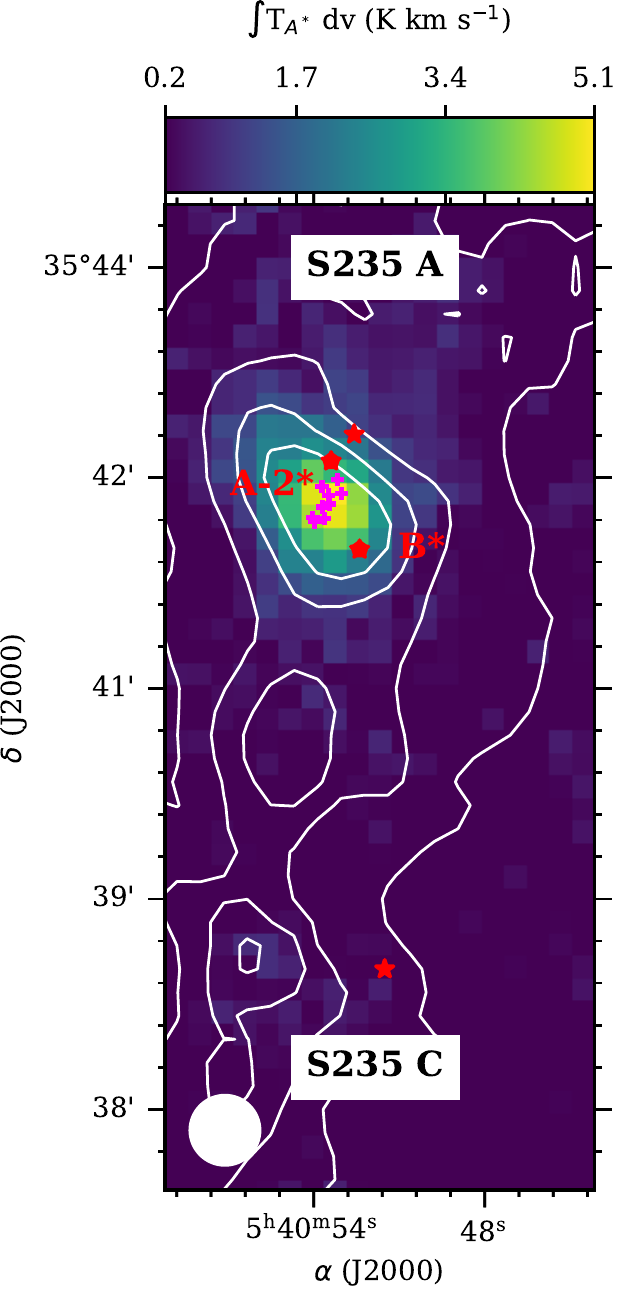}\\
    \vspace{0.5cm}
     \includegraphics[height=10.0cm]{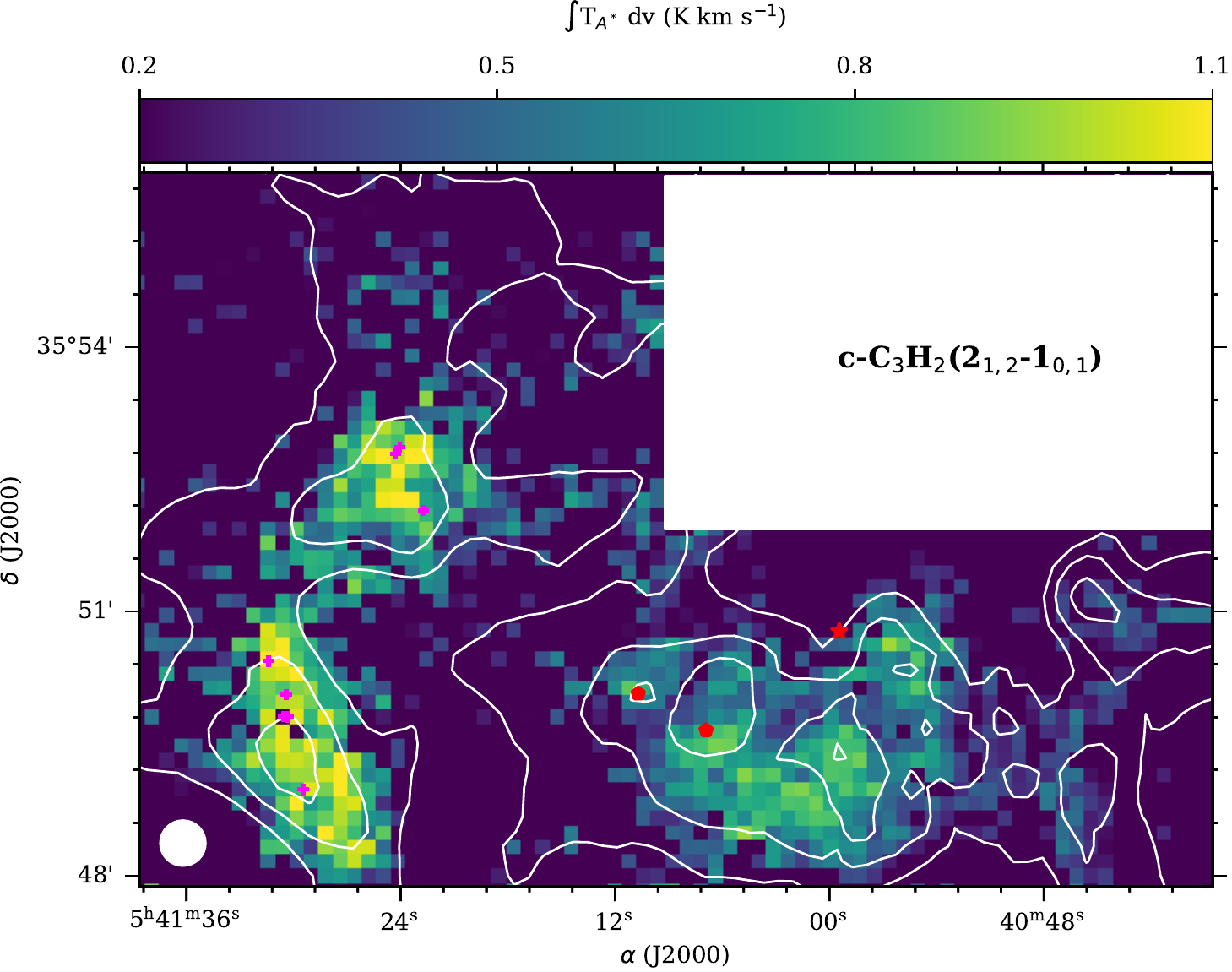}
    \includegraphics[height=10.0cm]{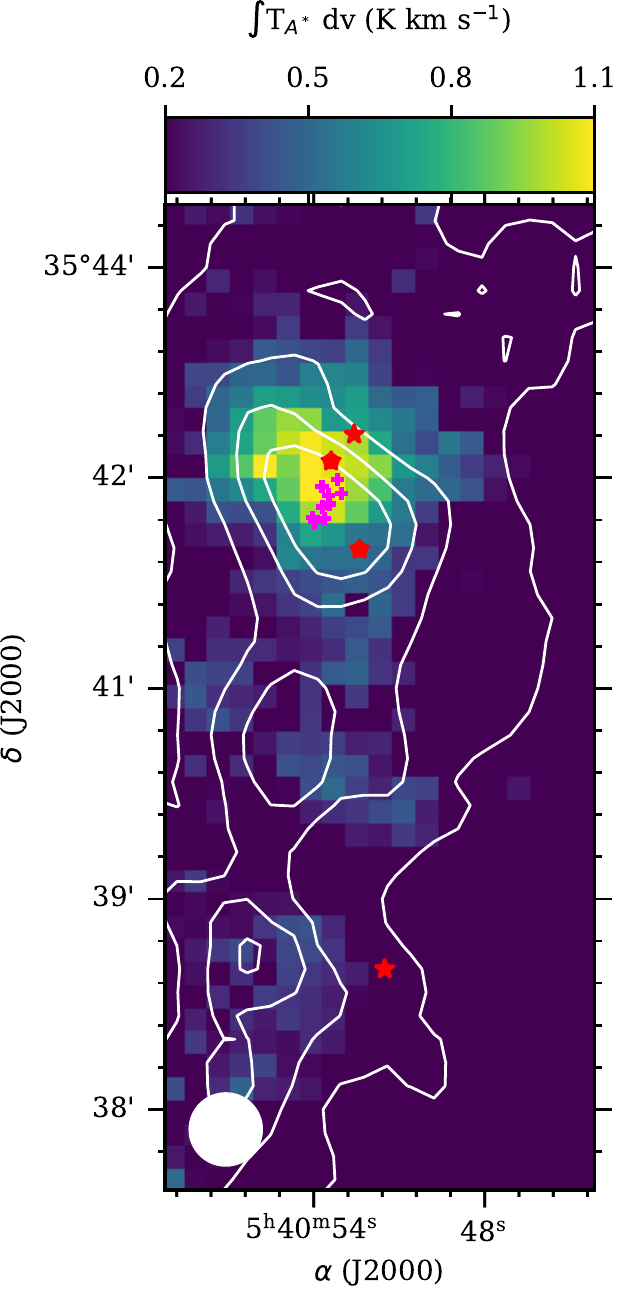}\\
    \caption{Integrated intensities of the detected lines in S235 (left panels) and S235\,ABC observed area (right panels). The telescope beam is shown by white circle. Pixel size corresponds to one third of the beam size. The colorscale begins at the $3\sigma$ level for each transition. White contours show levels of hydrogen column density: 1, 2, 3 and $4\cdot 10^{22}$~cm$^{-2}$ based on the CO observations by \citet{Bieging_2016}. Mm-continuum emission from \citet{Klein_2005} is shown by the white contours on the H$_2$CO and CH$_3$OH $3_K-2_K$ emission maps. The ionizing sources are shown by the red stars. Red diamonds show three bright infrared sources: IRS1, IRS2 \citep{Evans_1981} and S235~B$^*$ \citep{Boley_2009}. Magenta crosses show IR~sources from \citet{Dewangan_2011}.}
    \label{fig:mom0}
\end{figure*}

\begin{figure*}
\includegraphics[height=10.0cm]{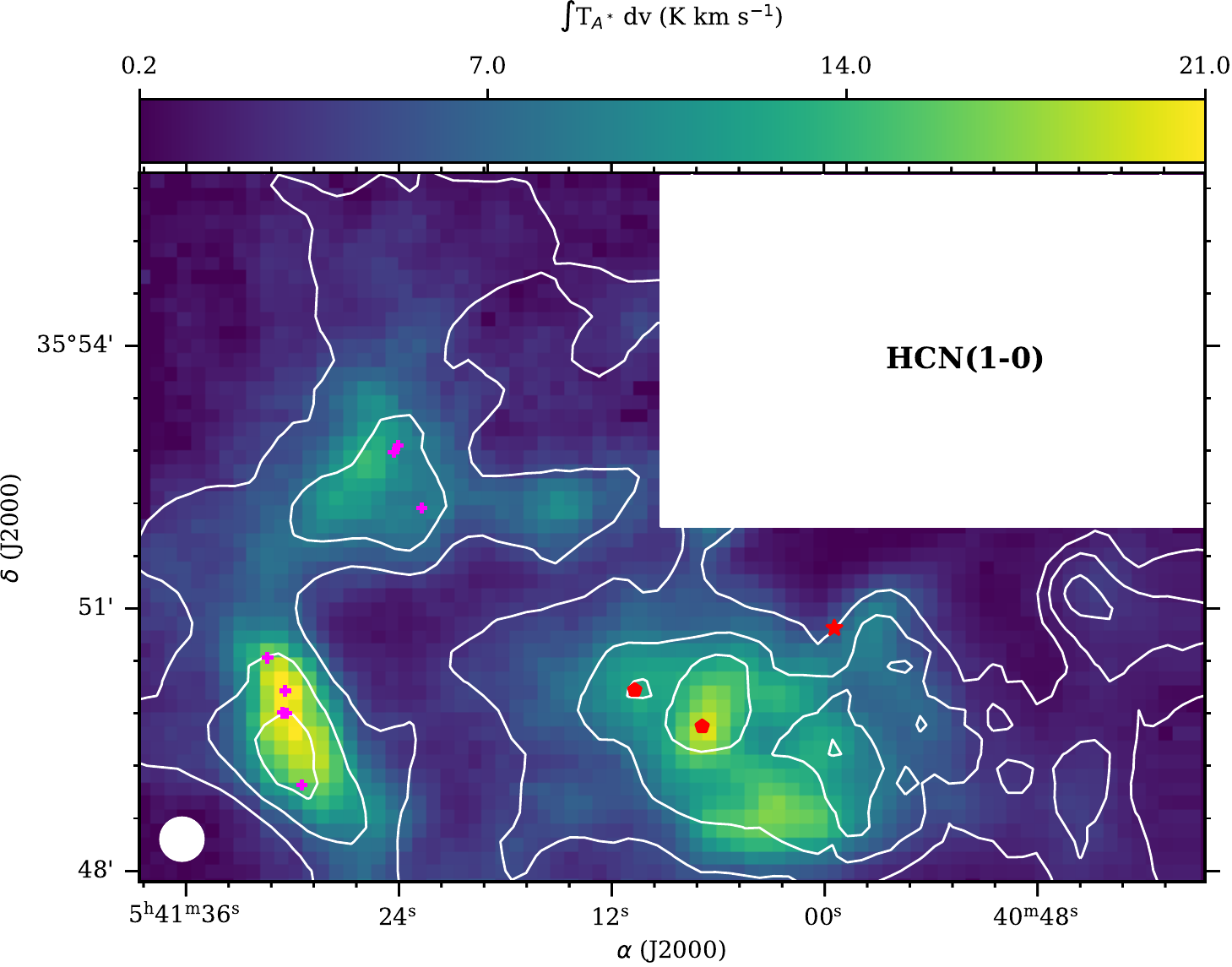}
\includegraphics[height=10.0cm]{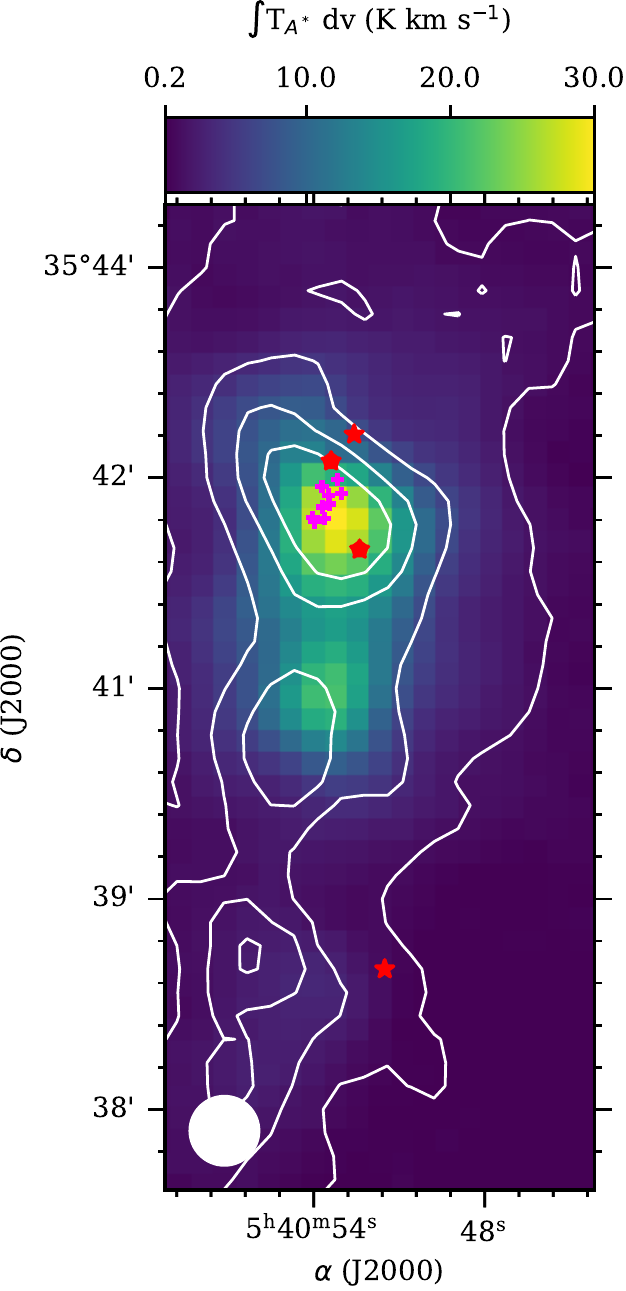}\\
\includegraphics[height=10.2cm]{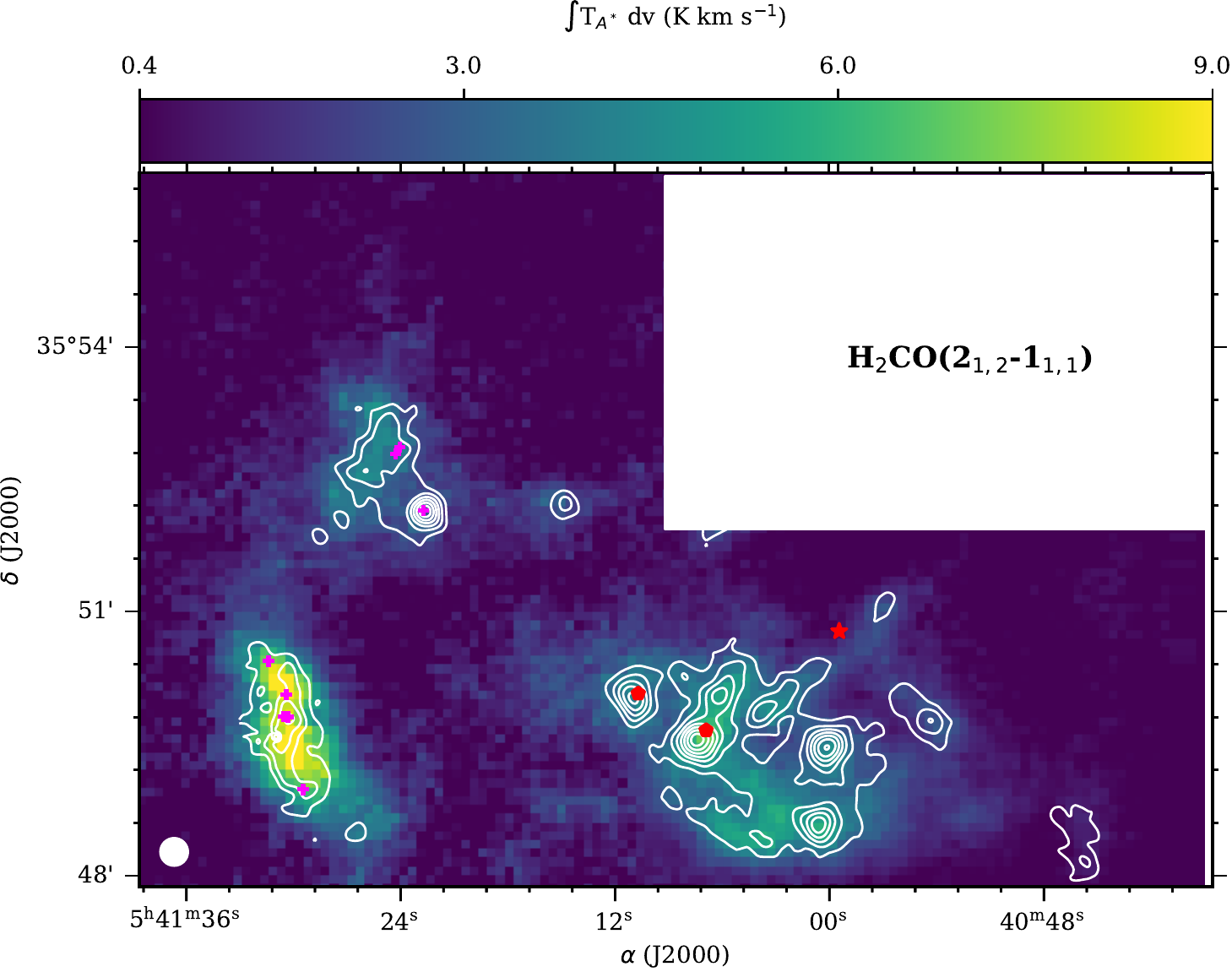}
\includegraphics[height=10.2cm]{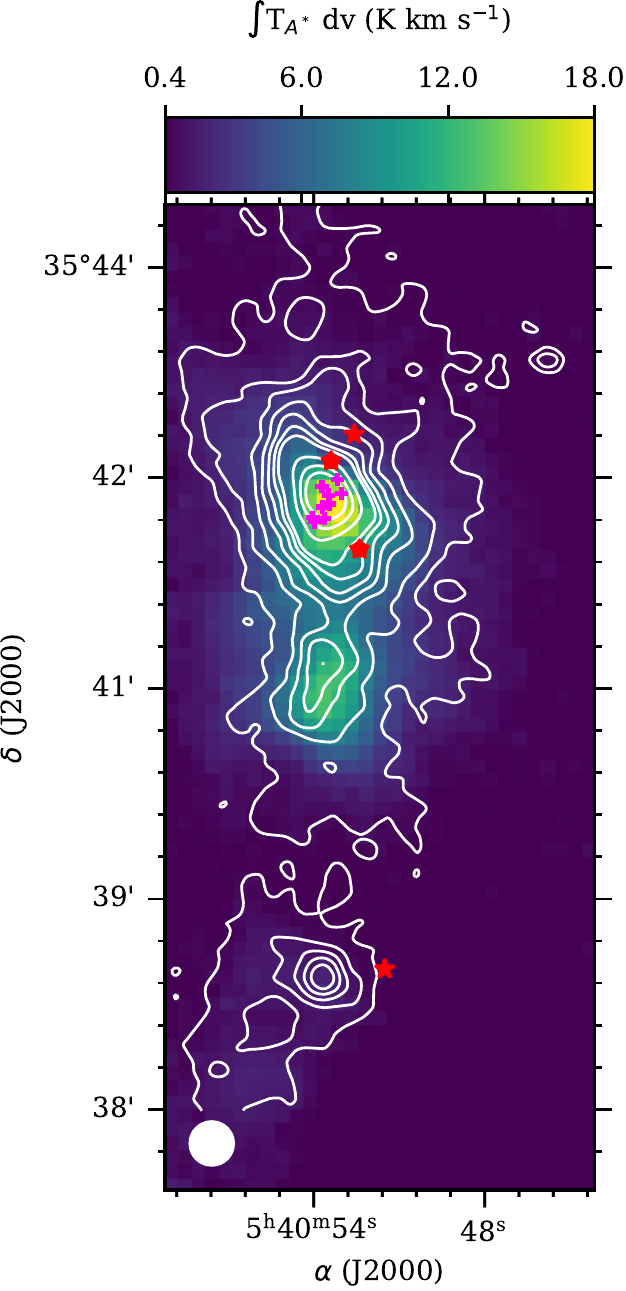}
\contcaption{. White contours in the map of the H$_2$CO(2$_{1,2}$--1$_{1,1}$) line emission show levels of SCUBA-850~\micron{} emission from 5 to 35~mJy~arcsec$^{-2}$ with a step 5~mJy~arcsec$^{-2}$ and additional contours at 55, 75 and 95~mJy~arcsec$^{-2}$ for S235\,ABC region based on \citet{Klein_2005}.}
\end{figure*}

\begin{figure*}
\includegraphics[height=10.0cm]{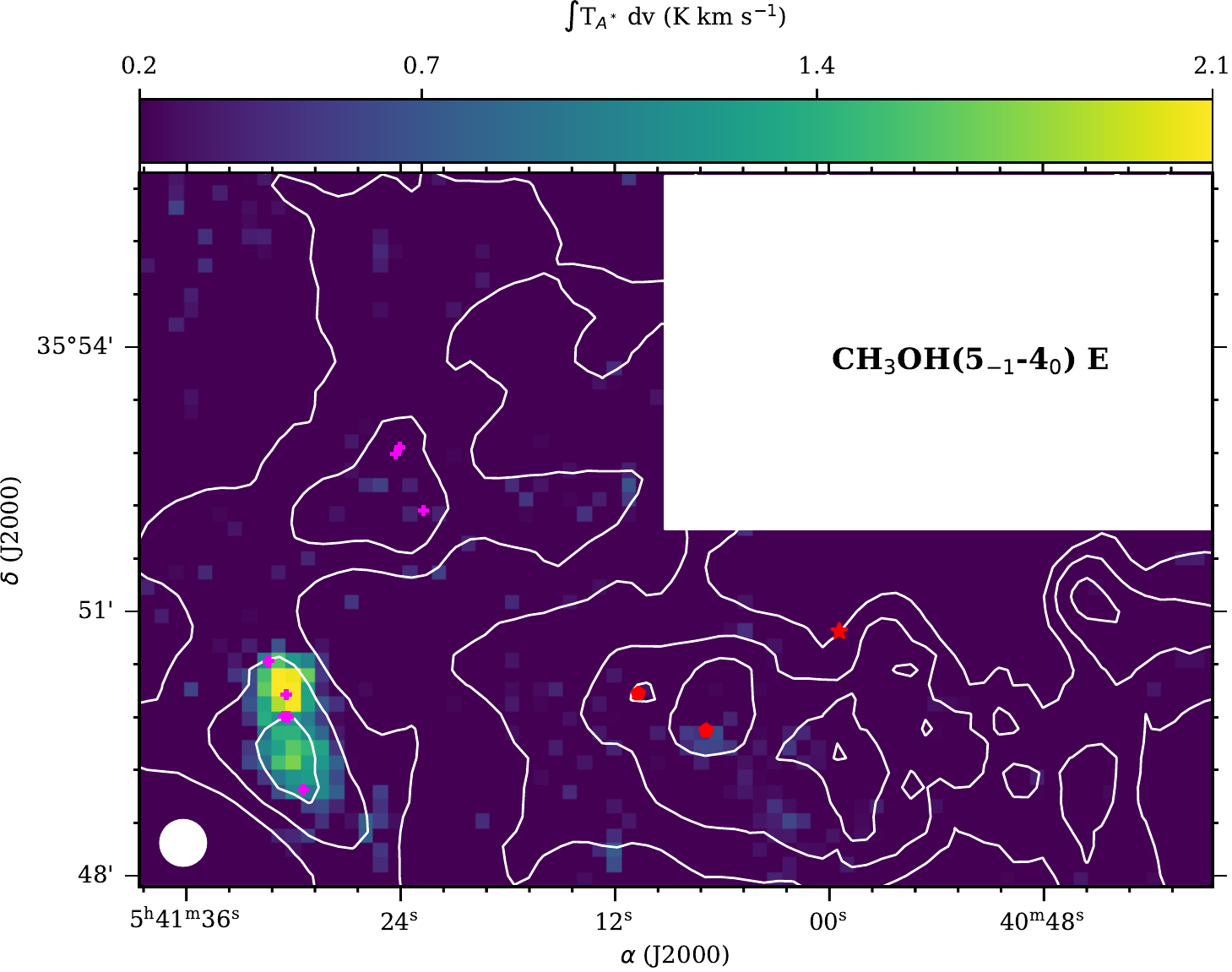}
\includegraphics[height=10.0cm]{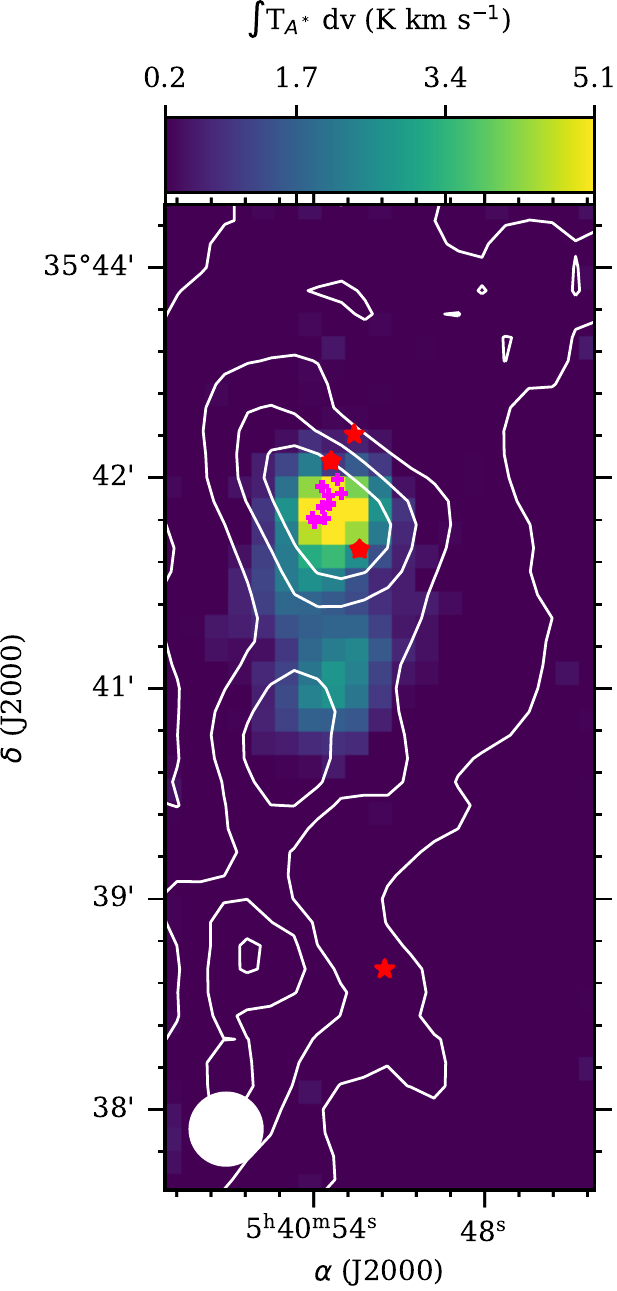}\\
\includegraphics[height=10.2cm]{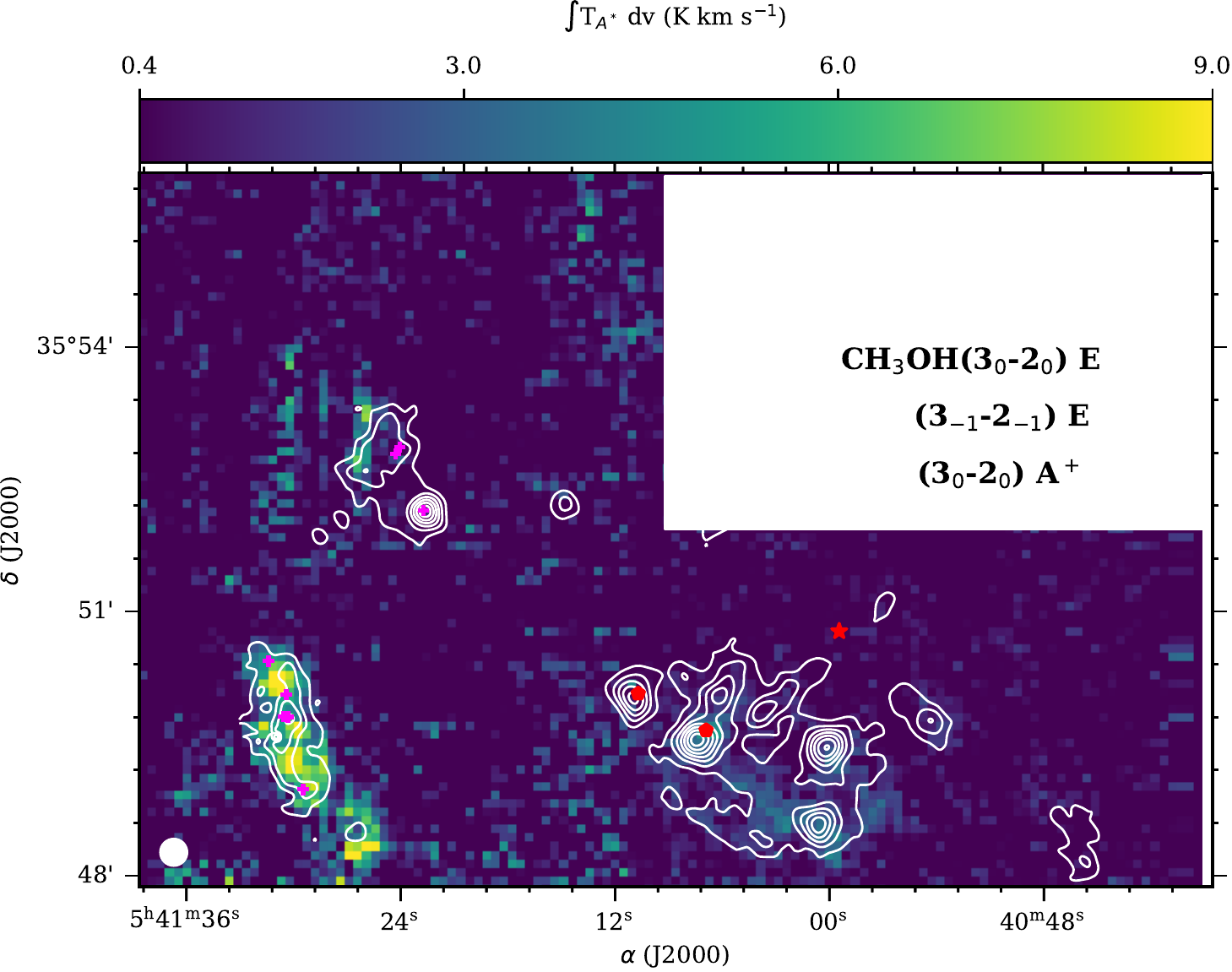}
\includegraphics[height=10.2cm]{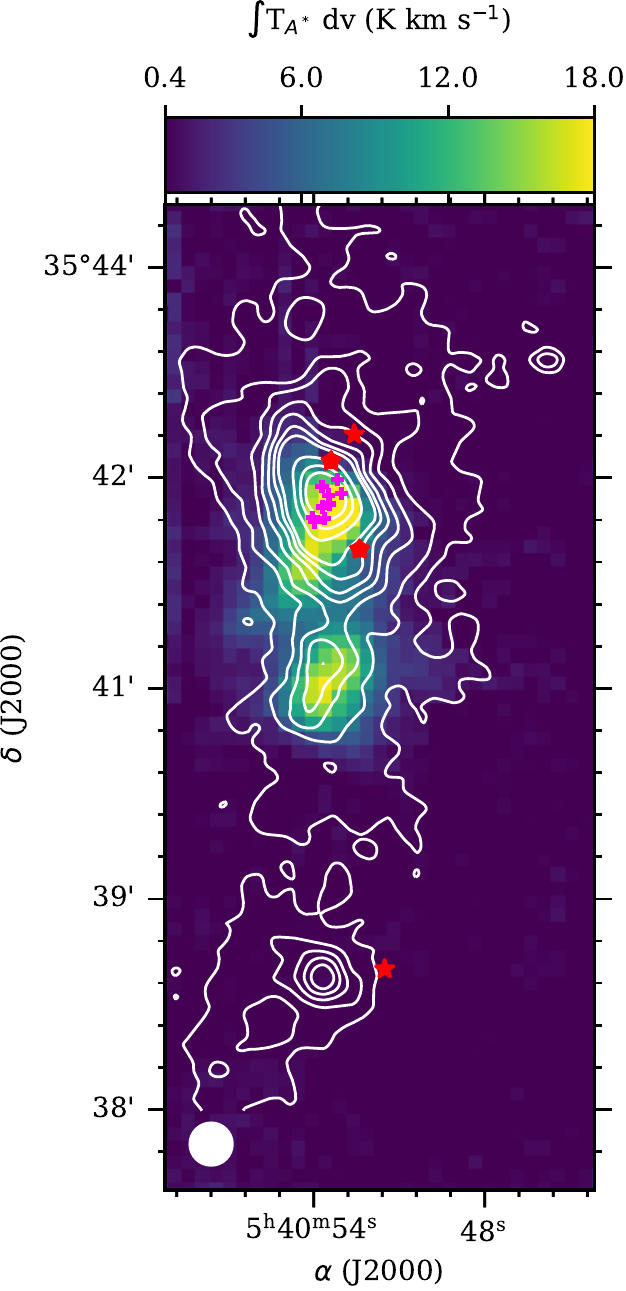}
\contcaption{. White contours in the map of the CH$_3$OH(3$_{K}$--2$_{K}$) line emission show levels of SCUBA-850~\micron{} emission from 5 to 35~mJy~arcsec$^{-2}$ with a step of 5~mJy~arcsec$^{-2}$ and additional contours at 55, 75 and 95~mJy~arcsec$^{-2}$ for S235\,ABC region based on \citet{Klein_2005}.}
\end{figure*}

\subsection{Column densities} 

For all detected species we calculated their column densities from the integrated line intensities in optically thin approximation, therefore the column density maps appear similar to the corresponding maps of the intensities. In this Section, we present a pixel-by-pixel comparison of the column densities of those pixels, where the S/N ratios are above 3 for both integrated intensity and column density.

\subsubsection{C$_2$H vs c-C$_3$H$_2$}

By fitting the C$_2$H(1--0) and HCN(1--0) hyperfine structures with the CLASS routine {\it hfs}, we find that the fitting does not converge in the the majority of the pixels in Fig.~\ref{fig:mom0} due to deviation from the Gaussian line shape. The fitting procedure converges in several pixels, and we find the median optical depth of those pixels to be $\tau_{\rm C_2H}=0.2$. The median value of the excitation temperature for those pixels is 6~K, which is not too far from the assumed 10~K value, see Sec.~\ref{sec:columndensity}. For all other pixels we find that the intensity ratio of the brightest hyperfine components $1_{3/2,2}-0_{1/2,1}$ and $1_{3/2,1}-0_{1/2,0}$ of the C$_2$H(1--0) line is in agreement with the optically thin case. For those pixels where we detected both the 85~GHz and 145~GHz lines of c-C$_3$H$_2$, we estimate the excitation temperature using rotational diagram method. We recognize the  uncertainty of this method when applied to the two line transitions only, and use it solely to test if the resulting excitation temperature value seems reasonable. The median value of this temperature is 14~K, which is not far from the assumed 10~K value. 

The pixel-by-pixel comparison of the C$_2$H and $c$-C$_3$H$_2$ column densities is presented in Fig.~\ref{fig:s235_c2h_ch3oh}. The values of $N_{\rm C_2H}$ and $N_{c-{\rm C}_3{\rm H}_2}$ are directly proportional to each other in S235, but the correlation is steeper in S235\,ABC. The maximum $N_{\rm C_2H} = 8\times10^{13}$~cm$^{-2}$ is observed in the East~1 clump, where $N_{\rm c-C_3H_2}= 5\times10^{12}$~cm$^{-2}$. We find that the values of both column densities are maximal in the range of $G=40-50$~Habings and become smaller at $G < 30$~Habings. In the majority of the pixels we find that the ratio falls into the interval $1 \leq N_{\rm C_2H}/N_{c-{\rm C}_3{\rm H}_2} \leq 20$ in S235 and S235\,ABC. The interesting feature of the $N_{\rm C_2H}/N_{c-{\rm C}_3{\rm H}_2}$ ratio in S235 and S235\,ABC is that the column densities are confined within the same range of values in the pixels with different FUV intensities. Therefore, even moderate UV~field unlocks carbon atoms from their carriers (e.g. CO or PAH or C-dust) which is then rapidly converted to hydrocarbons via gas-phase reactions. After that, the column densities do not increase with intensity of the UV~field.

Spatial distribution of the $N_{\rm C_2H}/N_{c-{\rm C}_3{\rm H}_2}$ ratio in S235 and S235\,ABC is shown in Fig.~\ref{fig:abundratio_S235} and \ref{fig:abundratio_S235A}, respectively. In the dense clumps around S235, the ratio is distributed almost uniformly around value of 8-12, but the higher ratio up to 20-24 is observed on the borders of the clumps and around YSOs IRS~1 and IRS~2. In the S235\,ABC area, the peak of the $N_{\rm C_2H}/N_{\rm c-C_3H_2} \approx 32$ appears towards S235~B (the color scale is saturated in Fig.~\ref{fig:abundratio_S235A} in order to use the same limits for both S235 and S235\,ABC area). The ratio becomes smaller ($\approx 8-12$) in the direction to and around the ionizing stars S235\,A$^*$ and S235\,C$^*$ and in the secondary density peak of the mm-continuum emission to the south from S235~B$^*$. Therefore,  the $N_{\rm C_2H}/N_{c-{\rm C}_3{\rm H}_2}$ ratios have maximum values towards young massive stars with emission lines in spectra (IRS~1, IRS~2 and S235~B$^*$), but become lower towards the ionizing stars and the dense molecular clumps.

In order to estimate how far our LTE calculations are from the non-LTE predictions, we calculate the line intensities of hydrocarbons using the RADEX software \citep{2007A&A...468..627V}. For that, we used appropriate gas temperatures and densities \citep[obtained with ammonia emission lines by][]{Kirsanova_2014} in several selected positions in the S235 and S235\,ABC areas. We find the excitation temperatures of about 4-6~K. These values are 2-3 times lower than the inferred gas temperatures. Therefore, we confirm the subthermal excitation in the region found by \citet{Bieging_2016} via analysis of the CO lines. The corresponding column densities of the hydrocarbons should be higher up to an order of magnitude to reproduce the observed line brightness. However, the optical depth of the C$_2$H(1--0) line becomes $>1$ in this case, which contradicts our observations. We conclude that the LTE-derived values can be considered as lower limits, especially for the low density gas around the ionizing star of S235, towards the East~2 cluster, to the north-west from the S235\,A$^{*}$ and in the S235\,C region.

\subsubsection{H$_2$CO vs CH$_3$OH}

Using rotational diagrams we estimated $T_{\rm rot}$ for the observed methanol lines in East~1 and Central clumps in S235 and in S235~ABC, see Appendix~\ref{app:rotdiag}. Insufficient signal-to-noise ratio of the data did not allow us to make the diagrams for the East~2 clump. The maximum $T_{\rm rot} \approx 40-50$~K is found in the northern part of East~1, while $T_{\rm rot} \approx 20$~K in the rest of the clump. Typical $T_{\rm rot} \approx 20$~K is observed also in the Central clump but it can be less than 10~K on its external border. The maximum $T_{\rm rot}$ value in S235\,ABC is 55~K found between S235A-2$^{*}$ and S235~B$^{*}$, and the value becomes $\approx 16-20$~K elsewhere. 

In contrast with column densities of the small hydrocarbons, the column densities of H$_2$CO and CH$_3$OH are weakly proportional to each other in S235, see Fig.~\ref{fig:s235_c2h_ch3oh}. We find $N_{\rm H_2CO} \approx 1-6 \times 10^{13}$ for the broad range of $10^{13} < N_{\rm CH_3OH} < 2 \times 10^{15}$ for positions with $20< G< 50$. \Nmeth{} becomes higher for lower values of the UV~field $G\approx 20-30$~Habings and $N_{\rm H_2CO}/N_{\rm CH_3OH} \approx 0.1$ there. The $N_{\rm H_2CO}/N_{\rm CH_3OH}$ becomes higher and $\approx 1$ for $G \geq 50$. We find $N_{\rm H_2CO}/N_{\rm CH_3OH} \approx 0.1$ in the majority of the observed positions in S235~ABC and more broad range of \Nform{} than in S235 for the same range of \Nmeth{} values. It is interesting that the similar values of the \Nform{} and \Nmeth{} are produced at different values of UV~fields in these two areas, namely, the less irradiated gas in S235~ABC produces as many precursors of COMs as the highly irradiated gas in S235.

Spatial distribution of the $N_{\rm H_2CO}/N_{\rm CH_3OH}$ ratios is different in two parts of S235: the low ratios $<0.2$ are observed in the East~1 clump, and the higher ratios up to $0.6-1$ are observed in the Central clump, see Fig.~\ref{fig:abundratio_S235}. In the S235\,ABC area, the highest $N_{\rm H_2CO}/N_{\rm CH_3OH} \approx 0.4$ ratios are observed on the external borders of the dense molecular clump. The region of the minima coincides with the peak of hydrogen column density.

\subsubsection{C$_2$H vs CH$_3$OH}

\begin{figure*}
    \centering
    \includegraphics[width=0.9\columnwidth]{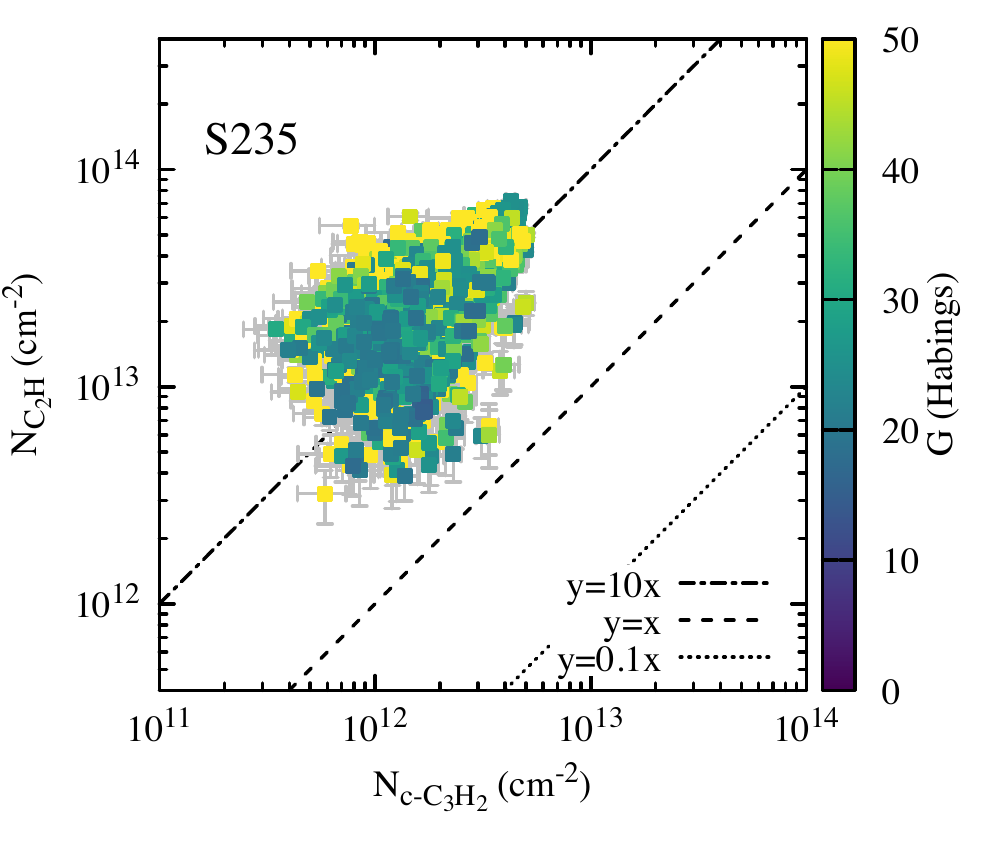}\hspace{0.5cm}
    \includegraphics[width=0.9\columnwidth]{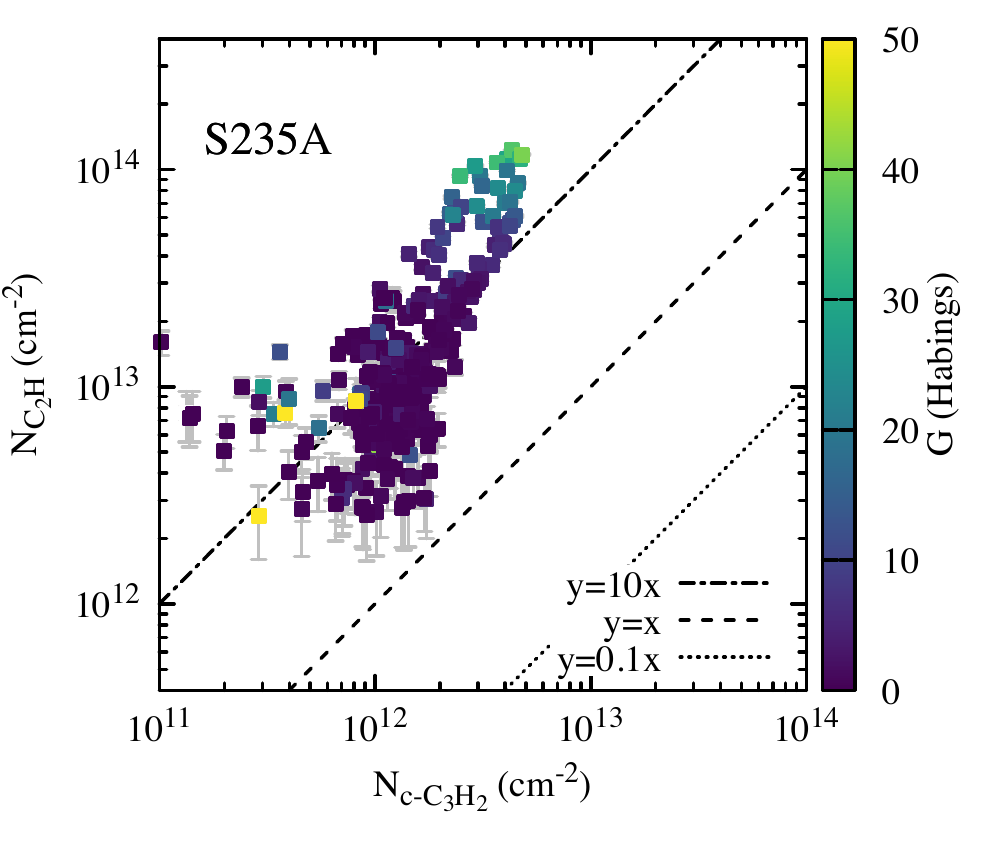}\\
    \includegraphics[width=0.9\columnwidth]{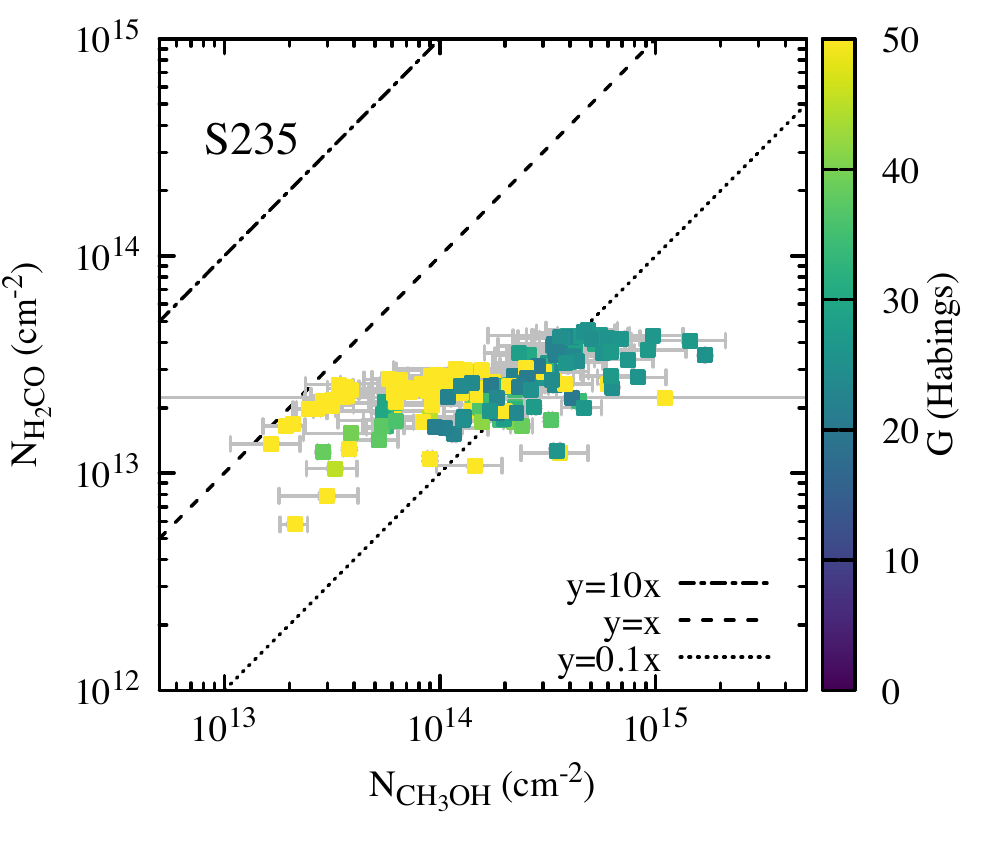}\hspace{0.5cm}
    \includegraphics[width=0.9\columnwidth]{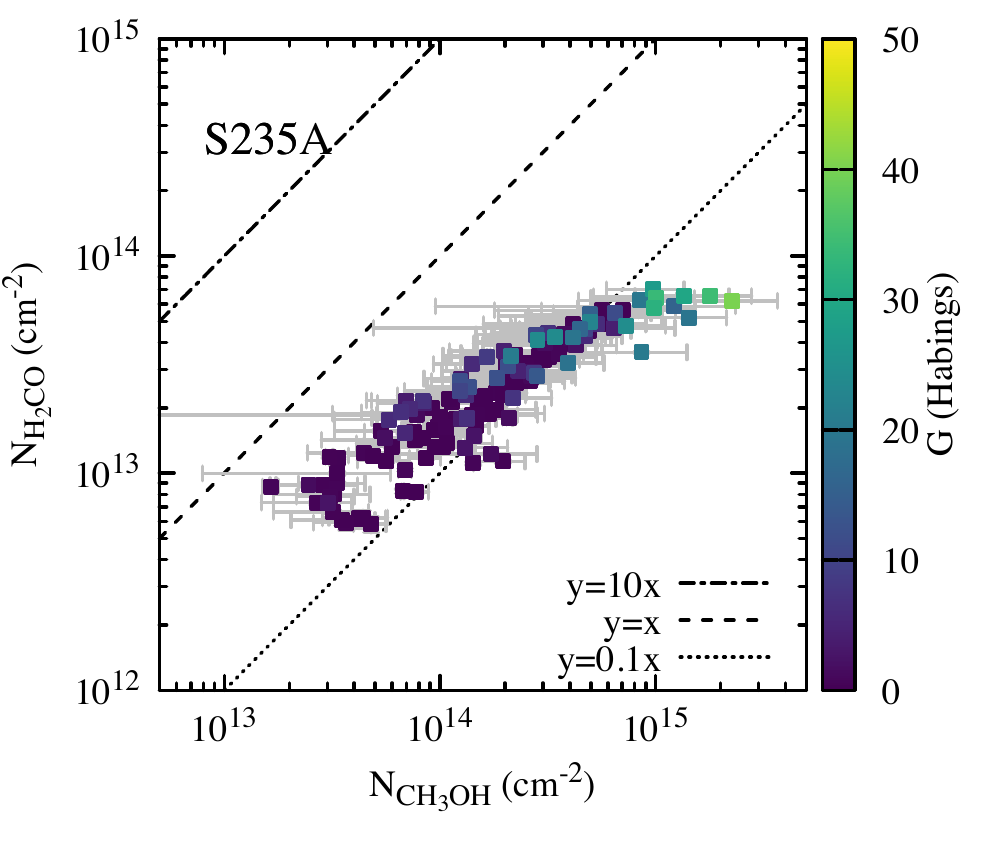}\\
    \includegraphics[width=0.9\columnwidth]{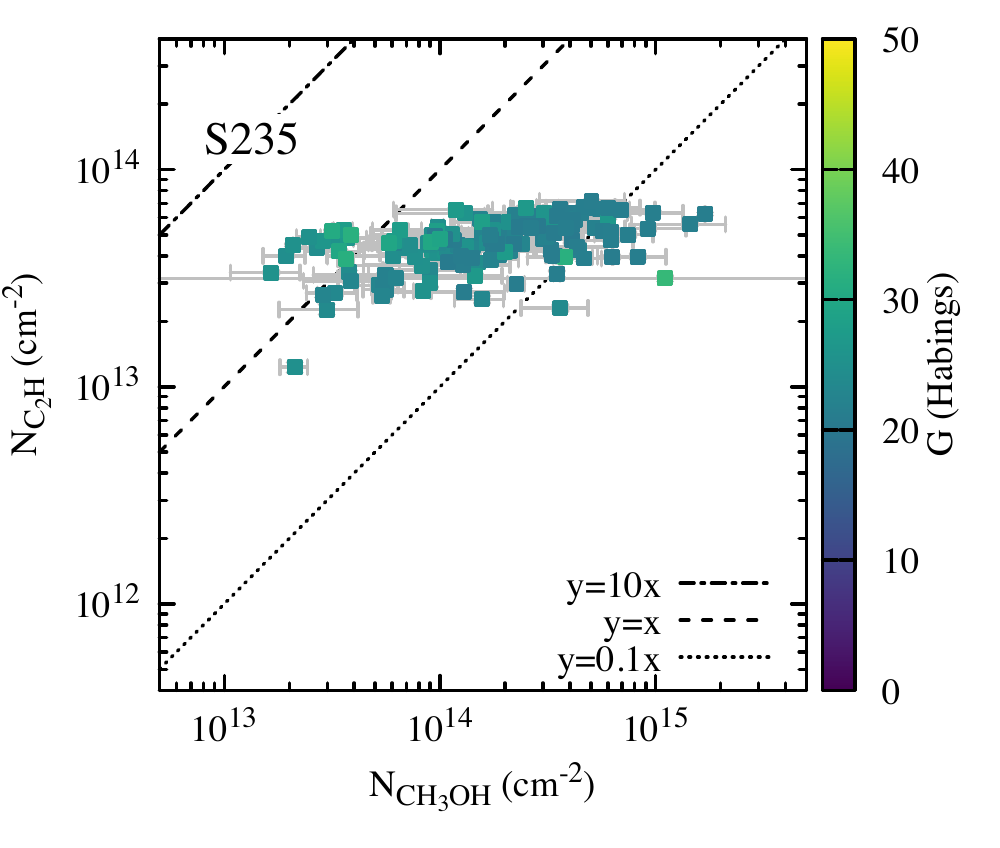}\hspace{0.5cm}
    \includegraphics[width=0.9\columnwidth]{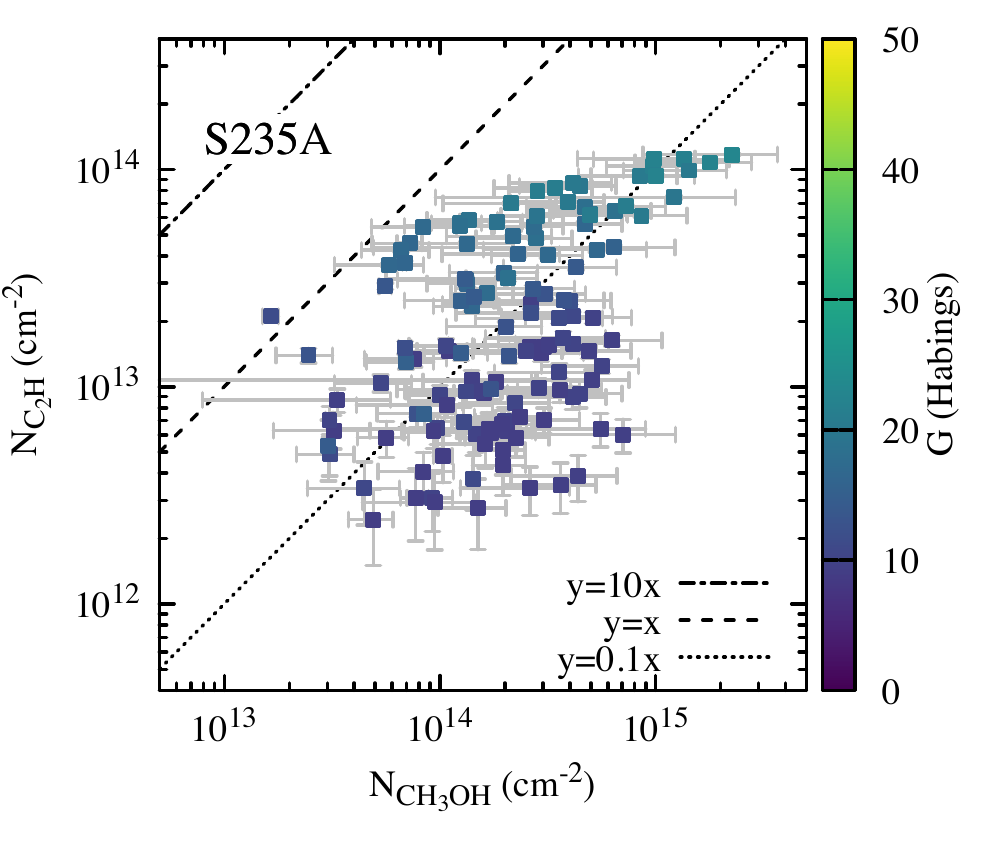}\\
    \caption{Top: pixel-by-pixel comparison of the C$_2$H and c-C$_3$H$_2$ column densities, middle: the same for H$_2$CO and CH$_3$OH, bottom: C$_2$H and CH$_3$OH. Positions where the column density values are smaller than the corresponding uncertainty multiplied by a factor of 3 are excluded from the plots. Color scale shows the UV-field intensities in Habing units. Straight lines show three different linear relations between the column densities.}
    \label{fig:s235_c2h_ch3oh}
\end{figure*}

The range of the C$_2$H column densities in the positions with known $N_{\rm CH_3OH}$ does not exceed one order of magnitude and is almost confined by the narrow range of $4-8\times 10^{13}$~cm$^{-2}$, while the methanol column densities vary by about two orders of magnitude from $10^{13}$ to $10^{15}$~cm$^{-2}$, see Fig.~\ref{fig:s235_c2h_ch3oh}. However, in S235\,ABC, the range of the $N_{\rm C_2H}$ values is almost as wide as the range of the methanol column densities: two orders of magnitude. There are no pixels with $G < 10$~Habings in the S235 region with known abundances of both molecules, as opposed to S235\,ABC, this may explain the narrow range of the C$_2$H column densities for S235. In the dark and dense molecular gas in S235\,ABC area, carbon is expected to be locked in CO. CO is formed from C$^+$ with small hydrocarbons being intermediates. Higher UV~field will unlock C from CO, resulting in higher abundances of small photostable hydrocarbons or C-ions. We conclude that the behavior of the C$_2$H~vs~CH$_3$OH column densities is similar to that of H$_2$CO~vs~CH$_3$OH. The minimum $N_{\rm C_2H}/N_{\rm CH_3OH}$ ratio is found for $G<10$~Habings and it becomes larger at higher and lower values of the UV-field.

\begin{figure}
    \centering
    \includegraphics[width=0.99\columnwidth]{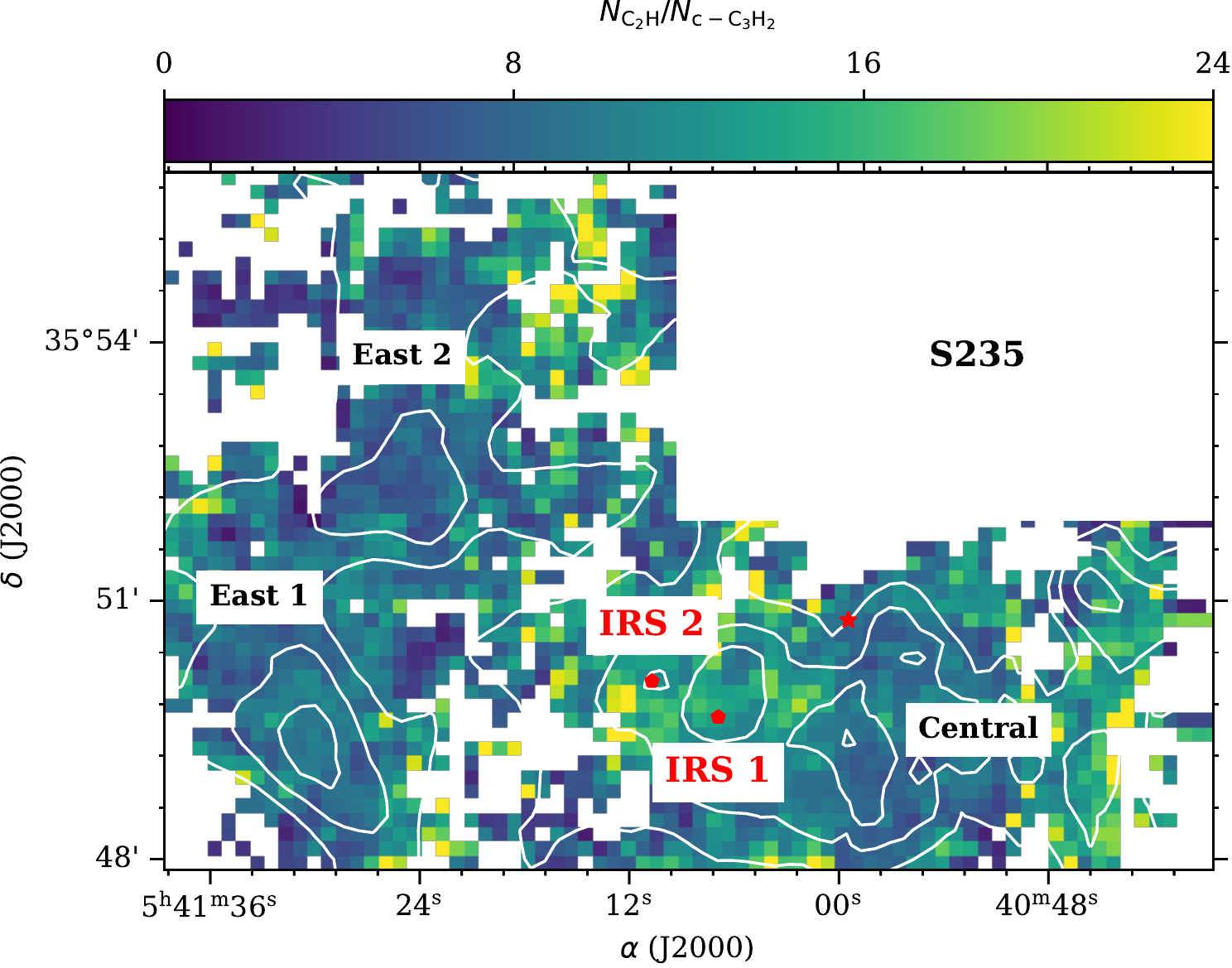}\\
    \vspace{0.5cm}
    \includegraphics[width=0.99\columnwidth]{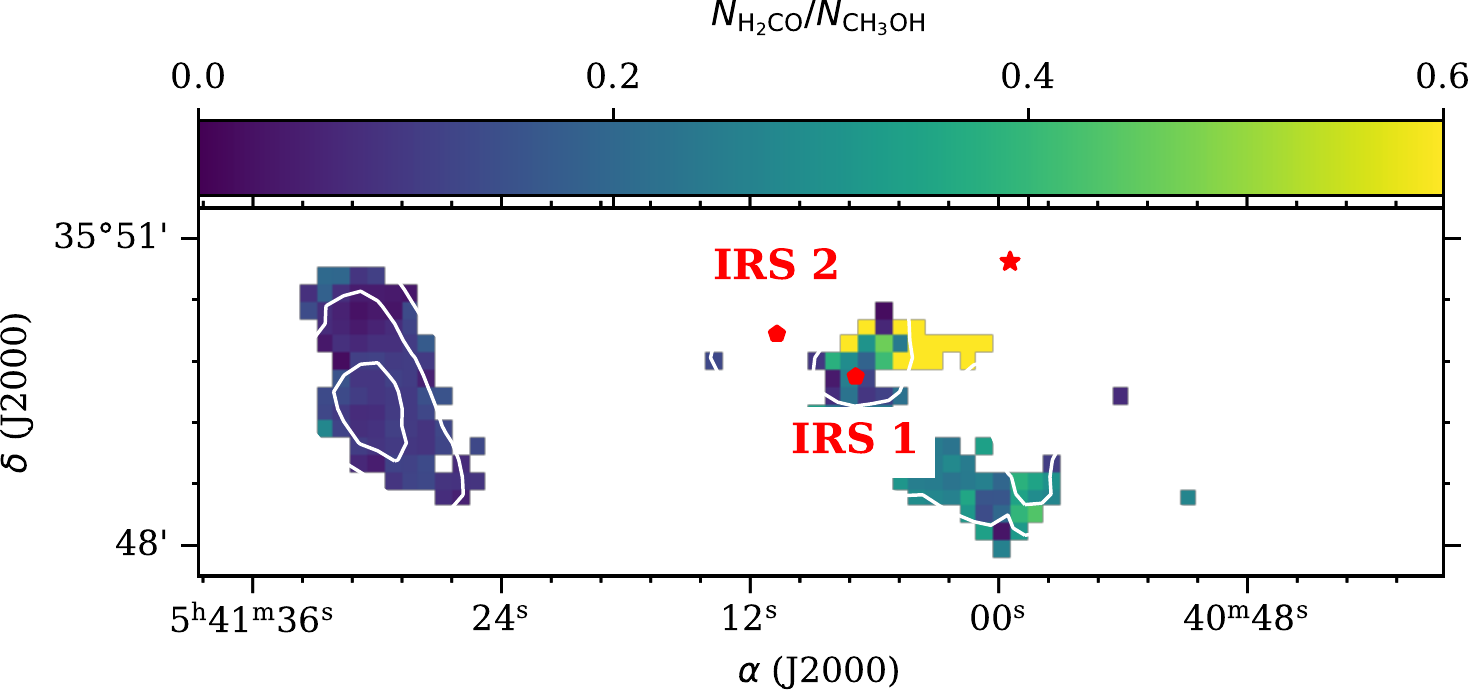}\\
    \vspace{0.5cm}
    \includegraphics[width=0.99\columnwidth]{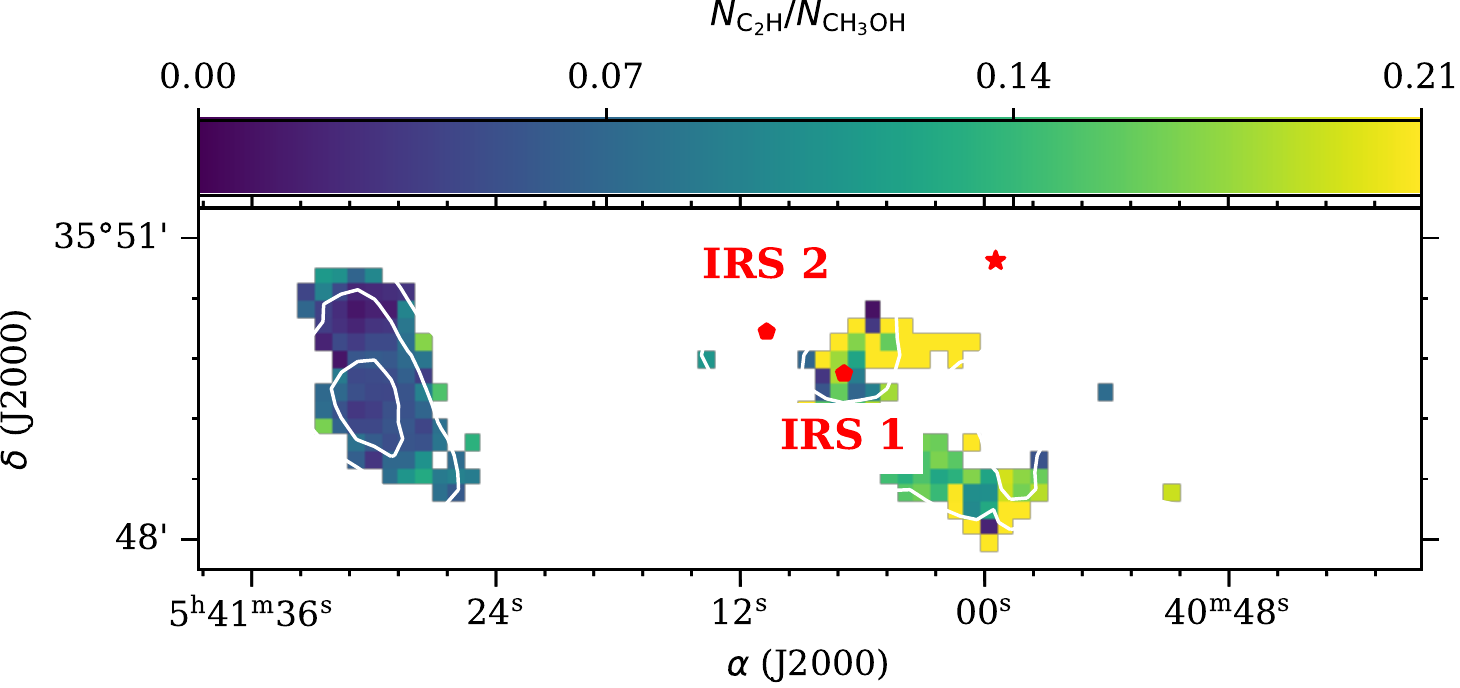}
    \caption{Top: Spatial distribution of the \Nethy{} to $N_{\rm c-C_3H_2}$ column density ratios in the S235 observed area, middle: the same for $N_{\rm H_2CO}$ to \Nmeth{}, bottom: the same for \Nethy{} to \Nmeth{} ratios. The hydrogen column density is shown by white contours. The IRS~1 and IRS~2 point sources are shown by red diamonds, the exciting star is shown by the red star. Only the pixels where the observed column density is at least three times higher than its uncertainty are shown.}
    \label{fig:abundratio_S235}
\end{figure}

\begin{figure*}
    \centering
    \includegraphics[width=0.65\columnwidth]{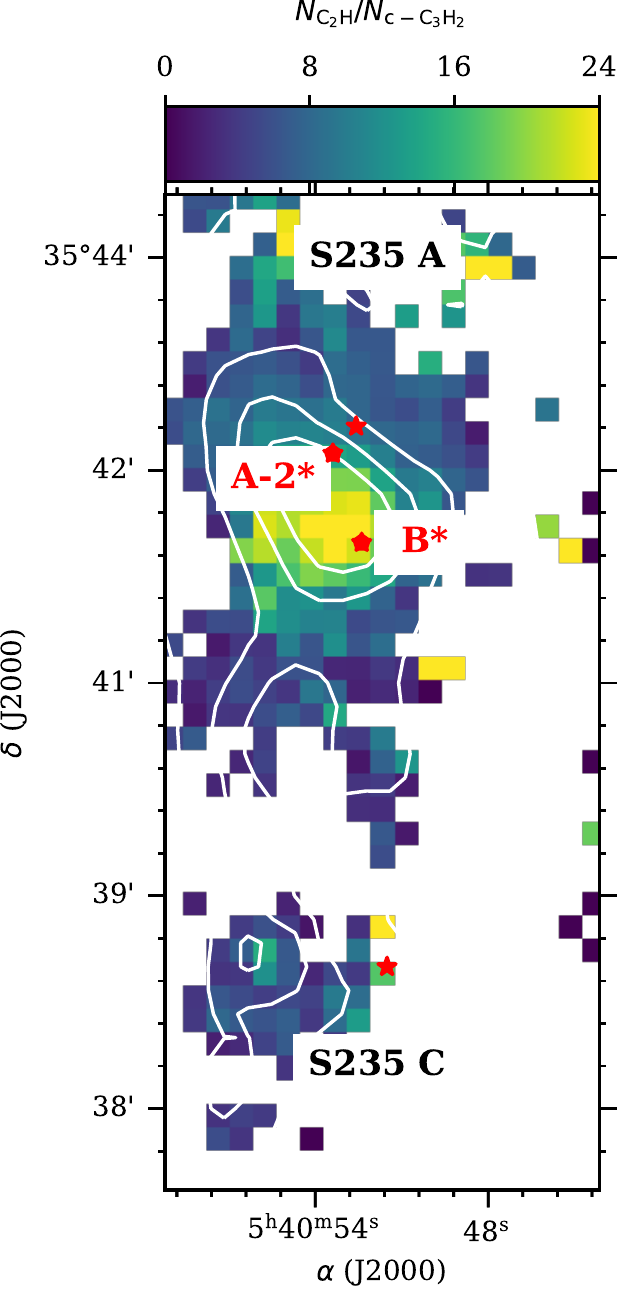}
    \includegraphics[width=0.65\columnwidth]{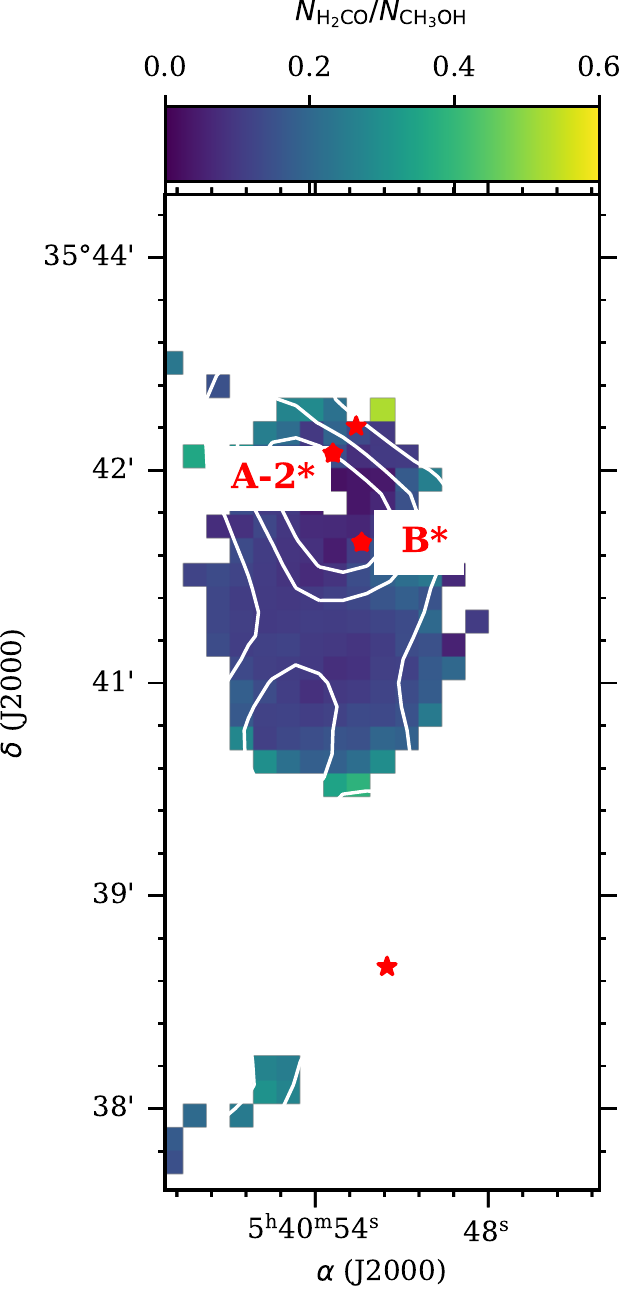}
    \includegraphics[width=0.66\columnwidth]{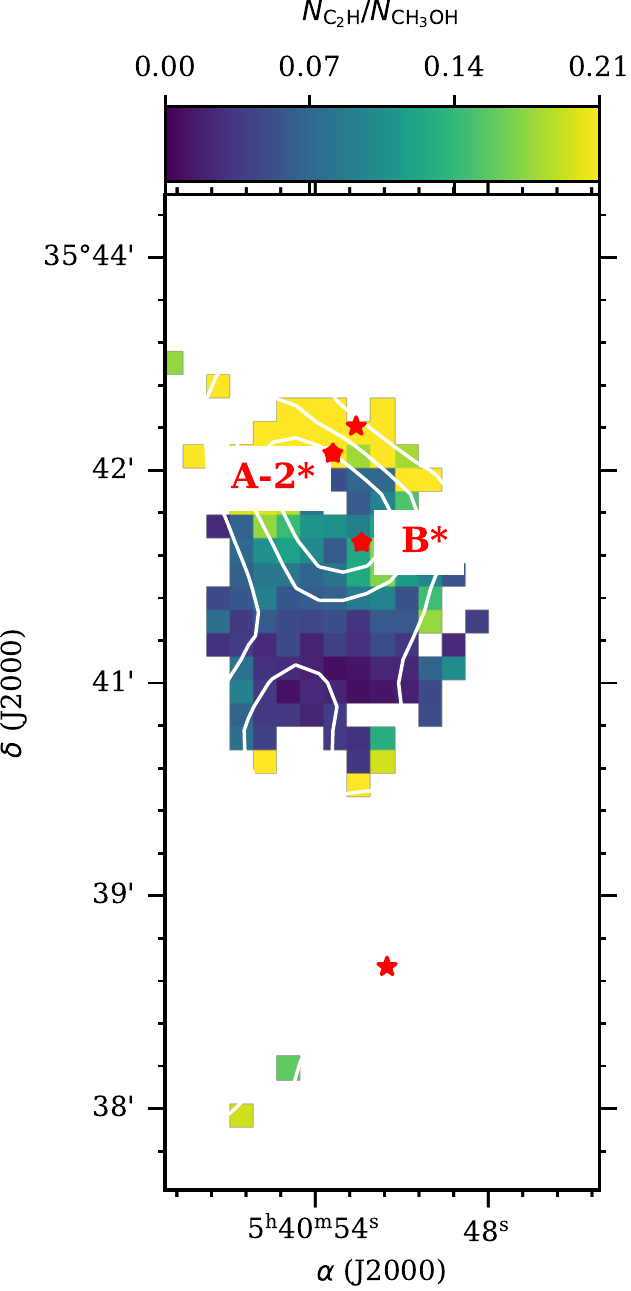}
    \caption{Left: Spatial distribution of the \Nethy{} to $N_{\rm c-C_3H_2}$ column density ratios in the S235\,ABC observed area, middle: the same  for $N_{\rm H_2CO}$ to \Nmeth{}, right: the same for \Nethy{} to \Nmeth{} ratio. The hydrogen column density is shown by white contours. S235~B$^*$ and S235\,A-2$^*$ point sources are shown by red diamonds, the exciting stars of the \hii{} regions S235\,A and S235\,C are shown by the red stars. Only the pixels where the observed column density is at least three times higher than its uncertainty are shown.}
    \label{fig:abundratio_S235A}
\end{figure*}

Spatial distribution of the $N_{\rm C_2H}/N_{\rm CH_3OH}$ ratios qualitatively similar to the $N_{\rm H_2CO}/N_{\rm CH_3OH}$ ratio distribution, see Fig.~\ref{fig:abundratio_S235} and Fig.~\ref{fig:abundratio_S235A}. We find values of $N_{\rm C_2H}/N_{\rm CH_3OH} < 0.1$ in the East~1 clump with relatively low UV-field. In the Central clump, where the UV~field is higher, the ratio reaches 0.2. In the S235\,ABC area, the ratio $N_{\rm C_2H}/N_{\rm CH_3OH} \approx 0.2$ in the vicinity of the ionizing stars S235\,A$^*$ and S235\,A-2$^*$ and it decreases systematically to the south of the \hii{} region.

We believe that using the LTE approach is well justified for those regions where we observed both the methanol and C$_2$H lines. The methanol emission probes dense gas, where physical conditions should be close to the LTE. Therefore, we do not expect any noticeable shift of the data points in the middle and bottom part in Fig.~\ref{fig:s235_c2h_ch3oh}.

\subsection{Abundances}

Spatial maps of the molecular abundances are shown in Fig.~\ref{fig:abundances}. The spatial distribution of the small hydrocarbons around S235 has a common appearance. The abundance values $x_{\rm C_2H}\approx 1-2\times10^{-9}$ and $x_{c-{\rm C}_3{\rm H}_2}\approx 3\times10^{-10}$ are higher up to a factor of 2-3 on the borders of the molecular clumps than in the inner parts of the clumps. Looking at the East~1 clump, we find that the abundances of small hydrocarbons are higher on the side of the clump that is illuminated by the stellar UV~field from the ionizing star of S235. The same is observed in the East~2 clump, especially for the $x_{\rm C_2H}$ distribution. Abundance of $c$-C$_3$H$_2$ increases modestly on the illuminated side of East~2 clump in contrast with the C$_2$H abundances. In the Central clump we also find $x_{\rm C_2H}$ up to $2\times10^{-9}$ on its edges and to the south of the IRS~1 and IRS~2 YSOs, and a uniform distribution of $x_{c-{\rm C}_3{\rm H}_2}\approx 1-1.5\times10^{-10}$. The Central (and probably partly East~2) clumps are situated in the back neutral wall of the \hii{} region \citep{Anderson_2019, Kirsanova_2020}, therefore the abundances of the hydrocarbons might not fully trace the changing UV~field with distance from the ionizing star. The minimum values of $x_{\rm C_2H}$ and $x_{c-{\rm C}_3{\rm H}_2}$ are observed in the inner and non-illuminated parts of the molecular clumps.

Abundance of C$_2$H reaches maximum values towards the ionizing star S235\,A$^{*}$ and IR~source S235\,A-2$^{*}$ in the S235~ABC area. The UV radiation from the deeply embedded ionizing star penetrates the molecular clump and leads to formation of the C$_2$H. The C$_2$H molecules are distributed from the north-west to south-east of S235\,AB region, their abundance decreases from $1.8$ to $0.1\times10^{-9}$ in the direction of the dense molecular clump to the south of S235~B$^{*}$, the secondary gas density peak in the region. The value of $x_{\rm C_2H}$ in S235\,C is only $\approx 4\times10^{-10}$, 5--6 times smaller than in the S235\,A region in spite of similar spectral type of the ionizing stars \citep[see also theoretical models of S235\,A and S235\,C by][]{Kirsanova_2020_PDR}. The peak value of $x_{\rm c-C_3H_2} = 2\times10^{-10}$ appears to the north-west of the S235\,A$^{*}$ where the low-density dust tail is observed (compare with Fig.~\ref{fig:general_view_spitzer}). General distribution of $c$-C$_3$H$_2$ is still similar to the C$_2$H distribution: decreasing to the south of S235~B$^{*}$ with the low values in S235\,C of the order of $10^{-11}$. Due to relatively low spatial resolution in comparison with compact size of the \hii{} region S235\,A, we did not resolve irradiated border of the molecular clump here and observe the highest C$_2$H and $c$-C$_3$H$_2$ abundances in the direction of the ionizing star S235\,A$^*$. In the S235 area we are able to resolve the irradiated borders of the molecular clumps.

There are three enhancements of the H$_2$CO abundances (up to $x_{\rm H_2CO} = 1.6\times10^{-9}$) around S235: the north-west illuminated border of the East~1 clump, the north-east of the East~2 clump and the area to the south-west of IRS~1. We assume above that the UV~radiation from the ionizing star of S235 leads to the enhancement of the abundances of the small hydrocarbons on the illuminated border of East~1. We find similar pattern in the distribution of the H$_2$CO abundances. The gas and dust density have irregular distributions in the Central clump, which is situated behind the \hii{} region. The high values of $x_{\rm H_2CO}$ are found in between the peaks of the mm-continuum emission. While $T_{\rm dust}$ value in the Central clump is 10~K ($\approx 50\%$) higher than in the East~1 and East~2 (see Fig.~\ref{fig:temperature_compsrison}), H$_2$CO abundances do not respond to that temperature enhancement. We only see that $x_{\rm H_2CO}$ is highest in the densest part of East~1 clump where $T_{\rm dust}$ drops to 15~K. The peak of $x_{\rm H_2CO} =1.8\times10^{-9}$ appears towards the secondary mm-continuum emission to the south of S235~B$^*$ where $T_{\rm dust} \approx 10-12$~K. The abundances towards the primary mm-continuum peak between S235\,A$^*$ and S235~B$^*$ ($T_{\rm dust} \geq 20$~K) are two times less than in the secondary one and are similar to the abundances in S235.

We find  the peak of $x_{\rm CH_3OH} = 3\times 10^{-8}$ in the northern part of the East~1 cluster, Fig.~\ref{fig:abundances}, i.~e. in the same direction to the bright methanol emission at 85~GHz, see Fig.~\ref{fig:noisemaps}. 
The methanol abundances drop by about one order of magnitude in the Central cluster, where $T_{\rm dust}$ is higher by about 10~K due to higher UV~field. There are two methanol abundance peaks in the S235\,ABC region: the first found towards the mm-continuum peak between S235\,A$^{*}$ and S235~B$^{*}$ and the second coincides with the H$_2$CO abundance peak to the south of S235~B$^{*}$. The primary peak of the formaldehyde abundance corresponds to the secondary peak of the methanol abundance in S235~AB area and vise versa.

\begin{figure*}
    \includegraphics[height=10.cm]{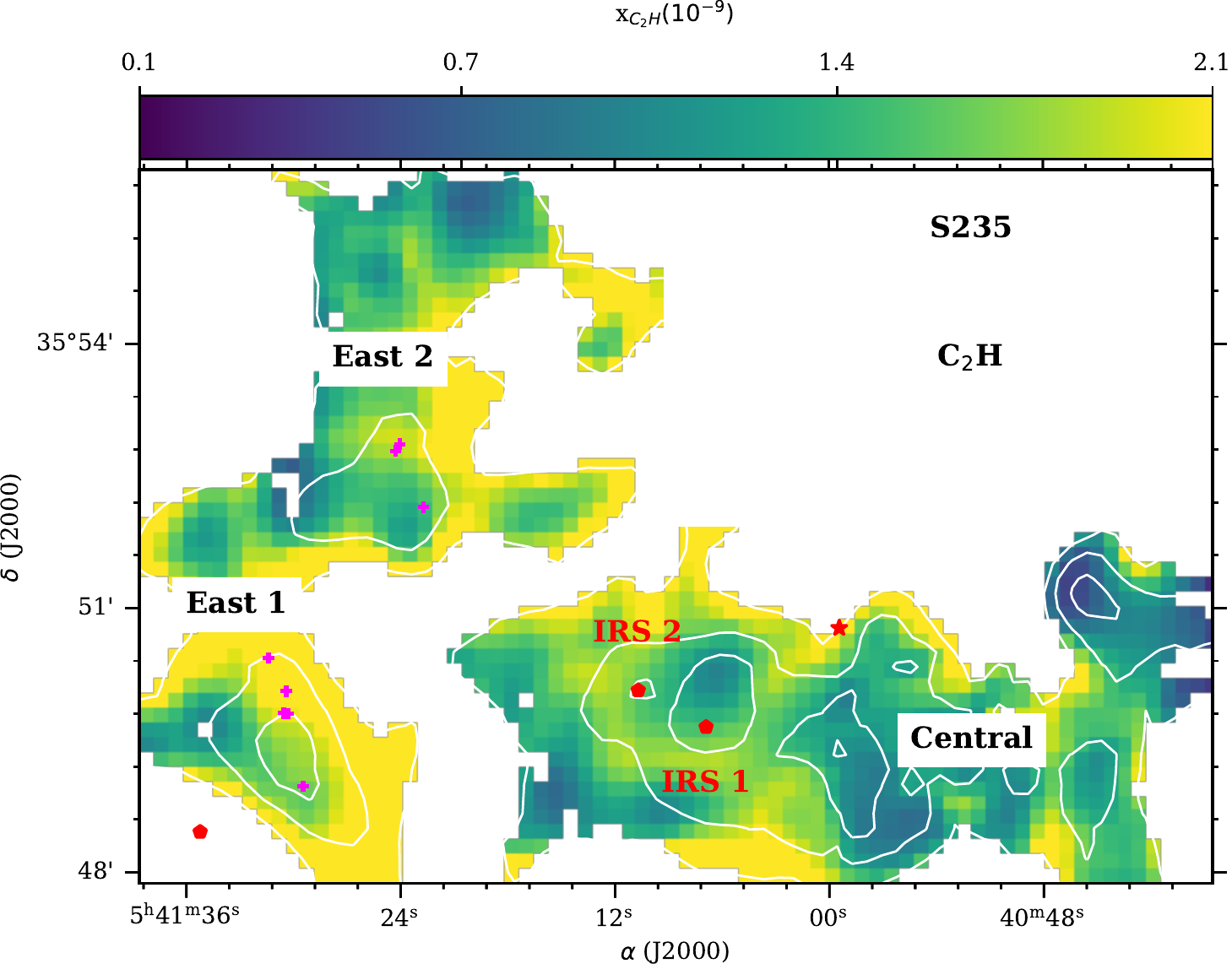}
    \includegraphics[height=10.cm]{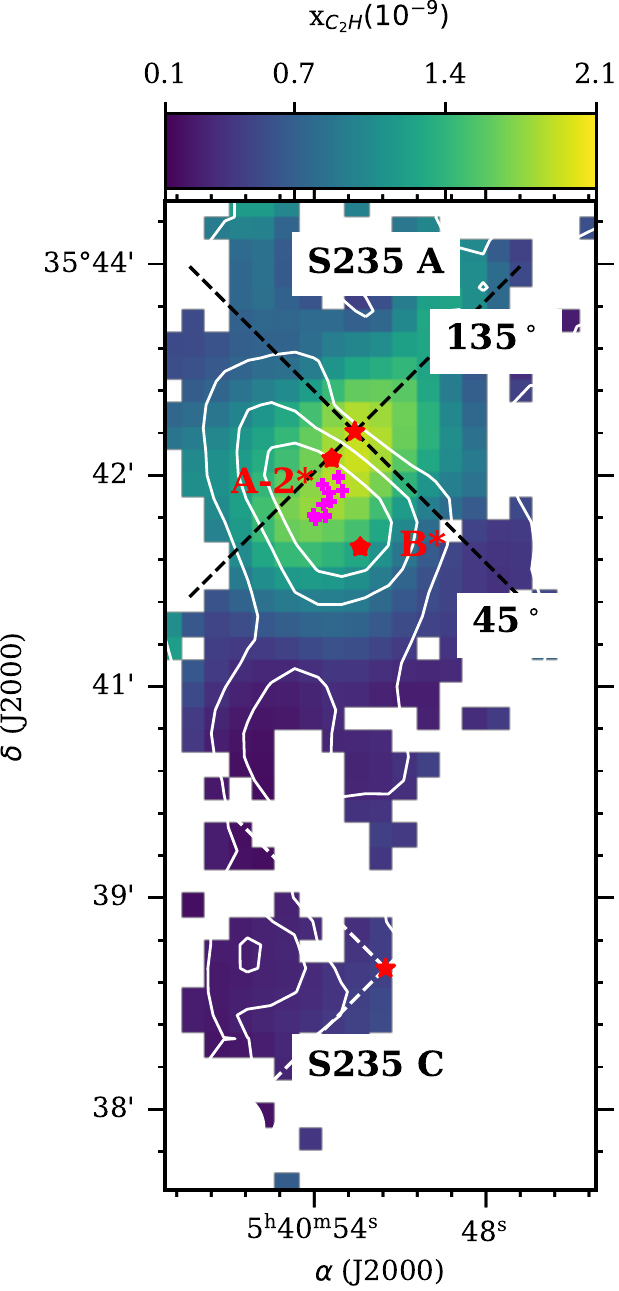}\\
    \vspace{3mm}
     \includegraphics[height=10.cm]{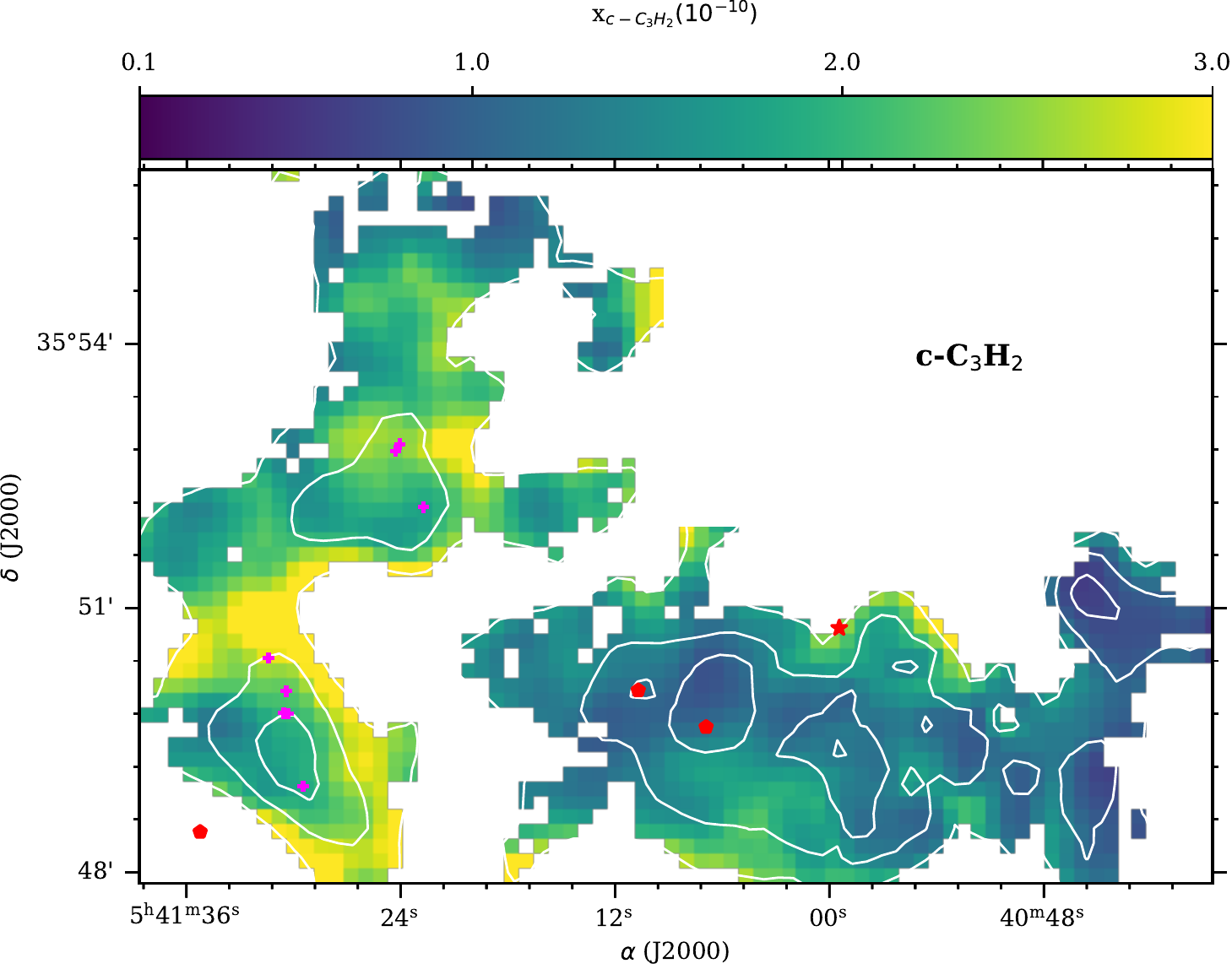}
    \includegraphics[height=10.cm]{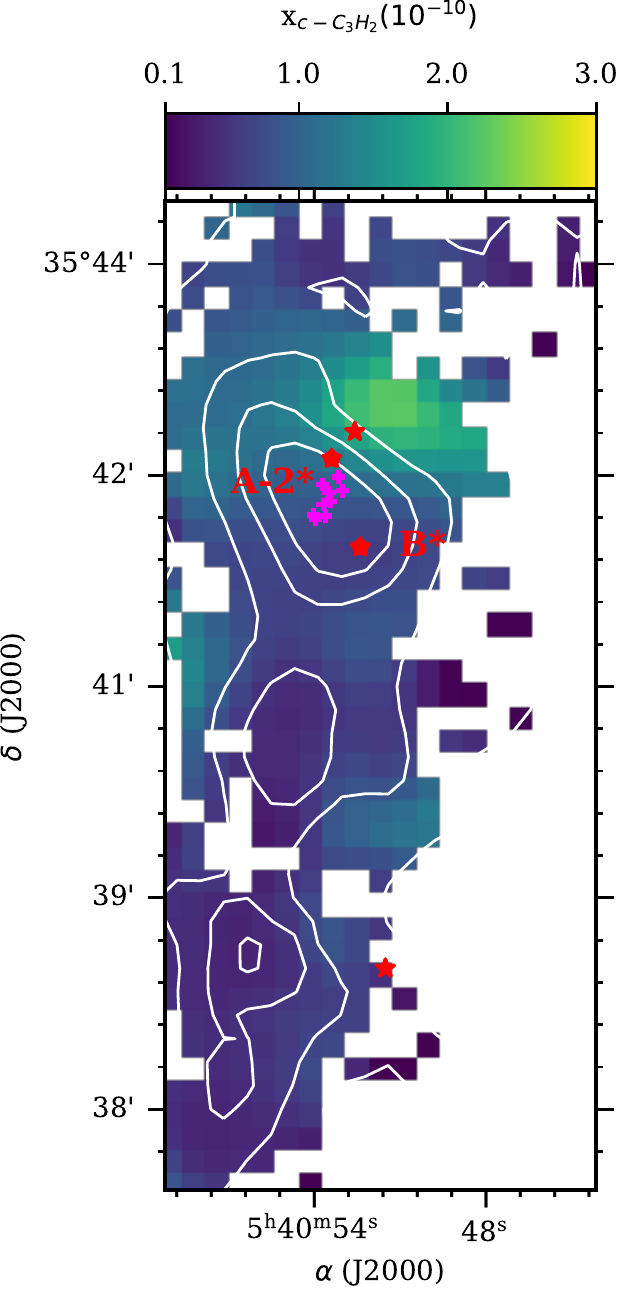}\\
    \caption{Molecular abundances in clouds belonging to S235 (left panels) and S235\,ABC regions (right panels). White contours show levels of hydrogen column density: 1, 2, 3 and $4\cdot 10^{22}$~cm$^{-2}$ based on the CO observations by \citet{Bieging_2016}. The ionizing sources are shown by the red stars. Red diamonds show three bright infrared sources: IRS1, IRS2 \citep{Evans_1981}, S235~East~IR \citep{Kirsanova_2020}, S235~B$^*$ \citep{Boley_2009}. Black dashed lines in the C$_2$H abundance map in the S235\,ABC area are the cuts for the pv~diagrams shown in Fig.~\ref{fig:hydrocarbons_pv}, where the numbers are the angles measured west of south, with the ionizing star at the origin. }
    \label{fig:abundances}
\end{figure*}

\begin{figure*}
\includegraphics[height=10.cm]{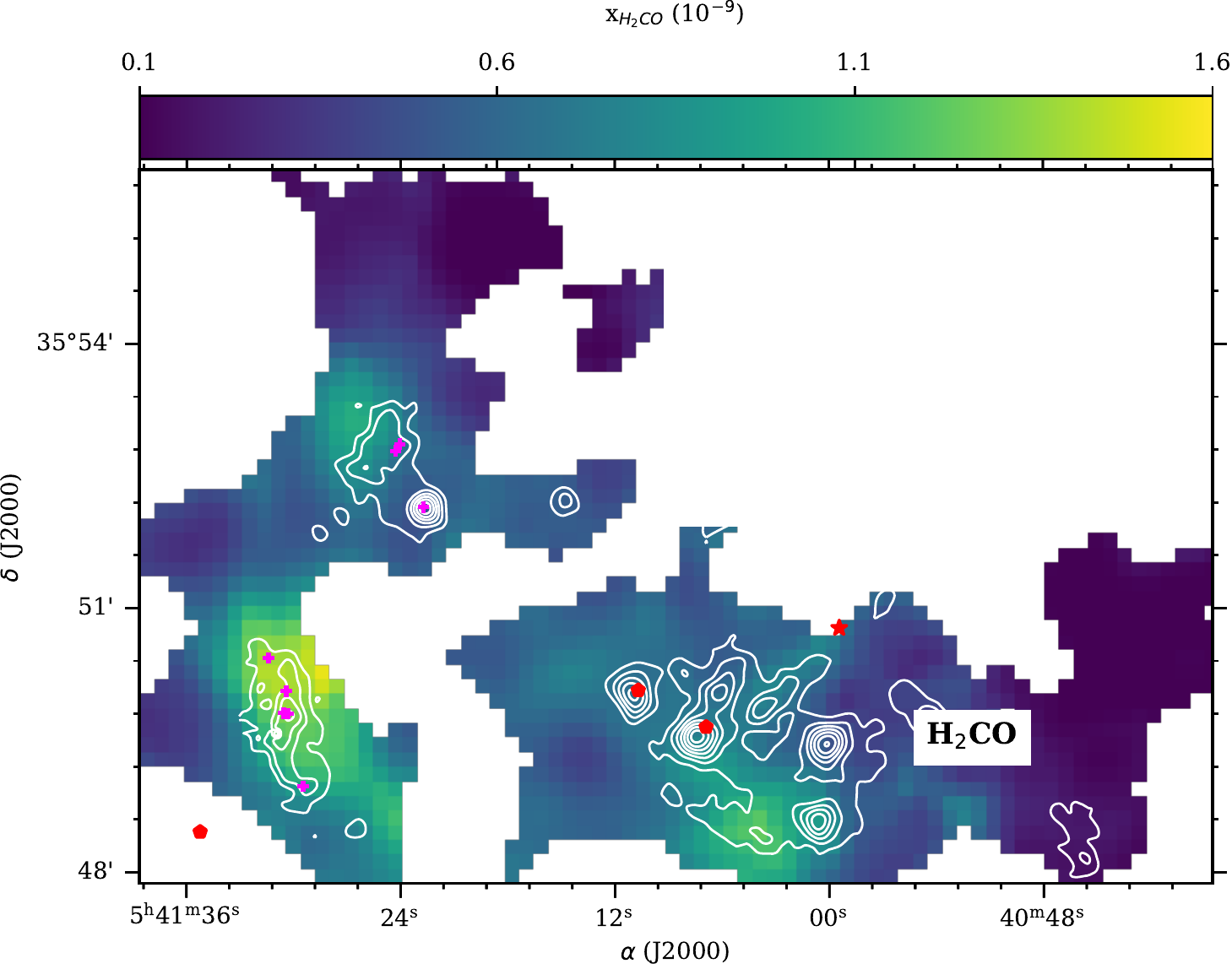}
\includegraphics[height=10.cm]{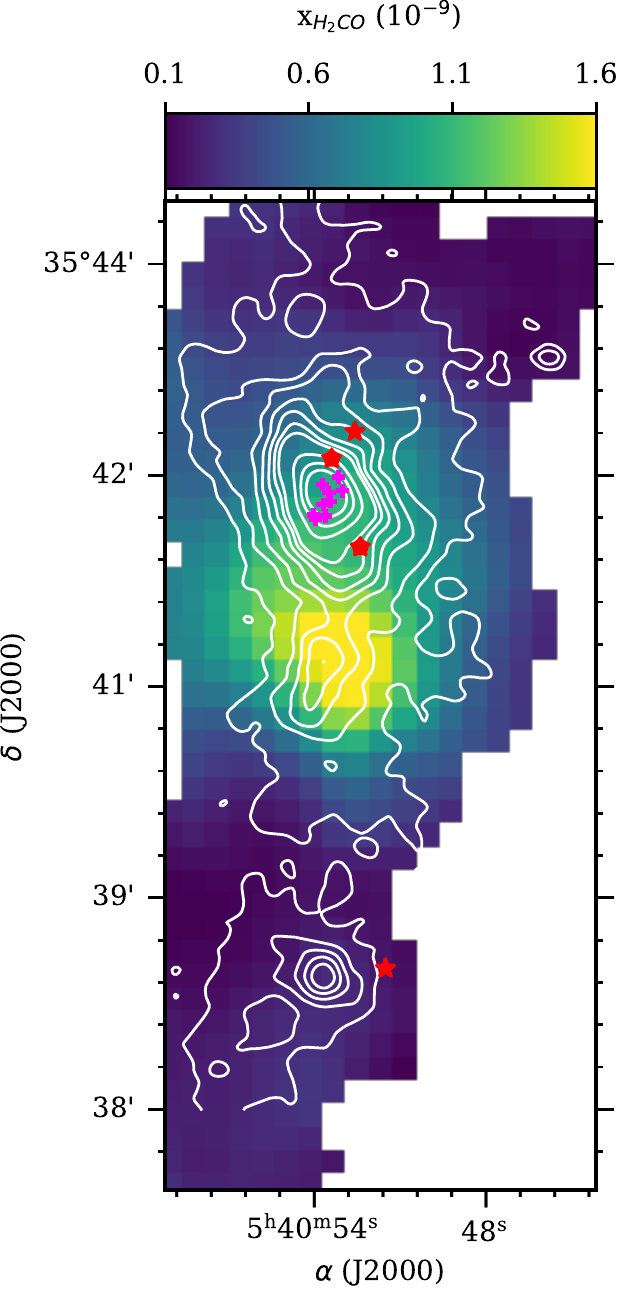}\\
\vspace{3mm}
\includegraphics[height=6.cm]{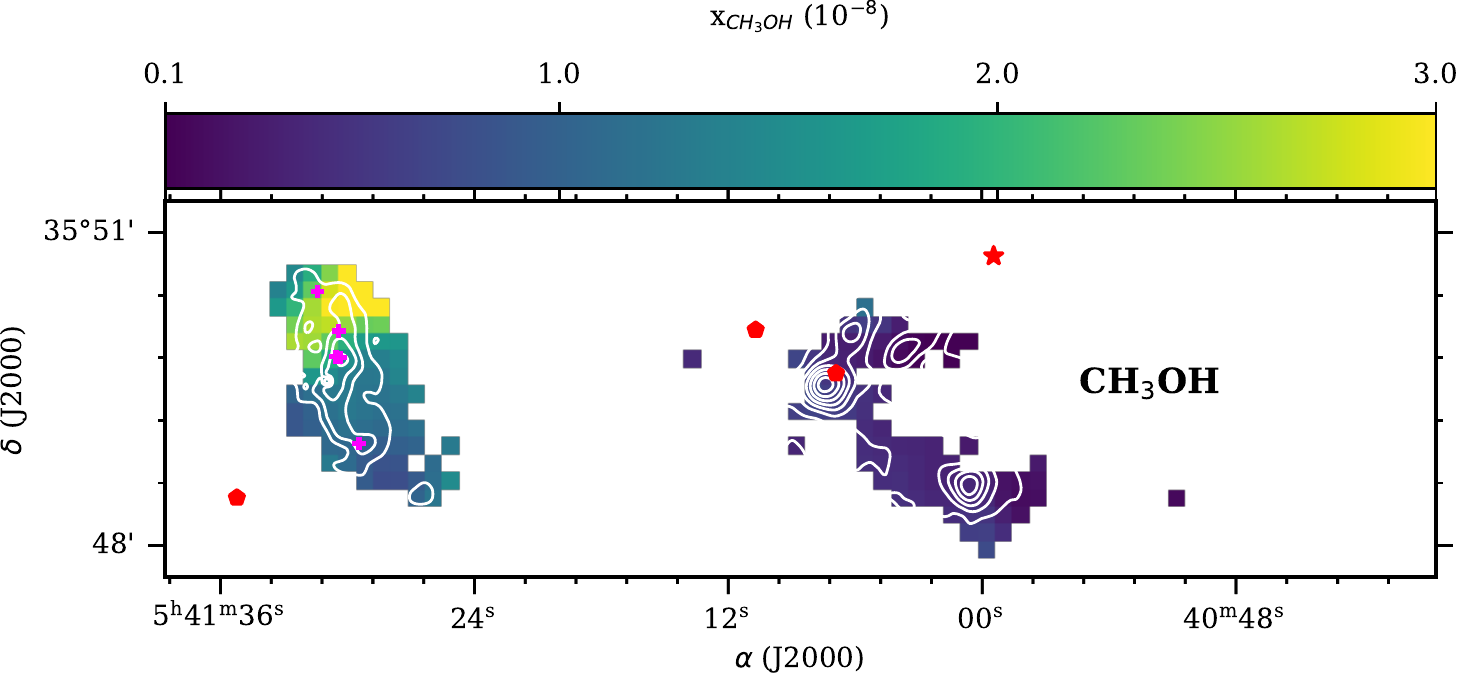}
\includegraphics[height=6.cm]{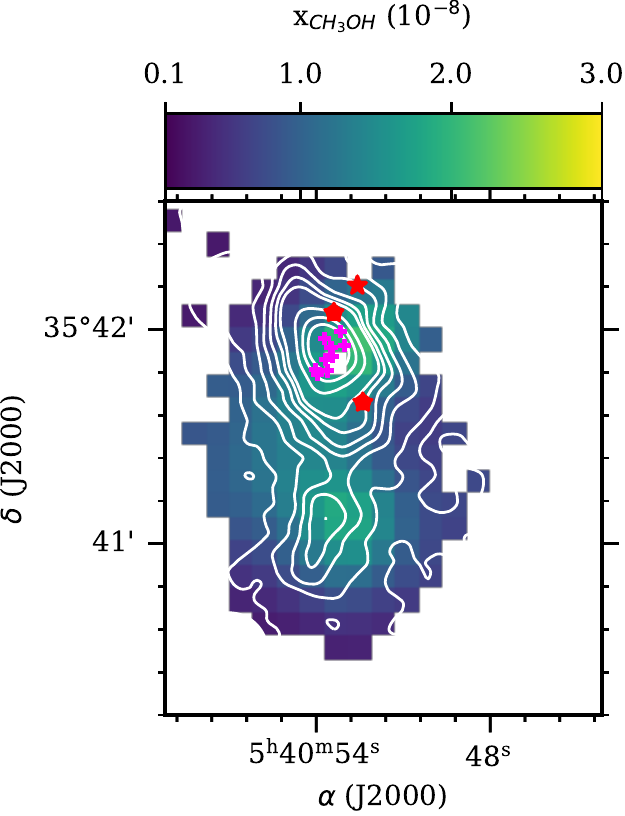}
\contcaption{. White contours show levels of SCUBA-850~\micron{} emission from 5 to 35~mJy~arcsec$^{-2}$ with a step of 5~mJy~arcsec$^{-2}$ and additional contours at 55, 75 and 95~mJy~arcsec$^{-2}$ for S235\,ABC region based on \citet{Klein_2005}.}
\end{figure*}

\section{Discussion}

\subsection{A lack of small hydrocarbons in S235 and S235\,ABC}

Comparing our results with the column densities of the small hydrocarbons and the organic molecules in the Orion Bar~\citep{Leurini_2010,Cuadrado_2015, Cuadrado_2017}, Horsehead PDRs~\citep{Teyssier_2004, Guzman_2013} and other PDRs, \citep[see][]{Bouvier_2020}, we see significant difference between the $N_{\rm C_2H}/N_{\rm CH_3OH}$ values in S235, S235~A and the other PDRs. The $N_{\rm C_2H}/N_{\rm CH_3OH}$ ratio in S235 and S235\,ABC regions (see Fig.~\ref{fig:s235_c2h_ch3oh}) lies in between straight lines $y=x$ and $y=0.1x$, but around $y=10x$ in many other PDRs. The difference arises due to relatively high \Nmeth{} and low \Nethy{} in S235 and S235~ABC in comparison with the other regions up to a factor of few or an order of magnitude. It is possible that relatively low $N_{\rm C_2H}$ arises due to moderate UV~field values in S235 and S235~ABC in comparison with other PDRs such as Orion Bar and Horsehead PDRs. On the other hand, \citet{Buslaeva_2021} found higher values of $N_{\rm C_2H} \approx 10^{14} - 10^{15}$~cm$^{-2}$ in PDRs with comparable UV~field to that of the regions under the study.

The $N_{\rm C_2H}/N_{\rm CH_3OH}$ ratios in S235 and S235~ABC coincide with the values observed towards protostars in Perseus~cloud \citep[see][]{Higuchi_2018} including known hot corinos there. \citet{Sakai_2009,Higuchi_2018,Bouvier_2020} analyzed the $N_{\rm C_2H}/N_{\rm CH_3OH}$ ratios in different types of ISM environments from molecular clouds to PDRs and found that it is a specific characteristic of a cloud illuminated by OB~stars. They assumed that if the starless phase was long ($\sim 10^5-10^6$~years), there would be enough time to form methanol from CO on the dust grains. On the other hand, if the cold phase was short, the protostellar phase would begin with high abundances of pristine chemical species typical of the diffuse gas, such as the small hydrocarbons. Following that model, S235 and S235\,ABC regions should have had enough time for developing the methanol-related surface chemistry. This conclusion is in agreement with the triggered star formation scenario proposed by \citet{Kirsanova_2008, Dewangan_2011, Kirsanova_2014}, who suggested that the star formation around S235 was initiated (accelerated) by the expansion of the \hii{} region on a timescale of $10^5-10^6$~years.

\citet{Cuadrado_2015} presented observed ratios of $N_{\rm C_2H}/N_{c-{\rm C}_3{\rm H}_2}$ vs dissociation parameter $G/n_{\rm H}$ for many of PDRs. Their compilation of the PDR observations predicts that $N_{\rm C_2H}/N_{c-{\rm C}_3{\rm H}_2}$ should be about 15--20 for $0.01 \leq G/n_{\rm H} \leq 0.1$ and around 10 for the lower values of the dissociation parameter. Using values of $n_{\rm H_2}$ from \citet{Kirsanova_2014} and assuming that all hydrogen is in H$_2$, we find the maximum dissociation parameter $0.01 \leq G/n_{\rm H} \leq 0.1$ in the Central clump and $0.001 \leq G/n_{\rm H} \leq 0.01$ in the East~1, East~2 clumps as well as in the S235\,ABC area. Therefore, $N_{\rm C_2H}/N_{c-{\rm C}_3{\rm H}_2}$ in the studied regions is in agreement with the results of \citet{Cuadrado_2015} and the behaviour of the ratio can be explained by the 4 times higher photodissociation rate of $c$-C$_3$H$_2$ than that of C$_2$H \citep[][]{2017A&A...602A.105H}.

\subsection{Photodesorption of methanol ice in PDRs}\label{sec:photodes}

We find in Fig.~\ref{fig:s235_c2h_ch3oh} that the values of $G=20-30$~Habings favour the appearance of methanol molecules in dense molecular clumps. Instead, amount of methanol decreases in the regions with the higher and lower UV~field due to interplay of the formation and destruction process. Methanol can be formed efficiently only via grain-surface reactions by the subsequent hydrogenation of CO molecules accreted from the gas phase~\citep[][]{Watanabe_2003}. \citet{Hasegawa_1992, Charnley_1997, Garrod_2007, Watanabe_2002, Vasyunin_2013} found that the most efficient methanol formation occurs at the dust temperature range of $\approx 10-20$~K in a number of experimental and theoretical studies. The grain-surface reactions and exothermic, and the fraction of released energy can be spent on braking the bond between newly formed methanol molecule and grain surface. Thus, methanol can be released to the gas phase even at low dust temperature via reactive (or chemical) desorption. Reactive desorption in cold and dense clumps is believed to be the most efficient way to deliver methanol to the gas phase \citep[see e.g.,][]{Garrod_2007, Vasyunin_2013, Vasyunin_2017, 2018ApJ...853..102C}. Reactive desorption can be responsible for the gas-phase methanol, observed in the regions with low $T_{\rm dust}$ and $G$: East~1 clump and the secondary density peak to the south from S235~B. The rate of methanol formation on the grains decreases at higher $T_{\rm dust}$ values because atomic hydrogen escapes from grain surfaces (it has low binding energy). Based on our analysis, we suggest that methanol cannot form efficiently on the grain surfaces in the Central clump ($G=50-60$~Habings which corresponds to $T_{\rm dust} \approx 25$~K for a grain radius of 0.1~\micron) or it is UV-dissociated in the gas phase there. We do not see a direct relation between the $G$ value and the column densities or abundances of methanol in the studied regions. This suggests that photodesorption of  methanol ice from the dust grains irradiated by a moderate UV~field is not the main pathway towards the gas-phase methanol.

Physical conditions favour methanol formation on grains surfaces and its survival in the gas phase in dense molecular gas between S235\,A and S235~B and also in the north of the East~1 clump. As stated above, the temperature range of $10 < T_{\rm dust} <20$~K is optimal for the methanol formation. The UV~field values $G=20-30$~Habings observed in these regions correspond to $T_{\rm dust} \approx 20$~K. Moreover, high number density found towards the mm-continuum peaks favours surface chemistry. After reactive desorption, methanol is able to survive in the gas phase due to the low UV~field during $10^9/n_{\rm H_2} \approx 10^4-10^5$ years. {Therefore, our conclusion about the efficient formation and survival of methanol in the dense clumps does not contradict the scenario of the triggered star formation in S235 because that chemical timescale is shorter/comparable than the estimated duration the star formation process \citep[][]{Kirsanova_2014}.}

Survival of methanol on grain surfaces in the cold regions and its evaporation due to chemical desorption around the young massive protostars suggest that the observed environments in S235 and S235\,ABC are more evolved than we thought before. The massive stars were formed in the region, resetting the chemical composition of the parental molecular cloud. We find "younger" chemical stage on the borders of the clumps irradiated by the massive stars, where the abundance of the small hydrocarbons increases. The "older" stage, where methanol is effectively formed on the grain surfaces and evaporates to the gas phase via chemical desorption, is still observed in the inner densest regions of the clumps.

\subsection{Gas kinematics traced by the small hydrocarbons}

\begin{figure*}
\includegraphics[width=0.99\columnwidth]{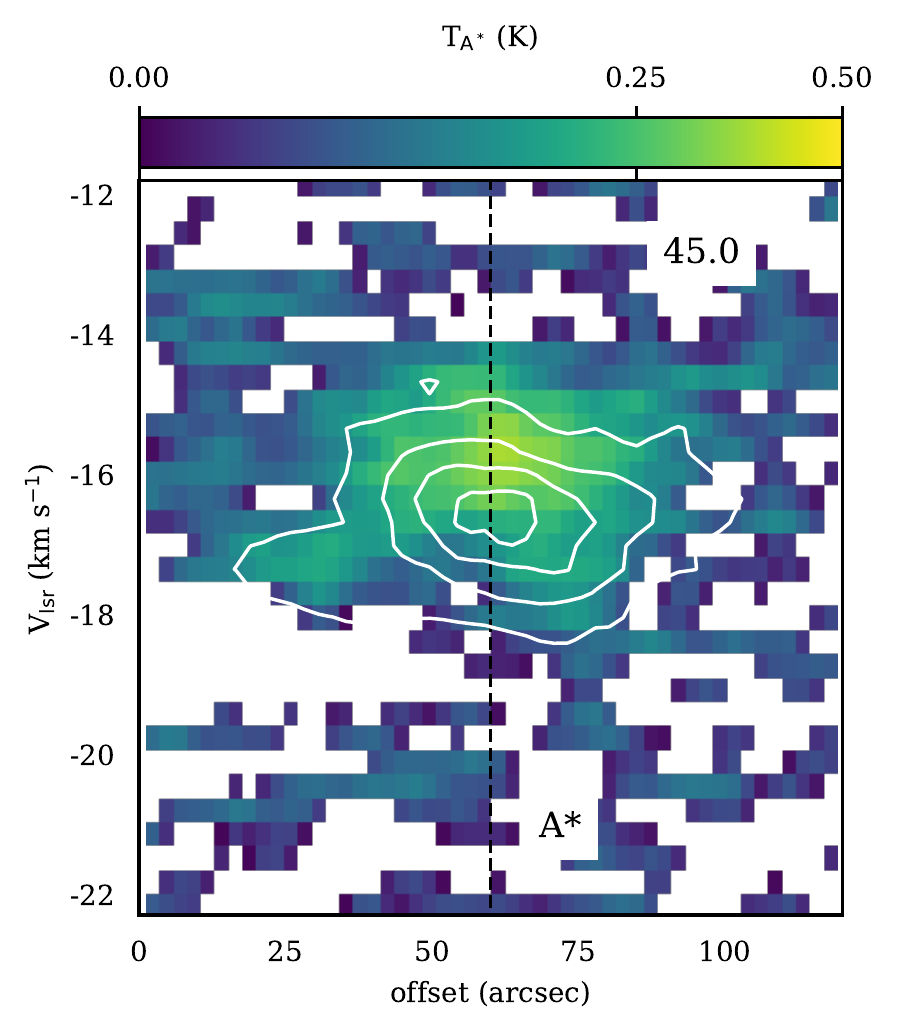}
\includegraphics[width=0.99\columnwidth]{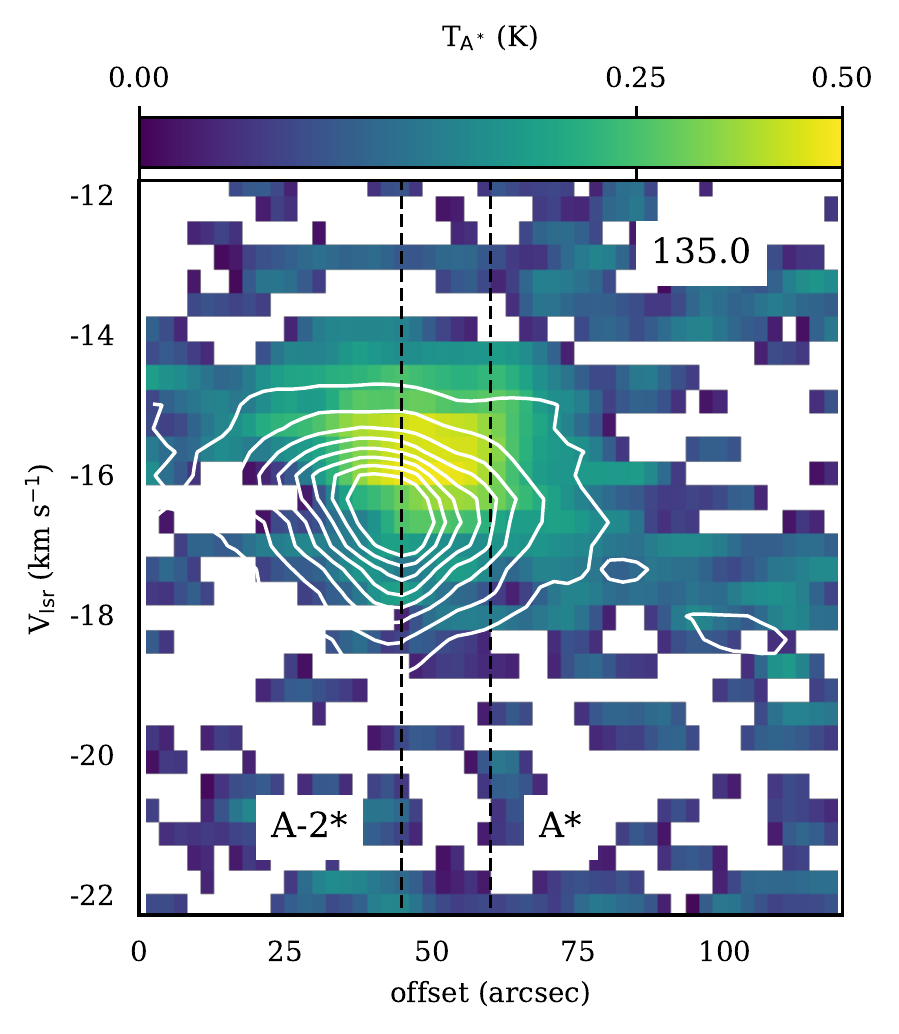}
\caption{The pv~diagrams of the $c$-C$_3$H$_2$ (colour) and C$_2$H (white contours) emission in S235\,A. The contours are from 0.25 to 2.0~K~\kms{} with a step of 0.25~K. Dashed vertical lines correspond to positions of the point sources S235\,A$^*$ and S235\,A-2$^*$.
}\label{fig:hydrocarbons_pv}
\end{figure*}

We find different spatial distribution of the C$_2$H and $c$-C$_3$H$_2$ abundances in the Sec.~\ref{sec:abund}, while both molecules are known as tracers of warm UV-irradiated gas. The C$_2$H abundance reaches its maximum in the dense region towards the ionizing star S235\,A$^*$, while the peak of $c$-C$_3$H$_2$ abundance is observed in the low-density part in the north-west of the S235\,A PDR, where the dense envelope breaks up, see Fig~\ref{fig:general_view_spitzer}. The broken envelope looks as a truncated arc in the 3.6~\micron{} image. The UV~radiation is expected to leak from the \hii{} region in this direction. In order to examine whether the molecular gas also escapes from the PDR there, we plot a position-velocity (pv) diagram in two directions shown by the dashed lines in Fig.~\ref{fig:abundances}. These directions cut the S235\,A PDR along and across the broken arc visible in the 3.6~\micron{} image. Looking to the 135~degrees pv~diagram in Fig.~\ref{fig:hydrocarbons_pv}, we find the broad C$_2$H and $c$-C$_3$H$_2$ lines in the dense part of the PDR between S235\,A-2$^*$ and S235\,A$^*$. In the low-density part of the PDR, at the offset~$\approx 100$\arcsec, we find double-component $c$-C$_3$H$_2$ lines while the C$_2$H(1--0) line remains with single component. The double-component line with a dip at $V_{\rm lsr}=-16$~\kms{} might be a signature of the expanding PDR, where the front and the rear walls move to the opposite directions. The expansion velocity $V_{\rm exp} \approx 2$~\kms{} can be obtained as a peak-to-peak value divided by a factor of two. The C$_2$H emission, while it is brighter than the $c$-C$_3$H$_2$, traces only the front side of the expanding envelope. The 45~degrees pv~diagrams, plotted through the dense part of the PDR, do not show the same signature of the expansion. We find a velocity gradient from the red in the east to the blue in the west. The same gradient was found for [C\,{\sc ii}],  [$^{13}$C\,{\sc ii}], and the HCO$^+$(3--2) line emission by \cite{Kirsanova_2020_PDR}.

It is interesting to compare the 135 degrees pv~diagram of $c$-C$_3$H$_2$ with the pv~diagrams of the [C\,{\sc ii}] and HCO$^+$(3--2) emission (see \citet{Kirsanova_2020_PDR} and Fig.~\ref{fig:pvdiag}.) While the dip at $-16$~\kms{} in the [C\,{\sc ii}] pv~diagram is related to the self-absorption effect in the dense part of the PDR (offsets around 50-75\arcsec), the [C\,{\sc ii}] pv~diagram at the offset~$\approx 100$\arcsec{} is similar to the C$_2$H(1--0) diagram: both of them have single-component line profiles at $V_{\rm lsr}=-18$~\kms. Therefore, while the [C\,{\sc ii}] pv-diagram does not trace the gas kinematics in the dense PDR due to high optical depth, it shows the real gas velocity in the less dense part of the PDR.

\cite{Kirsanova_2020_PDR} found that morphology of the S235\,A PDR is far from the spherically-symmetric shape. They compared velocities of the HCO$^+$(3--2) and [$^{13}$CII] lines and argued that the S235\,A PDR expands to the observer in its dense part with velocity up to 1~\kms. We propose that the low-density part of the PDR expands towards the observer and outwards with the velocity higher by a factor of two. The kinematic features of the high-density and the low-density part of the PDR are consistent with the model of a blister-type expanding \hii{} region. This type of \hii{} regions appear when a massive star forms close to the surface of a molecular cloud~\citep[see e.~g. studies of various \hii{} regions in][and many others]{1978A&A....70..769I, 1985A&A...152..387M, 1997A&A...320..972M, 2000Ap&SS.272..169F, 2005ApJ...627..813H, 2007A&A...464..995P, 2010AJ....140..985O, 2011MNRAS.410.1320R, 2017MNRAS.464.4835O}. We conclude that the pv~diagrams of the small hydrocarbon lines can be used to study gas kinematics in low-density PDRs along with the pv~diagrams plotted with the [C\,{\sc ii}] and CO lines \citep{Pabst_2019, Goicoechea_2020, Pabst_2020}.

The different spatial distribution and kinematic features of the C$_2$H and $c$-C$_3$H$_2$ lines were also found in deeply embedded low-mass protostars, \citep[e.g.][]{Murillo_2018}. They suggested that the distribution of the $c$-C$_3$H$_2$ line emission traces the outflow cavities, and C$_2$H might also trace surrounding envelope but not the cavity.  \citet{Cuadrado_2015} related the distribution of C$_2$H and $c$-C$_3$H$_2$ with different dissociation rates: the rate is higher for $c$-C$_3$H$_2$ than for C$_2$H. We also see the higher values of $x_{\rm c-C_3H_2}$ away from the ionizing star S235\,A$^*$, but the peak of the $x_{\rm C_2H}$ abundance is projected directly to this star. In our case, observations of the hydrocarbon lines with higher spatial resolution could help to clarify if $c$-C$_3$H$_2$ survives in the cavity away from the ionizing star.

\section{Conclusions}

We conclude our study of molecular emission in the PDRs around extended and compact \hii{} regions S235, S235\,A and S235\,C with the following results.

\begin{itemize}
    \item Emission of the small hydrocarbons C$_2$H and $c$-C$_3$H$_2$ appears in a broad environment of the PDRs over the whole observed areas in the plane of the sky. On the other hand, emission of H$_2$CO and CH$_3$OH molecules comes from the dense gas clumps irradiated by the PDRs. The methanol emission is less extended in comparison with the formaldehyde emission.
    
    \item Column densities of the small hydrocarbons are proportional to each other. The absolute values of \Nethy{} lie between $2\times10^{12}$ and 10$^{14}$~cm$^{-2}$, where $N_{c-{\rm C}_3{\rm H}_2} < N_{\rm C_2H} < 20\times N_{c-{\rm C}_3{\rm H}_2}$. In the compact and embedded PDR S235\,A we find that the column densities of the small hydrocarbons are proportional to the UV~field intensity. However, we do not see similar correlation between the column densities in the extended PDR S235 that has a stronger UV~field. We conclude that even a moderate UV~field can unlock carbon atoms from their reservoirs (e.g. CO or PAH or C-dust), those are then rapidly converted to hydrocarbons via gas-phase reactions. After that, the column densities do not scale strongly with an increasing intensity of the UV~field. The ratio of column densities $N_{\rm C_2H}/N_{c-{\rm C}_3{\rm H}_2}$ reaches its maximum values towards the UV sources. We explain this behaviour by a higher rate of the photodissociation for $c$-C$_3$H$_2$ than for C$_2$H. The peak abundance of $c$-C$_3$H$_2$ is observed in the low-density less irradiated part of the S235\,A PDR, where expanding motion of the neutral material is observed. Comparing abundances of the small hydrocarbons in the direction of the dense clumps and their irradiated edges, we found that the abundances of C$_2$H and $c$-C$_3$H$_2$ are higher towards the edges.
    
    \item The highest methanol column density up to $10^{15}$~cm$^{-2}$ appears in the directions with the UV~field intensities of $G\approx 20-30$~Habings. We propose that reactive desorption is responsible for the abundant gaseous methanol in the PDRs with moderate UV~fields. The ratio $N_{\rm H_2CO} / N_{\rm CH_3OH} \leq 1$ is consistent with the dark cloud chemistry in the majority of the observed positions. 
    
    \item The $N_{\rm C_2H}/N_{\rm CH_3OH}$ ratio in the studied PDRs is lower by a factor of 25 in comparison with the values measured towards the Orion Bar and the Horsehead PDRs, and are similar to the values observed in the hot corinos in Perseus. This means that the cold gas chemistry prevails over the irradiated-gas chemistry in both S235 and S235\,ABC regions. The PDRs likely inherited molecular abundances from the previous dark stage of molecular cloud evolution, which should have been long enough ($\gtrsim 10^5$~years) to form methanol ice from CO ice on the dust grain surfaces.
\end{itemize}

\section*{Acknowledgements}
We are thankful to S.~V.~Kalenskii and S.~V.~Salii for their helpful comments about molecular spectroscopy, to R.~Klein for providing the SCUBA mm-continuum emission map and also to anonymous reviewer for his/her comments and suggestions. This study is supported by the Russian Science Foundation, grant number 18-12-00351. This research has made use of NASA's Astrophysics Data System Bibliographic Services; SIMBAD database, operated at CDS, Strasbourg, France~\citep{Wenger_2000};  Aladin web page~\citep{2000A&AS..143...33B}; Astropy, a community-developed core Python package for Astronomy~\citep{Astropy_2013}; APLpy, an open-source plotting package for Python \citep[http://aplpy.github.com][]{APLpy_2012}. 

\section*{Data Availability}

The data underlying this article are available in Zenodo, at \url{https://doi.org/10.5281/zenodo.5176393}.

\bibliographystyle{mnras}
\bibliography{refs} 

\newpage
\appendix

\section{Noise level maps}\label{app:noise}

\begin{figure}
    \centering
    \includegraphics[height=4.7cm]{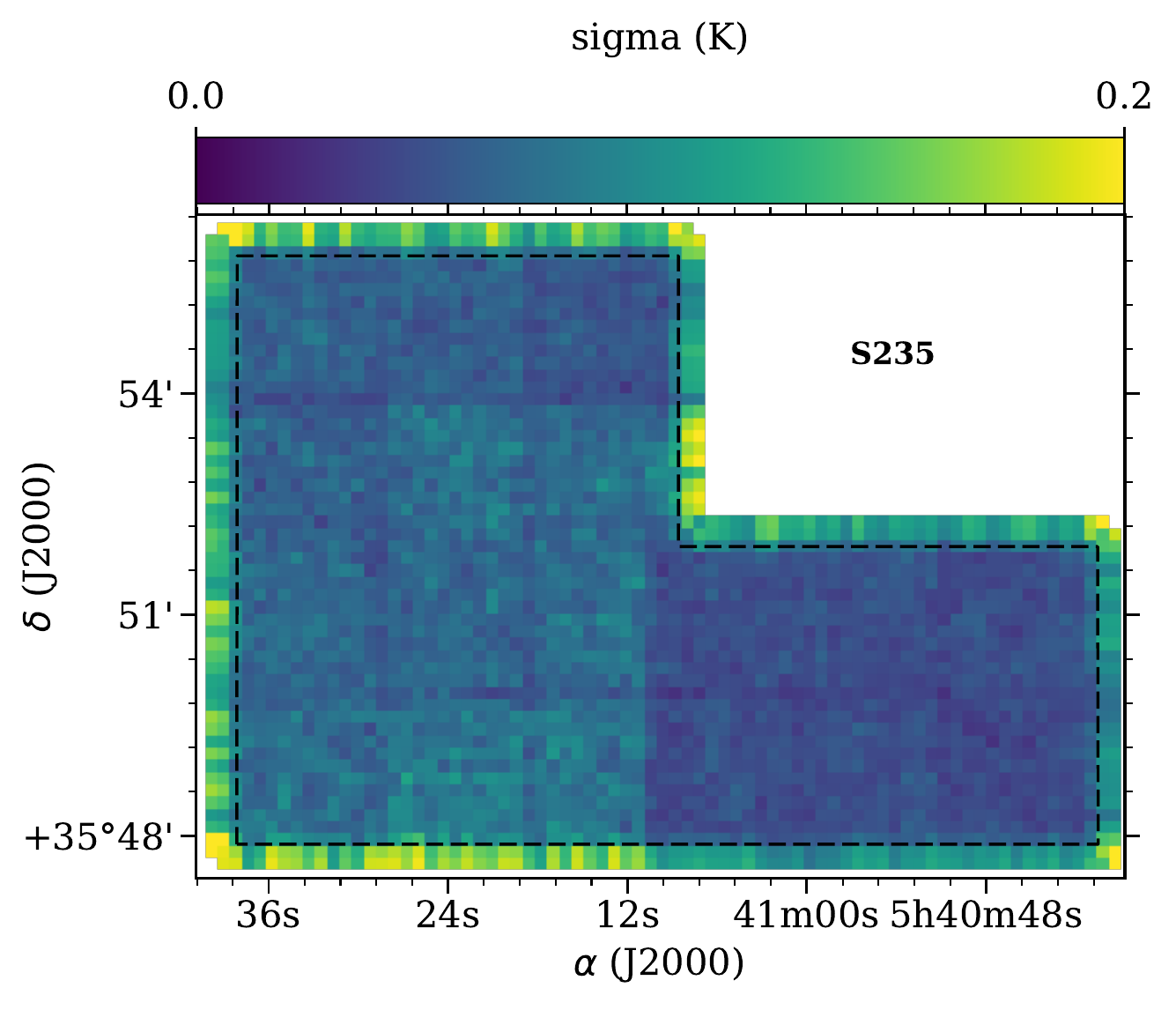}
    \includegraphics[height=4.7cm]{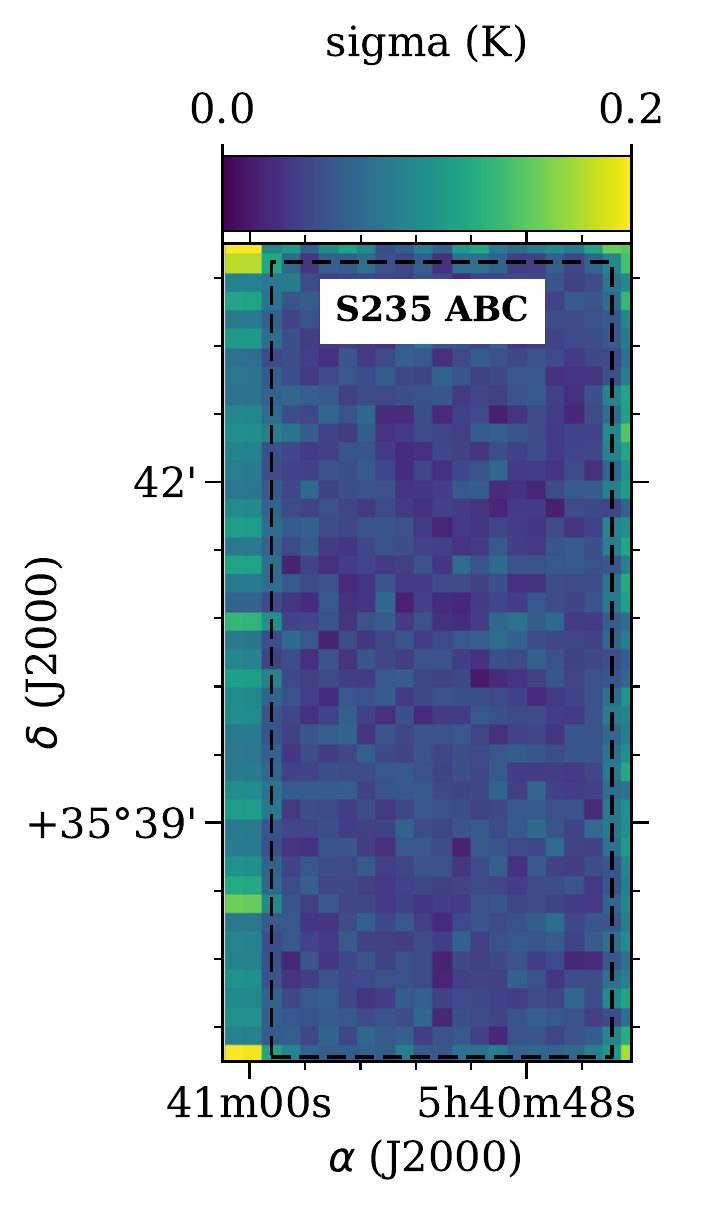}\\
    \includegraphics[height=4.7cm]{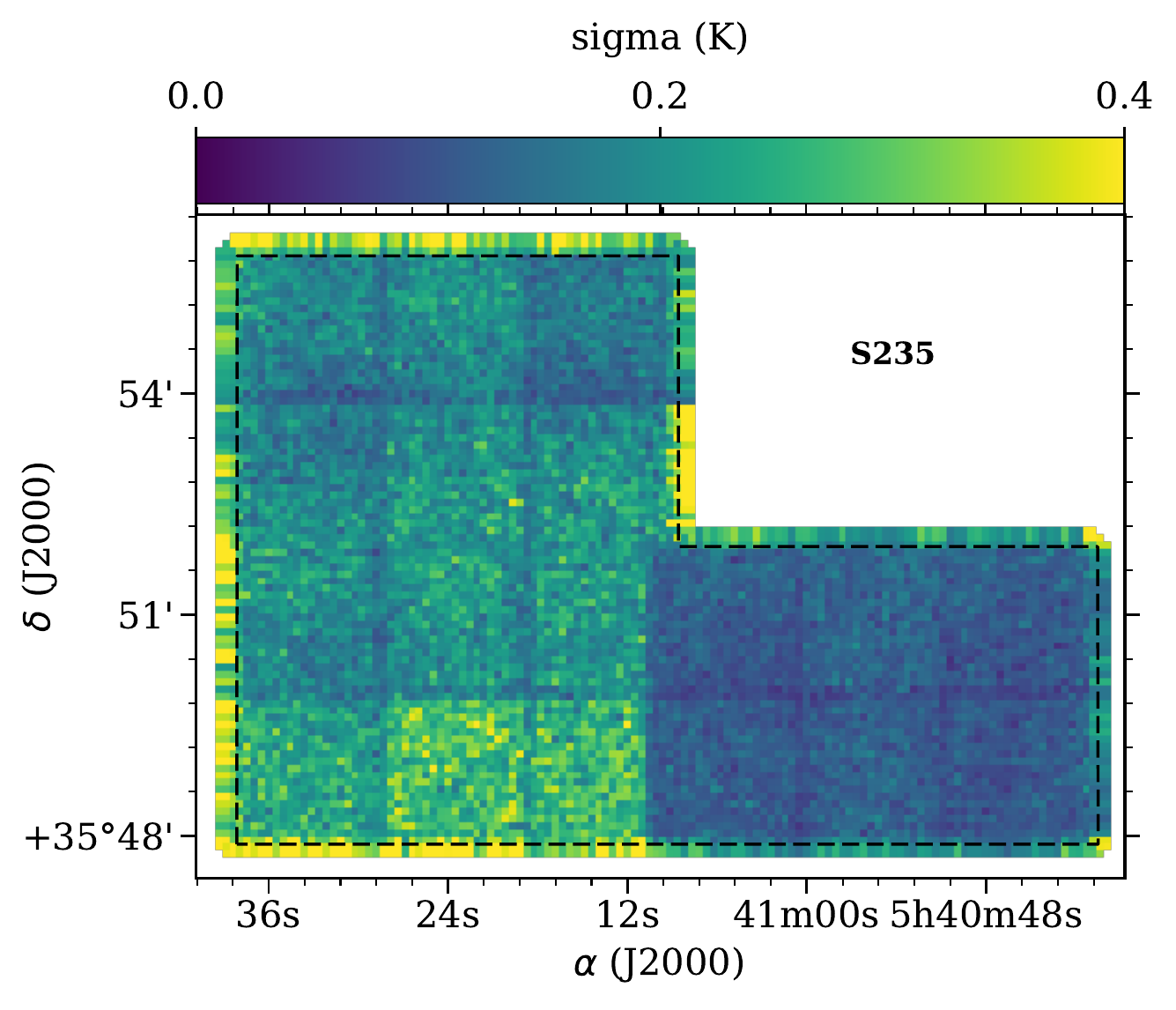}
    \includegraphics[height=4.7cm]{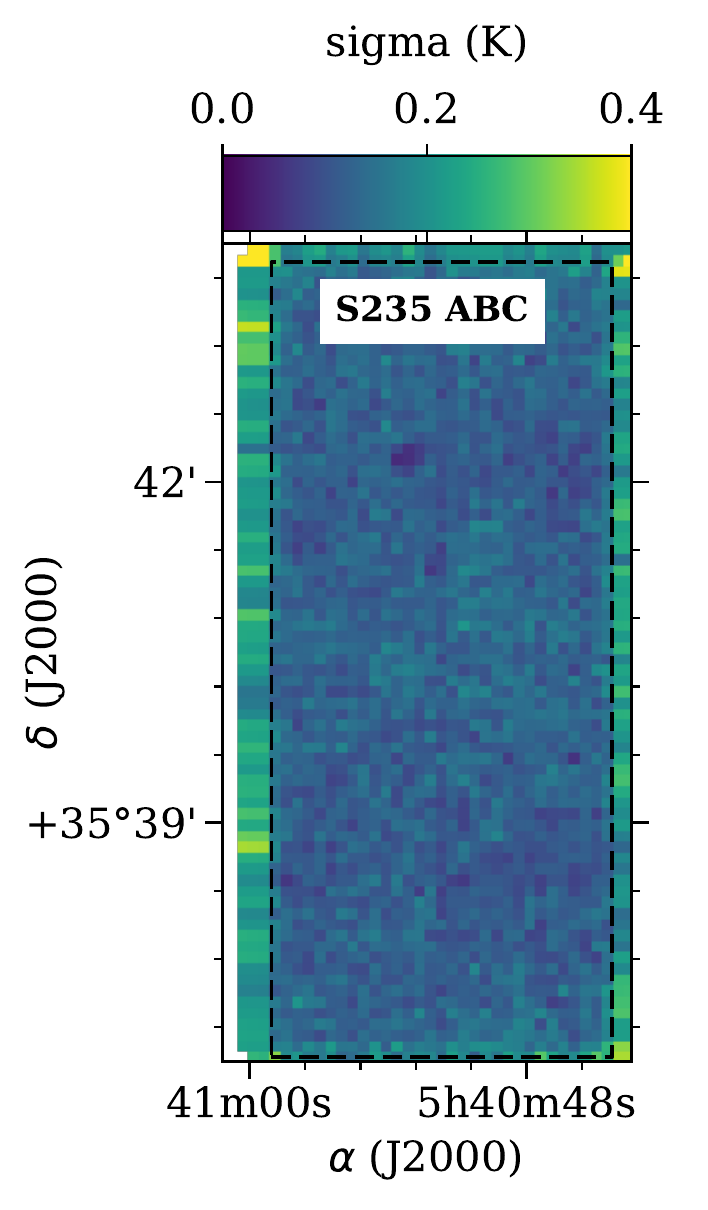}\\
    \caption{Noise maps ($1 \sigma$) at 3~mm (top) and 2~mm (bottom) in the S235 and S235\,ABC regions. Black dashed lines show the area mapped by the IRAM~30-m telescope. The temperature scale is in T$_{\rm A^*}$ units.}
    \label{fig:noisemaps}
\end{figure}

\section{Partition functions}\label{app:Qsprot}

Here we give partition functions from CDMS~\citep{Muller_2001}, which we used to calculate molecular column densities.

\begin{table}
    \centering
    \begin{tabular}{c|c|c|c}
    \hline
    $T$ (K) & \multicolumn{3}{c}{$Q$} \\
                & CH$_3$OH & H$_2$CO & c-C$_3$H$_2$ \\
                & A+E      & ortho   &  ortho   \\
    \hline
    300.0  & 37027.3292 & 2274.3609 & 3216.1633  \\ 
    225.0  & 20991.8100 & 1502.3796 & 2094.2004  \\ 
    150.0  & 9750.0398 & 846.3526 & 1146.1277  \\ 
    75.00  & 2924.3023 & 332.0760 & 412.1865  \\ 
    37.50  & 920.9637 & 144.4132 & 150.8781  \\ 
    18.75  & 274.9880 & 72.6702 & 57.1980  \\ 
    9.375  & 78.1736 & 40.5840 & 23.2568  \\ 
    5.000  & 26.7190 & 26.2131 & 11.5647  \\ 
    2.725  & 11.8899 & 19.5632 & 7.0058  \\ 
    \hline 
    \end{tabular}
    \caption{Rotation-spin partition functions taken from CDMS}
    \label{tab:partition_functions}
\end{table}

\section{Rotational diagrams for the observed methanol lines}\label{app:rotdiag}

\begin{figure*}
    \centering
    \includegraphics[width=0.65\columnwidth]{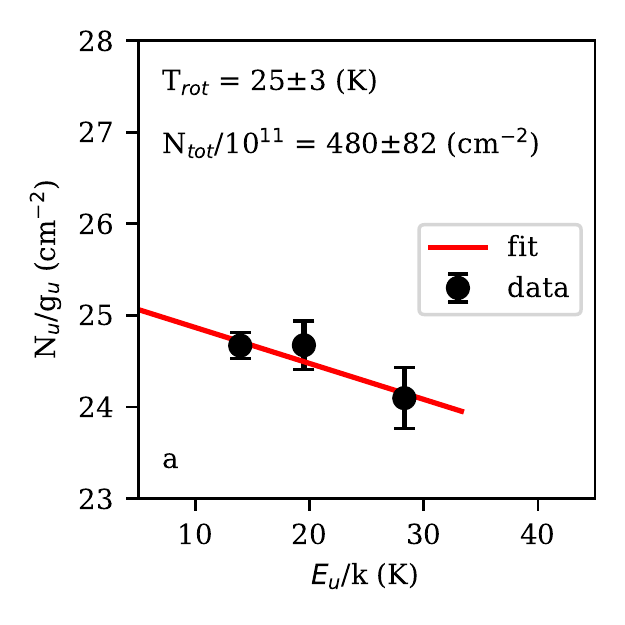}
    \includegraphics[width=0.65\columnwidth]{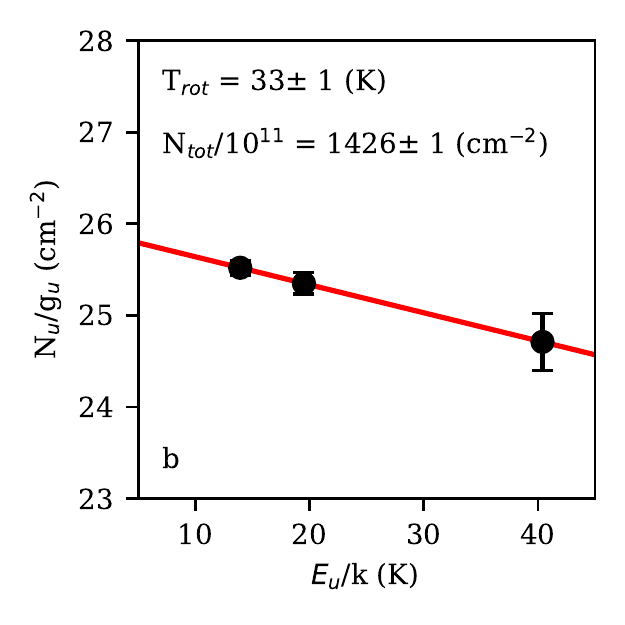}
    \includegraphics[width=0.65\columnwidth]{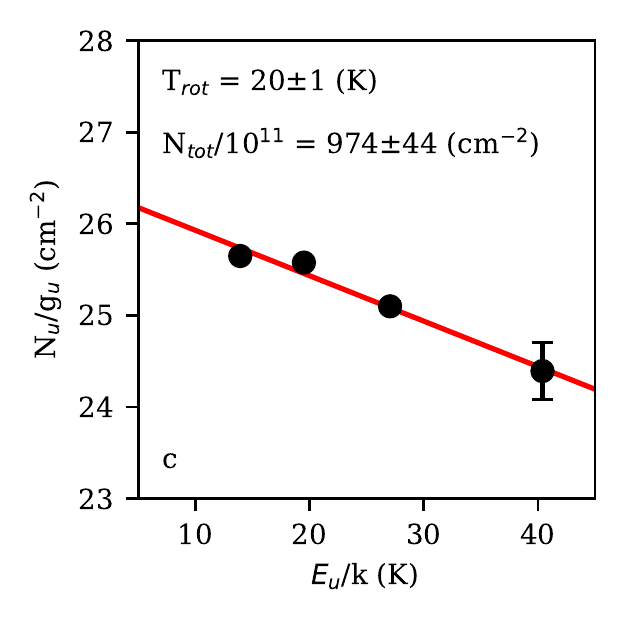}\\
    \includegraphics[width=0.65\columnwidth]{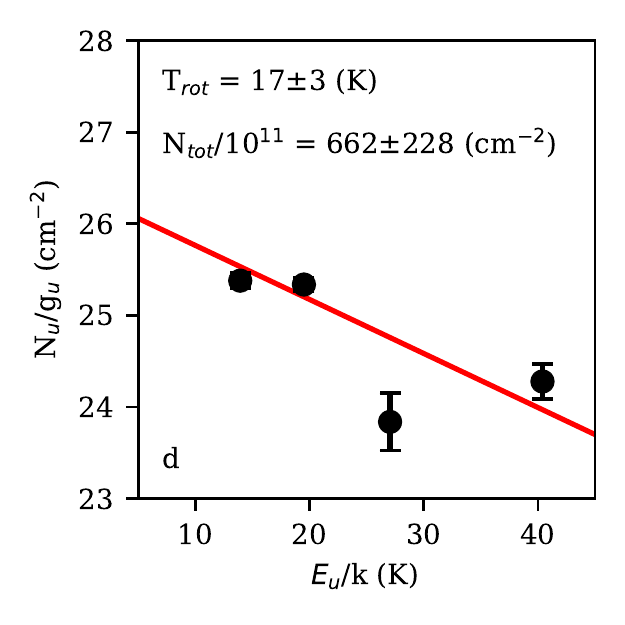}
    \includegraphics[width=0.65\columnwidth]{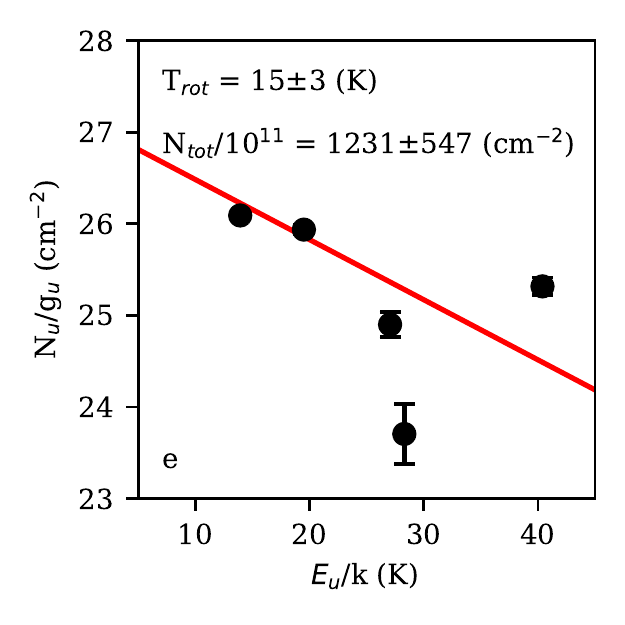}
    \includegraphics[width=0.65\columnwidth]{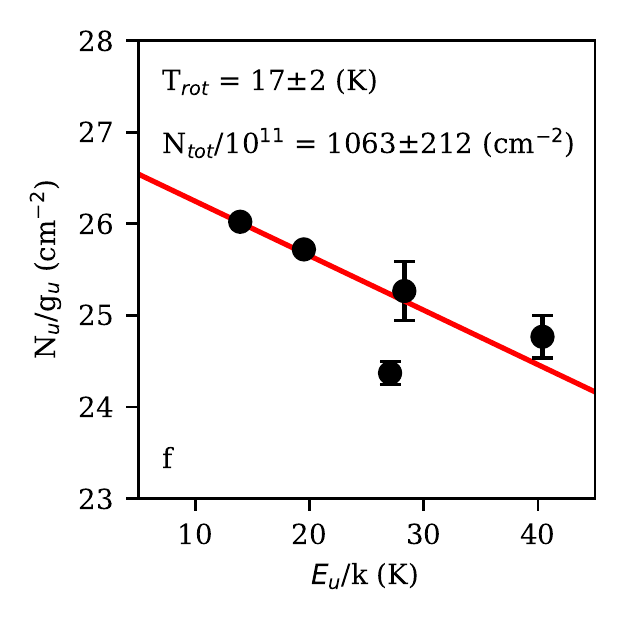}
    \caption{Rotation diagrams with a linear fit over the CH$_3$OH(3$_K$--2$_K$) and (5$_{-1}$--4$_0$) lines. The integrated intensities of the lines with the S/N ratio $<3\sigma$ are eliminated.}
    \label{fig:methanol_rot_diag}
\end{figure*}

\begin{figure*}
    \centering
    \includegraphics[height=9.cm]{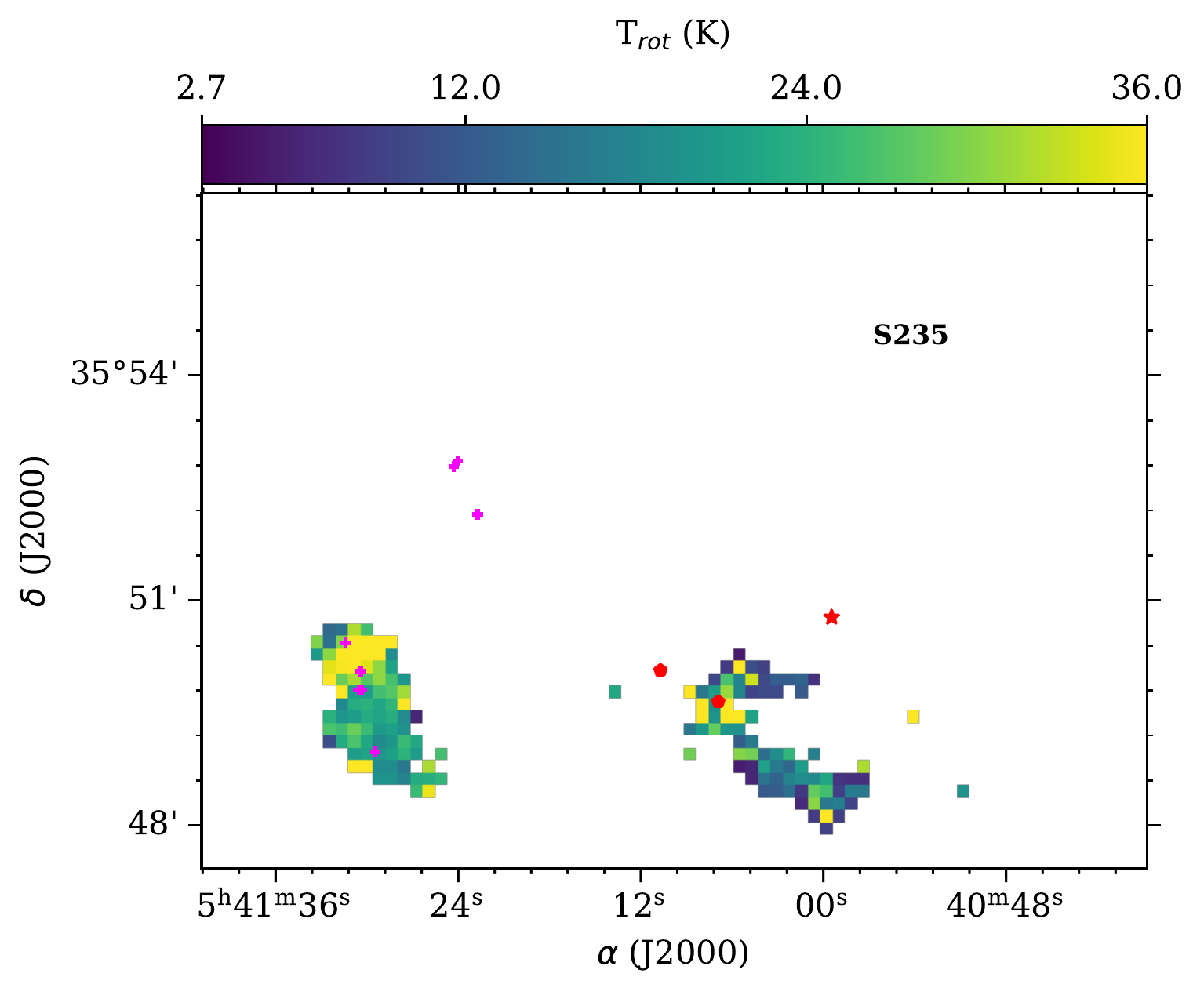}
    \includegraphics[height=9.cm]{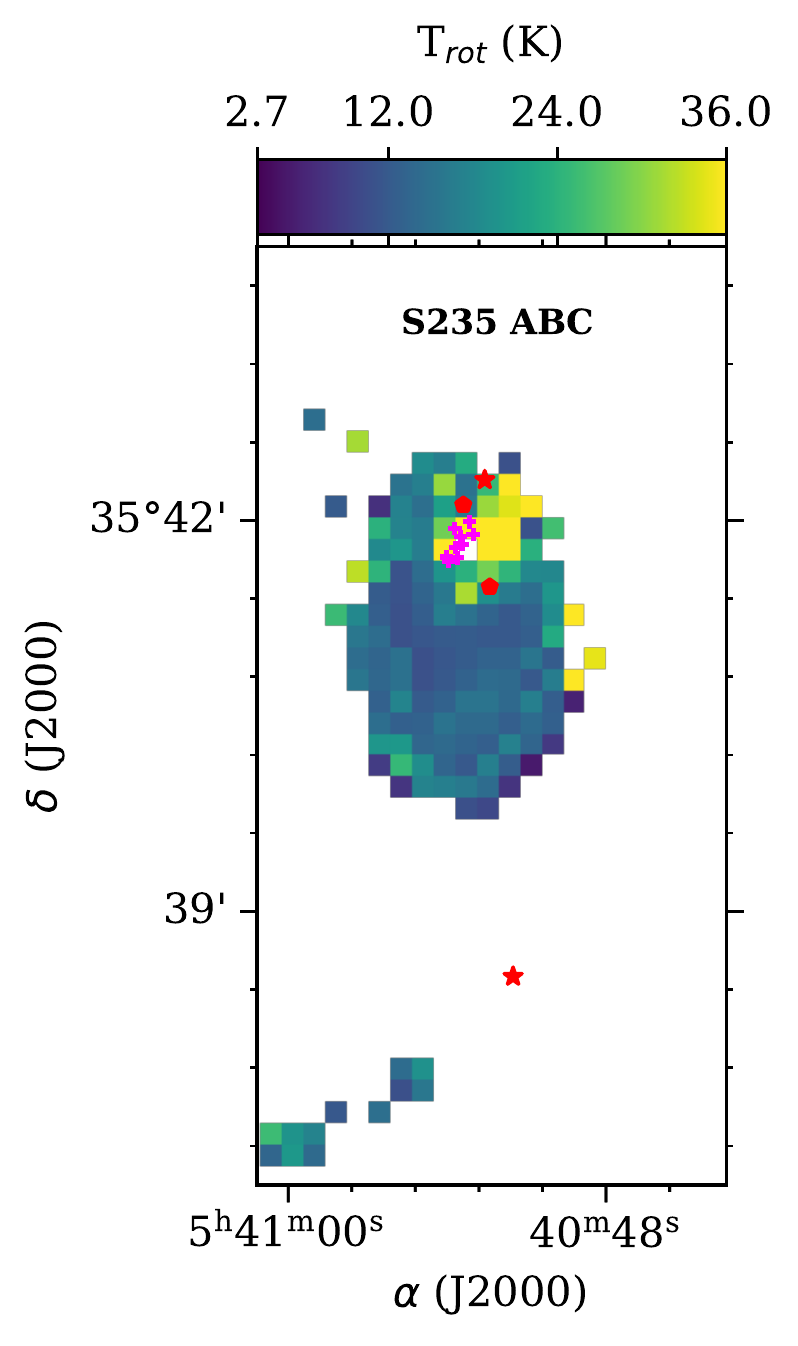}\\
    \includegraphics[height=9.cm]{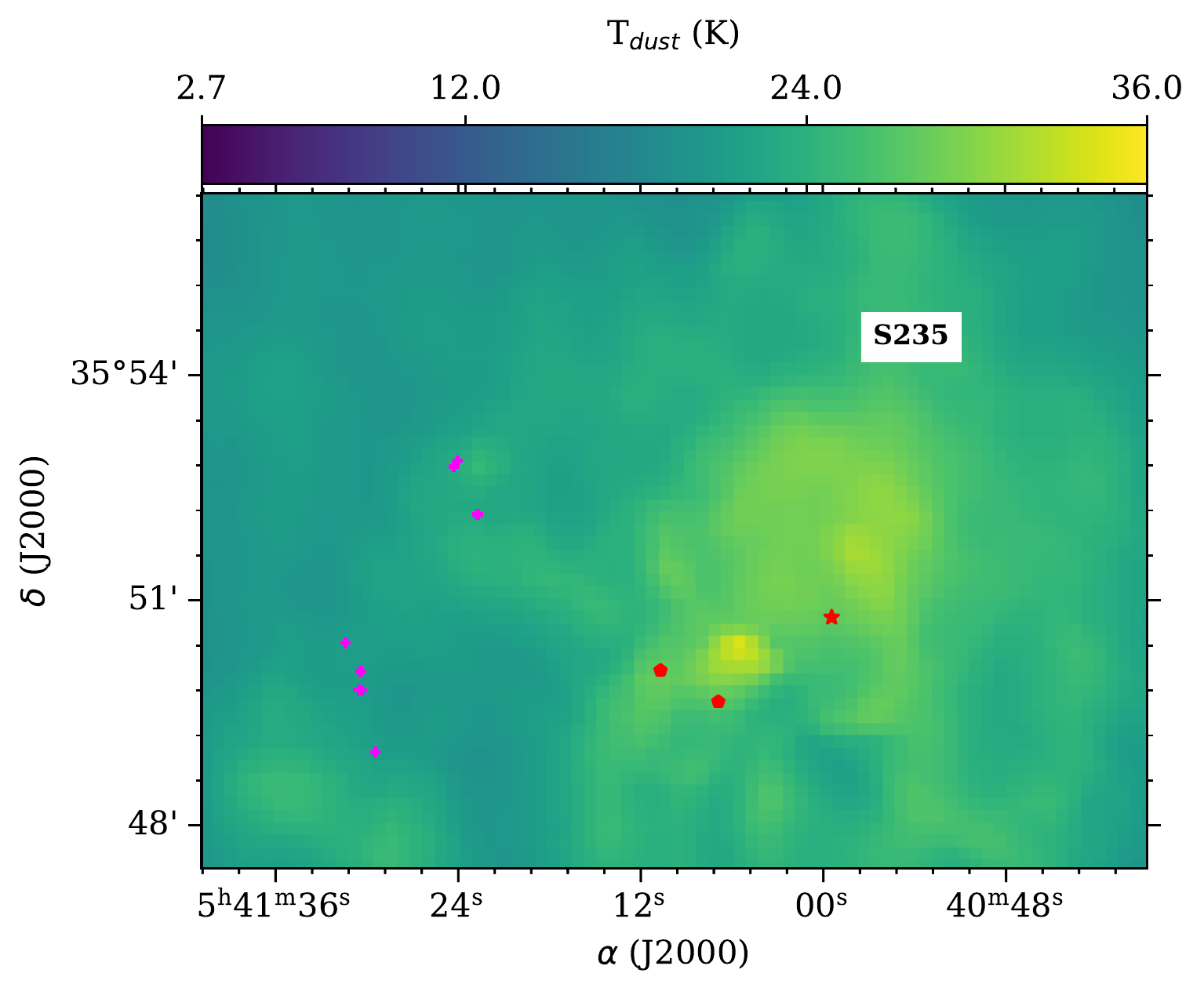}
    \includegraphics[height=9.cm]{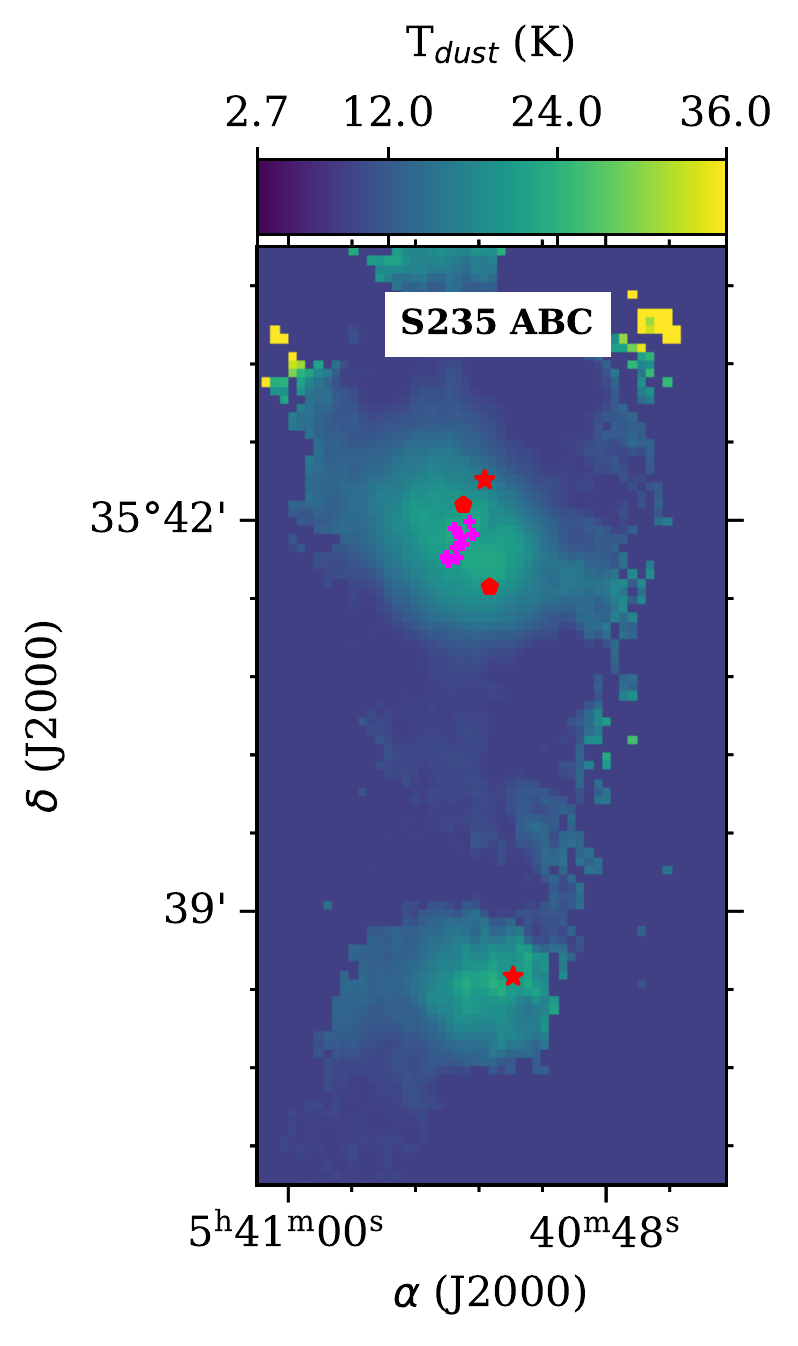}
    \caption{The $T_{\rm rot}$ (top row) and the $T_{\rm dust}$ (bottom row) values. The ionizing sources are shown by the red stars. Red diamonds show three bright infrared sources: IRS1, IRS2 \citep{Evans_1981} and S235~B$^*$ \citep{Boley_2009}. Magenta crosses show IR~sources from \citet{Dewangan_2011}.}
    \label{fig:temperature_compsrison}
\end{figure*}

We present selected rotational diagrams for the observed methanol lines (see  Fig.~\ref{fig:methanol_rot_diag}) and maps of the $T_{\rm rot}$ (determined in the present study) and $T_{\rm dust}$ values \citep[][]{Kirsanova_2020_PDR, Kirsanova_2020} in the observed regions (see  Fig.~\ref{fig:temperature_compsrison}).

\section{Additional spectra}\label{app:addspec}

\begin{figure}
\includegraphics[width=0.8\columnwidth]{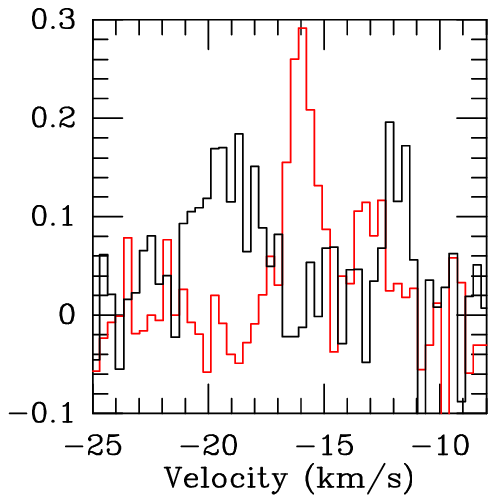}
\caption{Examples of the double-component spectra of the c-C$_3$H$_2$ emission line.}\label{fig:add_spec}
\end{figure}

\section{Additional pv~diagrams}\label{app:pvdiag}

\begin{figure*}
\includegraphics[width=0.99\columnwidth]{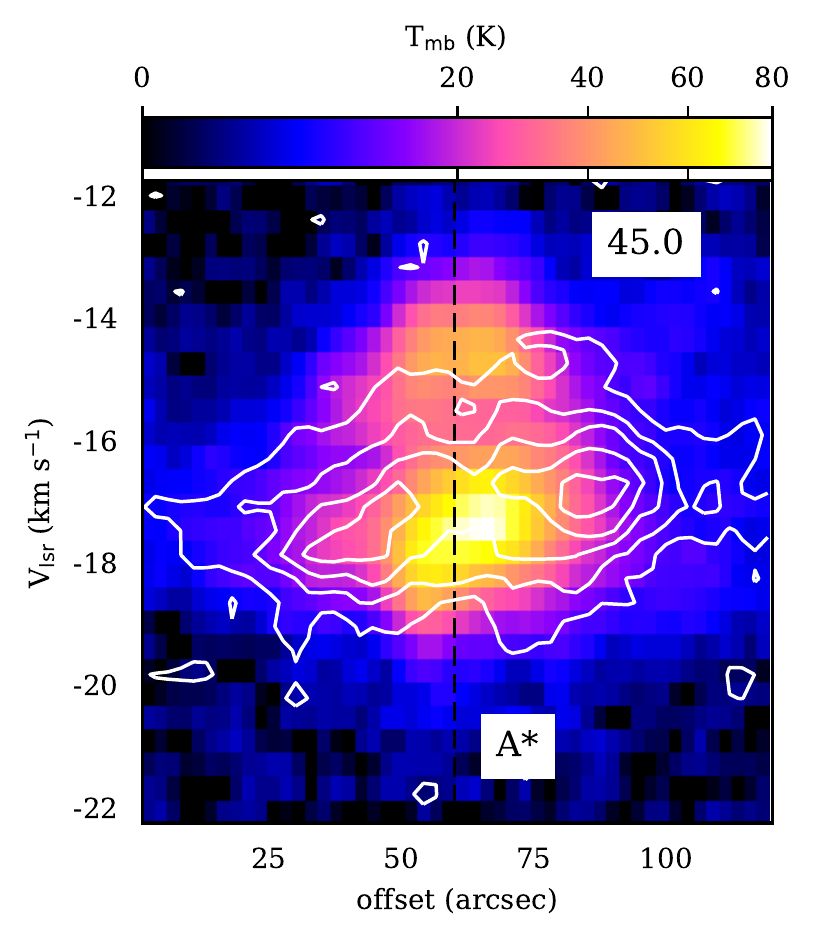}
\includegraphics[width=0.99\columnwidth]{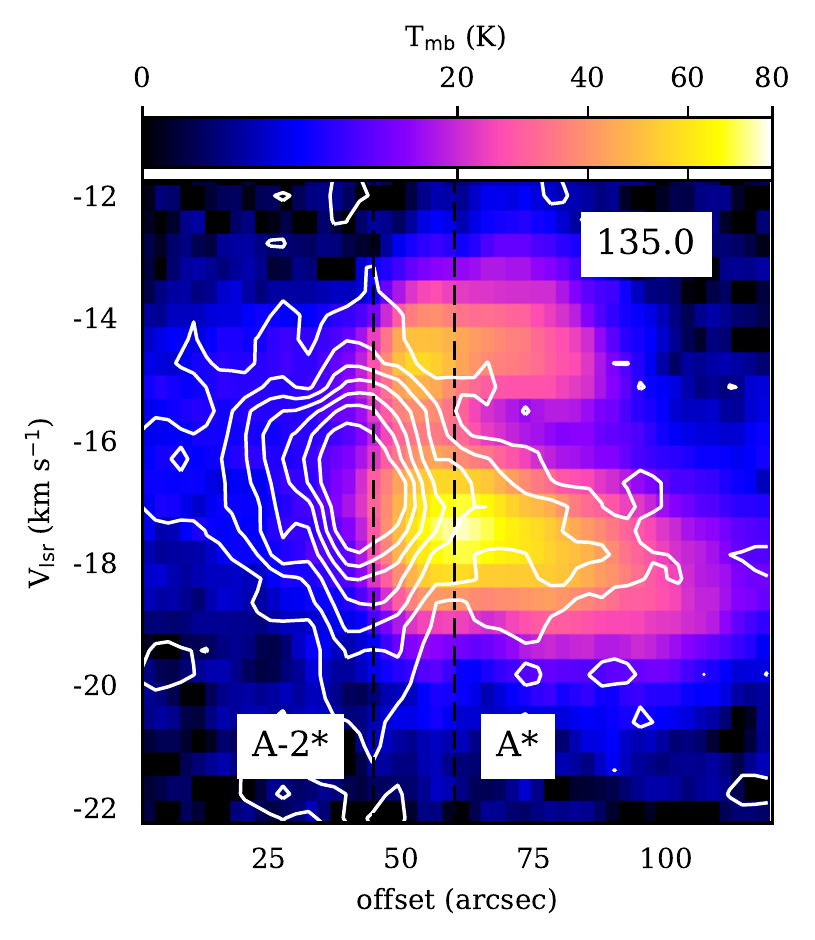}
\caption{Position-velocity diagrams of the \CII{} (colour) and HCO$^+$(3-2) (white contours) emission in S235\,A taken from \citet{Kirsanova_2020_PDR}. Dashed vertical lines correspond to positions of the point sources S235\,A$^*$ and S235\,A-2$^*$.}\label{fig:pvdiag}
\end{figure*}

\bsp	
\label{lastpage}
\end{document}